\documentclass[nohyper,12pt,letterpaper]{JHEP}  
\usepackage{epsfig}

\newfont{\frak}{eufm10 scaled 1200}

\newfont{\Bbb}{msbm10 scaled 1200}     
\newcommand{\mathbb}[1]{\mbox{\Bbb #1}}
\DeclareSymbolFont{AMSa}{U}{msa}{m}{n}
\DeclareSymbolFont{AMSb}{U}{msb}{m}{n}
\let\Box\relax
\DeclareMathSymbol{\Box}{\mathord}{AMSa}{"03}

\def\IR{{\mathbb R}}

\def\IQ{{\mathbb Q}}
\def \eqn#1#2{\begin{equation}#2\label{#1}\end{equation}} 
\def \rut{2/5 transformation}
\def \Rut{{\mathbb G}}             
\def \lp{L_{p}}
\def \ls{L_{s}}

\def\hacek{\accent20}                           

\title{M~Theory and Cosmology }

\author{Tom Banks\\
  Department of Physics and Astronomy\\
  Rutgers University, Piscataway, NJ 08855-0849\\
E-mail: \email{banks@physics.rutgers.edu}}

\abstract{This is a series of lectures on M-theory for cosmologists.
After summarizing some of the main properties of M-theory and its
dualities I show how it can be used to address various fundamental and 
phenomenological issues in cosmology.
}


\received{???????? ?st, 1998}
\accepted{???????? ?th, 1998}

\preprint{\hepth{9911067}\\RUNHETC-99-34}
\begin{document}


\def\mth{M-theory}
\section{Introduction}

This is a series of lectures on superstring/M-theory for 
cosmologists.  It is definitely
{\it not} a technical introduction to \mth\ and almost all technical 
details will be omitted.
A secondary aim of these lectures (or rather the lecture notes -- for 
there will probably
not be many professional string theorists at the actual sessions) is 
to proselytize
for a certain point of view about \mth, which is not the conventional 
wisdom.  A crude statement
of this point of view is that many of the key questions of \mth\ can 
be asked only in the
cosmological context, in particular the central phenomenological
question 
of vacuum selection.
I also believe that some of the fundamental structure of \mth, and 
the relation between
quantum mechanics and spacetime geometry is obscured when one tries 
to study only Poincar\'e
invariant vacuum states of the theory, and ignore cosmological questions.
The 
latter ideas are very speculative however, and I will not discuss them here.

The classic justification of string theorists for studying states 
of \mth\ with $d \geq 4$
Poincar\'e invariance, in a world which is evidently cosmological, is 
that the universe we observe
is locally approaching a Poincar\'e invariant vacuum.  Many of the 
properties of the world
should be well approximated by those of a Poincar\'e invariant state.
It is a philosophy rooted in particle physics, and we shall see that 
it has been quite
successful in \mth\ as well.  One of the key features of such states is 
that they can have
{\it superselection sectors} (a special case of which is the 
phenomenon of spontaneous
breaking of global symmetries).  There can be different 
Poincar\'e invariant states in the
same theory which ``do not communicate with each other'' 
in the following sense:
Certain finite energy excitations of Poincar\'e invariant vacua can be 
classified as
asymptotic states of a number of species of particles.  States with 
any finite number of
particles differ from the vacuum only in a local vicinity of the 
particles' asymptotic 
trajectories (this is more or less the cluster property).   We can 
construct the scattering 
matrix for particle excitations of a given vacuum state and it is 
unitary.  No initial
multiparticle excitation of a given vacuum ever scatters to produce 
excitations of another
\footnote{I have taken pains here not to use arguments from local 
field theory, which can
be only approximate in \mth.}.
In supersymmetric (SUSY) theories it is quite common to have 
superselection sectors that
are not related by a symmetry.  A single theory can produce 
many different kinds of
physics, which means that it does not make definite predictions.

In the early days of string theory, when the vast vacuum degeneracy of 
string perturbation
theory was discovered, it was hoped that nonperturbative effects would 
either lift
the degeneracy or show us that many of the apparent classical ground 
states were 
inconsistent (as {\it e.g.} an $SU(2)$ gauge theory with an odd number 
of isospin one half
fermions is inconsistent).  From the earliest times there were 
arguments that this
was unlikely to be true for those highly supersymmetric ground 
states which least resemble
the real world.  Recent discoveries in string duality and \mth\ make 
it virtually
impossible to believe that these archaic hopes will be realized.

{}From the very beginning I have argued (mostly in private) that 
the resolution of this
degeneracy would only come from the study of 
cosmology\footnote{The earliest conversation
of this type that I remember was with Dan Friedan and took place in 
1986 or 1987.}.  That
is, the physics that determines the correct Poincar\'e invariant vacuum  
took place in
the very early history of the universe.  To understand it one 
will have to understand
initial conditions, and not just stability criteria for possible 
endpoints of cosmological
evolution.  Not too much progress has been made along these lines, 
but there are not
many people thinking about the problem (I myself have probably 
devoted a total of no more
than two years since 1984 to this issue.).  Nonetheless, I hope 
to convince you that it
is a promising area of study.

Associated with the vacuum degeneracy, there are massless excitations.  
I do not have an argument for this which does not depend on an 
effective field 
theory approximation.  In effective field theory, the vacuum degeneracy
is parametrized by the zero modes of a collection of scalar 
fields (which we will call
the moduli fields or simply the moduli\footnote{The term {\it moduli space}
is used by mathematicians to describe multiparameter families of solutions
to some mathematical equations or conditions.  Thus one speaks of 
``the moduli space of Riemann surfaces of genus $g$'' or ``the moduli
space of solutions to the {\bf X} equation''.  Physicists have adopted
this language to describe spaces of degenerate ground states of 
certain supersymmetric theories.  We will be making a further abuse of
the terminology in our discussion of cosmology.}) with no 
potential.  Fields like this spell trouble for phenomenology.  
It is difficult 
to find arguments that they couple significantly more weakly than 
gravity (see however 
\cite{damourpolyakov}), 
and there is no reason for them to couple universally.  Thus, 
they should affect the
orbits of the planets and Eotvos-Dicke experiments.    

On the other hand, we know that SUSY is broken in the real world, 
and then there is no reason
for scalar fields to remain massless.  This however does not 
eliminate all of the problems
and opportunities associated with the moduli.  First of all, one 
can argue that the potential
for the moduli vanishes in many different, phenomenologically 
unacceptable, extreme regions
of moduli space, where supersymmetry is restored.  Examples of such 
regions are weakly coupled SUSY
string compactifications and regions where the world has more than 
four large dimensions.
A quite general argument \cite{ds} shows that one cannot find a stable
minimum of the system by any systematic expansion in the small parameters
which characterize those extreme regions.  Either one must accept the
possibility of different orders in an asymptotic expansion being equally 
important in a region where the expansion parameter is small, or one
is led to expect that the moduli vary with time on cosmological time
scales.  The latter option typically leads to unacceptable time
variation of the constants of nature.
Of course, it might also provide interesting models of the 
fashionable ``quintessence''
\cite{steinhetal}, if these difficulties can be overcome.

Even if one finds a stable minimum for the modular potential 
there are still difficulties.
These are a consequence of additional assumptions about the nature 
of SUSY breaking.
It is usually assumed that SUSY has something to do with the 
solution of the gauge
hierarchy problem of the standard model.  If so, the masses of 
superpartners of quarks
should not be more than a few TeV and one can show that this 
implies that the
fundamental scale of SUSY breaking cannot be larger than 
about $10^{11}$~GeV.
One then finds that the moduli typically have masses and lifetimes 
which are such
that the universe is matter dominated at the time nucleosynthesis 
should have been
occurring.  This is the {\it cosmological moduli problem}.  
There have been several solutions
proposed for it, which are discussed below.  On the other hand, 
there has been
much recent interest in models with very low scales of SUSY breaking.  
These include
gauge mediated models and models with 
TeV scale Planck/String mass
and large extra dimensions\footnote{The latter are often discussed 
without reference to SUSY,
since the hierarchy problem is ``solved'' by the low 
Planck scale.  However,
if they are to be embedded into \mth\ they must have SUSY, broken at 
the TeV scale.}
Here the cosmological moduli problem is more severe, though a recent
paper claims that it can be solved by thermal inflation \cite{rectherm}.  

Another potential problem with moduli was pointed out by \cite{brustein}.
Since the SUSY breaking scale is smaller by orders of magnitude than the 
natural scale of the
vacuum energy during inflation (in most models) one must find an explanation
of the discrepancy.  A favored one has been that the true vacuum 
lies fairly deep 
in an extreme region of moduli space, typically the region of weak 
string coupling.
The universe then begins its history at an energy density many orders 
of magnitude
larger than the barrier which separates the true vacuum from the 
region of extremely
weak coupling where time dependent fundamental parameters and 
unwanted massless particles
destroy any possibility of describing the world we see.  Why doesn't 
it ``overshoot'' the true vacuum and end up in the weak 
coupling regime?  We will discuss a cosmology at the end of these
lectures that resolves this problem.

Not all the news is bad.  One of the things I hope to convince you of 
in these lectures
is that \mth\ moduli are the most natural candidates in the world for 
inflaton fields.
The suggestion that moduli are inflatons was first made in \cite{binetry}.
The word natural is used here more or less in its technical field 
theoretic sense;
that large dimensionless constants in an effective Lagrangian require 
some sort of
dynamical explanation.  For moduli, in order to get $n$ $e$-foldings of 
slow roll
inflation one needs dimensionless parameters of order $1/n$ in 
the Lagrangian.
Another interesting point is that with moduli as inflatons, the scale 
of the vacuum energy 
that is required to explain the amplitude of primordial density 
fluctuations is the same
as the most favored value of the unification scale for couplings and
close to the
scale determining the dimension five operator that gives rise to 
neutrino masses.
These numbers fit best into an \mth\ picture similar in gross detail 
to that first
proposed by Witten \cite{horwitb} in the context of the 
Ho\v rava-Witten \cite{horwita}
description of strongly coupled heterotic strings.
In such scenarios, $10^{16}$~GeV is the fundamental length scale 
and the fields of the
standard model live on a domain wall 
in an eleven dimensional
space with 7 compact dimensions of volume $\sim 10^4$ fundamental
units. 
The four dimensional Planck scale is an artifact of the large volume. 
The SUSY breaking
vacuum energy responsible for inflation must also come from effects 
confined to a
(perhaps different) domain wall.

To summarize, \mth\ has a number of features which require cosmological 
explanations
and a number of potentially interesting implications for cosmologists.  
The two
subdisciplines have very different cultures, but they ought to 
see more of each other.
The plan of these lectures is as follows.  I will first introduce 
the elements of
string duality and \mth, starting from the viewpoint of 11D SUGRA, 
which involves
the smallest number of new concepts for cosmologists.  The key ideas 
will be the
introduction of the basic  half SUSY preserving branes of 11D SUGRA 
and the demonstration of how various string theories arise as 
limits of compactified versions
of the theory.  We will see that the geometry and even the topology 
of space as seen by low energy observers
can change drastically in the course of making smooth changes of parameters.
Another key concept is that of the moduli space of vacua which
preserve a certain amount of SUSY, and the various kinds of 
nonrenormalization 
theorems which allow one to make exact statements about the 
properties of these spaces. 

{}From this we will turn to a discussion of the fundamentals of 
quantum cosmology.
This discussion will be incomplete
since the material is still under development.  We will review the
problem of time in quantum cosmology and a standard resolution of
it based on naive semiclassical quantization of the Wheeler-DeWitt
equation.  We will argue that \mth\ promises to put this argument
on a more reliable basis, and in particular that the peculiar Lorentzian 
metric on the space of fields, which is the basis for the success of
the Wheeler-DeWitt approach to the problem of time, can be derived 
directly from the duality group of \mth\ (at least in those highly
supersymmetric situations where the group is known).  This leads 
naturally into a discussion of whether cosmological singularities can
be mapped to nonsingular situations via duality transformations (it turns
out that some can and some can't and that the distinction between them
defines a natural arrow of time).  We also present a weak anthropic
argument which attempts to answer the question of why the world we see
is not a highly supersymmetric stable vacuum state of \mth.  Finally,
as an amusement for aficionados of heterodoxy, we present some 
suggestions  for M-theoretic resolutions of certain cosmological
conundra without the aid of inflation.  

The remainder of the lectures will be devoted to more or less standard
inflationary models based on moduli and will examine in detail the
properties of these models adumbrated above.  Compared to much of the
inflation literature, these sections will be long on general properties 
and short on specific models which can be compared to the data.  
\mth\ purports to be a fundamental theory of the universe, rather than a
phenomenological model.  Inflaton potentials are objects to be calculated
from first principles rather than postulated in order to fit the data.
There is nothing wrong with phenomenological models of inflation,
but they are not the real province of \mth\ cosmology. Unfortunately,
our understanding of the nonperturbative properties of \mth\ in the regime
where the supersymmetry algebra is sufficiently small to allow 
for a potential on the moduli space (the alert reader may have already
noted that the preceding phrase contains an oxymoron) is too limited
to allow us to build reliable models of the potential.  Thus, if we
are honest, we must content ourselves with general observations 
and the definition of a set of goals.

\section{\mth, Branes, Moduli and All That}

\subsection{The story of M}

Once upon a time there were six string theories.  Well, actually 
there were five 
(because one, the Type IA theory,
was an ugly duckling without enough Lorentz invariance) and actually there 
were an infinite number, or rather continuous families $\ldots$.   
What's going on here?
The basic point is the following: what string theorists called a string 
theory in the
old days was a set of rules for doing perturbation theory.  What was 
perhaps misleading
to many people is that these rules were usually given in terms of a 
Lagrangian, more generally
a superconformally invariant $1+1$ dimensional quantum field theory, 
(with some extra 
properties).  We are used to think of Lagrangians as defining theories.  
The better way to think of the world sheet Lagrangians of string theory 
is by imagining
a quantum field theory with many classical vacua.  Around each vacuum 
state we can construct
a loop expansion.  The quadratic terms in the expansion around a 
vacuum state define 
a bunch of differential operators, whose Green's functions are the 
building blocks 
of the perturbation expansion.  Using Schwinger's proper time 
techniques 
we can describe these Green's functions in terms of an auxiliary 
quantum mechanics, and if
we wish we can describe this quantum mechanics in terms of a Feynman 
path integral with
a Lagrangian.  The world sheet path integrals of string theory are the 
analogs of these
proper time path integrals.  One of the beautiful properties of string 
theory is
that, unlike field theory, the Lagrangian formulation of the propagator 
completely determines
the perturbation expansion.  To compute an $n$ particle scattering 
amplitude in tree level string
theory one does the path integral on a Riemann surface with no handles 
and (for theories
whose perturbation expansion contains only closed strings) 
$n$ boundaries.  The boundary
conditions on the boundaries are required to be superconformally 
invariant and carry fixed
spacetime momentum.  The Lagrangian itself is superconformally 
invariant, and the allowed
boundary conditions are generated by acting on a particular 
boundary condition which defines 
the ground state of the single string with a set of {\it vertex operators} 
which represent small
perturbations of the action which preserve superconformal 
invariance\footnote{Actually the 
situation is a bit more complicated, and to do it justice one must use 
the BRST formalism.
For our purposes we can ignore this technicality.}.  A given vertex 
operator creates 
a state of the string which propagates as a particle with given 
mass and quantum numbers.
The higher orders in perturbation theory just correspond to computing 
the same path integral
on Riemann surfaces of higher genus.  One sums over all Riemann 
surfaces, or in some cases
only over oriented ones.

The conditions of superconformal invariance have many solutions.  
Classically (in the sense of
two dimensional classical field theory -- this should not be confused with
tree level string theory which corresponds to summing all orders in the 
semiclassical expansion 
of the world sheet field theory, on Riemann surfaces with no handles), 
for the particular
case of Type II string theories, the bosonic terms in the most general 
superconformal
Lagrangian have the form
\eqn{classsucon}{{\cal L} = (G_{\mu\nu}(x) + i B_{\mu\nu}(x))
\partial x^{\mu} \bar{\partial} x^{\nu} + h.c. + \Phi (x) \chi }
where the derivatives are taken with respect to complex coordinates on 
a Euclidean world sheet.
$G_{\mu\nu}$ is symmetric and $B_{\mu\nu}$ is antisymmetric.  
$\chi$ is the Euler density
of the world sheet, a closed two form (for more on forms, closed and 
otherwise, see below) whose integral is the Euler Character.
Quantum mechanically, there are restrictions on the functions, 
$G,B$, and $\Phi$.
To lowest order in the world sheet loop expansion the condition of 
superconformal invariance
coincides with the equations of motion coming from a spacetime 
Lagrangian
\eqn{stlag}{{\cal L}_{st} = \sqrt{-g}e^{-2\Phi}
[R + 4 (\nabla\Phi)^2 + (d B)^2]}
This fact, combined with the fact that vertex operators are allowed 
perturbations of these
equations, shows us that string theory is a theory of gravity.  
One cannot choose the background metric arbitrarily; 
it must satisfy an equation of motion.

In classifying consistent solutions of all these rules, string 
theorists found a number of
discrete choices depending on the number of fermionic generators 
in the world sheet 
superconformal group and on the types of Riemann surfaces allowed.  
This led to
the five different types of string theory.
Once these discrete
choices were made, there still seemed to be a multiparameter 
infinity of choices.
However, it was soon understood that the continuous infinity 
corresponded to expanding the
same basic theory around different solutions of its classical equations
of motion (approximately
the equations generated by (\ref{stlag})).  This was a little surprising, 
because one
of the rules for the perturbation expansion was that there be 
some number of flat
Poincar\'e invariant dimensions (I explained in the introduction 
why string theorists insisted
on this).  So each of these solutions was a static classical vacuum state.
Why are there so many vacua?  The answer is spacetime SUSY.  
Indeed almost every known
perturbatively stable vacuum state of string theory is 
supersymmetric\footnote{There are
no known exceptions.  Recently however \cite{evashamit} some 
constructions which appear to
be stable at least through two loops have been found.  This is 
the reason for the
word almost in the text.}.   It is well known that spacetime 
SUSY often leads to
nonrenormalization theorems
which prevent the existence of potentials for scalar fields.  
The strongest theorems of
this type come when there is enough SUSY to guarantee that the scalars are in
supermultiplets with gauge or gravitational fields, but there are 
other examples.
As noted in the introduction, we will call the space of classical 
vacua the moduli space.
It should be noted that the moduli space is not connected 
({\it e.g.} the branches with
different amounts of SUSY are generally disconnected from each other), 
nor is it a manifold.
The reason for the latter property is that often,new massless states 
can appear 
on submanifolds of the moduli space.  Often these include scalars, 
which, as long as the original
moduli are restricted to the submanifold, have no potential.
One can then define a new branch of moduli space on which the original
moduli are restricted to the submanifold, but the new massless scalar
fields have expectation values.  
Thus the moduli space has 
several disconnected components, each of which is a bunch of 
manifolds of different
dimension, glued together along singular submanifolds.

Thus, circa 1994-95 we had five discrete classes of string theory, 
Type $II_{A,B}$ ; ${\rm Heterotic}_{A,B}$
(A refers to the $E_8 \times E_8$ heterotic string theory and B to 
the $SO(32)$ version)
and Type $I_B$.  The Type $I_B$
theory has the same symmetries in spacetime
as the $Het_B$ theory although the world sheet theories are completely
different.  In Type $I_B$ the gauge quantum numbers are carried on the
ends of open strings (like flavor quantum numbers on QCD strings) 
and nonorientable world sheets appear in the perturbation expansion.
The heterotic theory has only closed strings, orientable world sheets
and gauge quantum numbers carried by the body of the string.
There is also a Type $I_A$ theory which has a similar relation to 
$Het_A$.  This theory does not have ten dimensional Lorentz invariance
because it has two $8+1$ dimensional domain walls at the ends of a finite
or infinite $9+1$ dimensional ``interval''.  It has $SO(16) 
\times SO(16)$ gauge symmetry carried by the ends of open strings
which can only propagate on the domain walls. 

The labels A and B refer to theories which are different in 10 (the
maximal dimension for perturbative strings) dimensions but are actually
equivalent to each other when compactified on a circle.  The equivalence
is due to a stringy symmetry called T duality.  The momentum of
a string on a circle is the integral of the time derivative of its 
coordinate;
{\it i.e.} the time derivative of the center of mass position.
\eqn{mom}{P = \int d\sigma \partial_t \theta.}
Strings on a circle carry another quantum number called winding number,
which is defined by
\eqn{wind}{w = \int d\sigma \partial_{\sigma} \theta.}
The Euclidean world sheet Lagrangian for the string coordinate $\theta$ is
\eqn{wslagcirc}{{\cal L}_{ws} = (\partial_t \theta)^2 + 
(\partial_{\sigma} \theta)^2}
Instead of $\theta$ we can introduce a new coordinate by the two 
dimensional analog of an electromagnetic duality transformation
\eqn{dual}{\partial_a \theta = \epsilon_{ab} \partial_b \bar{\theta}.}
It turns out that when one performs this transformation one automatically
takes a Type A theory to a Type B theory.

We learn two things from this:
First, there are only half as many different string theories as we 
thought, and second, to see that theories are the same we may have
to compactify them.  Decompactification loses important degrees of
freedom (in this case string winding modes, which go off to infinite
energy) which are necessary to see the equivalence.  
There are no more such equivalences which can be seen in perturbation
theory, but we might begin to suspect that there are further equivalences
which might only appear nonperturbatively.  How can we hope to realize
this possibility in a theory which is formulated only as a perturbation
series?  

The key to answering this question is the notion of SUSY preserving
or BPS states.  To explain what these are, let me introduce
the SUSY algebra
\eqn{susalg}{\{ \bar{Q}_a, Q_b \} = \gamma^{\mu}_{ab} P_{\mu}}
Actually, this is only the simplest SUSY algebra one can have in
a given dimension.  We will see more complicated ones in a
moment.  If we look at particle representations of the SUSY algebra,
then $P_{\mu}$ is either a timelike or a lightlike vector.
In the timelike case, the matrix on the right hand side of (\ref{susalg})
is nondegenerate, while in the lightlike case it is 
degenerate -- half of the states in the representation are annihilated by 
it.  This means that in the lightlike case half of the SUSY generators
annihilate every state in the representation.  Thus massless
supermultiplets are smaller than massive ones.  This means, that
in general in a supersymmetric theory, small changes in the parameters
will not give mass to massless particles.  In order to do so
one must have a number of massless multiplets which fit together to
form a larger massive multiplet (the super Higgs mechanism) in order
for states to be lifted.  If this is not the case for some values
of the parameters, then small changes cannot make it so and the
massless particles remain.  Of course, we did not really need SUSY
to come to this conclusion for massless particles of high enough
spin.  In that case it is already true that the Lorentz group 
representations of massless and massive particles are of different
multiplicity.

The new feature really comes if we compactify the theory in a way
which preserves all the SUSY generators.  This can be done by
compactifying on a torus with appropriate boundary conditions.
The SUSY algebra remains the same, but now some of the components
of the momentum are discrete.  Also, the Lorentz group is broken
to the Lorentz group of the noncompact dimensions, so the spinor
representation breaks up into some number of copies of the
lower dimensional spinor.  The algebra now looks like
\eqn{susalgcomp}{{\{ \bar{Q}_a^i, Q_b^j \} = \gamma^{\mu}_{ab} P_{\mu} 
\delta^{ij} + \delta_{ab} Z^{ij}.}}
 Spinor and vector indices now run over their lower dimensional
values, and $i,j$ label the different copies of the lower dimensional
spinor in the higher dimensional one.  The generators $Z^{ij}$
are scalars under the lower dimension Lorentz group.  They are 
combinations of the toroidal momenta and are examples of what are called
central charges.  

Now consider a state carrying nonzero values of the central charge
in such a way that the higher dimensional momentum is lightlike.
It represents a massive Kaluza-Klein mode of the massless particle
in the higher dimension.
In a nonsupersymmetric theory the masses of
Kaluza Klein modes of higher dimensional massless fields are
renormalized by quantum corrections.  But in a theory with the
extended SUSY algebra (\ref{susalgcomp}) we can ask whether 
the representation is annihilated by half the supercharges (other
fractions are possible as well).  If it is, we get a computation of
the particle's nonzero mass in terms of its central charges.
{\it This mass cannot change as parameters of the theory are varied
in such a way as to preserve extended SUSY, 
or rather the formula for its variation with parameters may be
read directly from the SUSY algebra.}  We will see below that it
is possible to realize the central charges $Z^{ij}$ of  
the extended lower dimensional SUSY algebra in other ways.  
Instead of representing KK momenta in a toroidal compactification, 
they might arise as winding numbers of extended objects, called branes,
around the compact manifold.  The italicized conclusion will be valid for 
these states as well. 

The argument for the statement in italics above, is again based
on the smaller dimension of the representation.
To see it more explicitly, work in the frame where the spatial momentum
is zero, and take the expectation value of the anticommutator in states
of a single particle with mass $M$ (actually we mean a whole SUSY multiplet
of particles).  Then (\ref{susalgcomp}) reads
\eqn{rest}{M \delta_{ij} + \gamma^0 Z^{ij} = \ < [Q_a^i, Q_b^j ]_+ >\\ \geq 0.}
The last inequality follows because we are taking the expectation 
value of a positive operator.  It says that the mass is bounded
from below by the square root of the sum of the squares of the
eigenvalues of the matrix $Z$, which are also called the 
central charges. 

Equality is achieved only when the expectation value vanishes, which, since
the SUSY charges are Hermitian, means that some of the charges annihilate
every state in the representation.  These special representations of the
algebra have smaller dimension and cannot change into a generic 
representation, which satisfies the strict inequality, as parameters are
continuously varied.

Thus, in theories
with extended SUSY, certain masses can be calculated exactly from the
SUSY algebra.  These special states are called Bogolmony Prasad
Sommerfield or BPS states, since these authors first encountered this
phenomenon in their classical 
studies of solitons \cite{bps}.  The connection
to SUSY, which makes the classical calculations into exact
quantum statements, was noticed by Olive and Witten \cite{ow}.

Notice that although we motivated this argument in terms of Kaluza-Klein
states, it depends mathematically only on the structure of
the extended SUSY algebra.  Thus if we can obtain this algebra in another
way, we will still have BPS states.  An alternative origin for
central charges and
BPS states comes from ``wrapped branes'' of a higher
dimensional theory.

To understand this notion, note that, strictly from the point of
view of Lorentz invariance, the SUSY algebra could contain terms like
\eqn{susalgbrane}{\{ Q_a, Q_b \} = \gamma^{\mu}_{ab} P_{\mu} + 
\gamma^{\mu_1 \ldots \mu_p}_{ab} Z_{\mu_1 \ldots \mu_p}}
The multiple indices are antisymmetrized.
The famous Haag-Lopusanski-Sohnius \cite{hlscm} generalization
of the Coleman Mandula theorem, tells us that this pth rank
antisymmetric tensor charge, must vanish on all finite energy
particle states.  On the other hand, the purely spatial components
of it have precisely the right Lorentz properties to count the
number of infinite energy $p$-branes, or $p$-dimensional domain walls,
oriented in a given hyperplane.  We will have more to say about
these brane charges when we talk about branes and gauge theories
below.

Now suppose we have compactified $p$ or more dimensions, and the resulting
compact space has a topologically nontrivial $p$-dimensional
submanifold, or $p$-cycle.  To see what we mean, consider the two torus

\epsfig{file=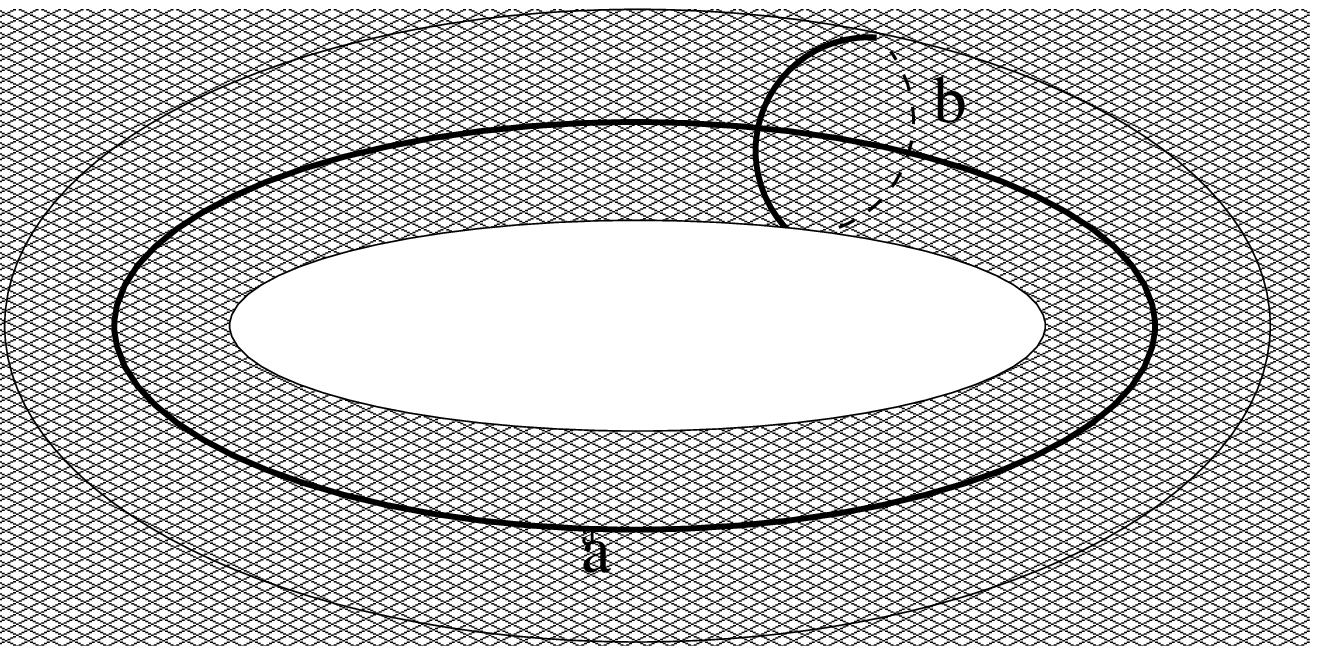}

\centerline{{\bf Figure 1:} A Two Torus With Nontrivial Cycles Labelled.}
\vskip.1in

\noindent
It has two different kinds of nontrivial 1 cycle, labelled a and b
in the figure.  The whole torus is a nontrivial two cycle.  The
word nontrivial cycle or just cycle implies that the submanifold cannot
be contracted to a point 
``because it wraps around a hole in the manifold''.
If we wrap a $p$-brane around the $p$-cycle, we get a finite energy
particle state.  The tensor charge with all indices pointing
in the compact directions is a scalar charge in the remaining noncompact
directions and is allowed to appear as a central charge in an extended
SUSY algebra.  Often, the corresponding particles have the BPS property.

With this background, we can get on with the story of M.  Practitioners
of string duality realized that BPS states gave them a powerful handle
on nonperturbative physics.  For example, consider a weakly coupled
string theory and a solitonic state, whose mass in string units is
proportional to $g_S^{-r}$\footnote{In string theory both $r=1,2$ are 
realized. $r=2$ corresponds to a conventional soliton, arising as 
a solution of the classical equations of motion.  $r=1$ corresponds
to Dirichlet brane or D brane states.}. If 
it is a BPS state then as the coupling becomes infinitely
strong, it becomes infinitely lighter than the string scale (if not
for the BPS property, we could not trust the weak coupling formula
at strong coupling).  In all the cases which have been studied
one can, by thinking about the lightest BPS states in the strong
coupling limit, realize that they are just the elementary states
of another weakly coupled theory.  In most cases, this is another
string theory, but there is a famous exception.

If one considers Type IIA string theory in ten dimensions, it
contains a single $U(1)$ gauge field.  None of the perturbative
states are charged under this field; they have only magnetic
moment couplings.  If one considers hypothetical BPS states
charged under this $U(1)$, then it is easy to show that their
spectrum in the strong coupling limit is precisely that of the
supergravity multiplet in 11 dimensions.  Thus one is led to
conjecture \cite{witvar} that the strong coupling limit of Type IIA
string theory has eleven flat dimensions and a low energy limit
described by SUGRA.

None of this was much of a surprise to the SUGRAistas \cite{dht}.  It had long been known that the low energy limit of
IIA string theory was a ten dimensional SUGRA theory which was the
dimensional reduction of 11D SUGRA, with the string coupling appearing
as the ratio of the three halfs power of the radius of the reducing
circle to the eleven dimensional Planck mass.  The SUGRAistas
even had a correct explanation of where the strings come from.
As we will see 11D SUGRA couples naturally to a membrane, the M2 brane.
If we wrap one leg of the M2 brane around the circle whose radius
is being shrunk to zero we get a string whose tension is going to
zero in Planck units.

String theorists get a C for closed mindedness for
ignoring the message of the SUGRAistas for so long.
Behind their resistance lay the feeling that
because both 11D SUGRA and the world volume theory
of membranes are nonrenormalizable, one could not
trust conclusions drawn on the basis of these
theories.  It was only with the advent of 
an unambiguous, 
string theoretic construction of the KK gravitons
of 11D SUGRA as bound states of D0 branes 
\cite{jopo} \cite{witbound} that the last bastions
of resistance fell.  What one should have realized
{}from the beginning was that conclusions about BPS
states, based as they are only on the symmetry
structure of the theory, can be extrapolated from
effective theories far beyond the limited range of
validity these low energy approximations.  

The picture as we understand it today\footnote{or 
rather a cartoon of it, for moduli space is much
more complicated than a two dimensional deerskin.}
is illustrated by the famous ``modular deerskin''.

\epsfig{file=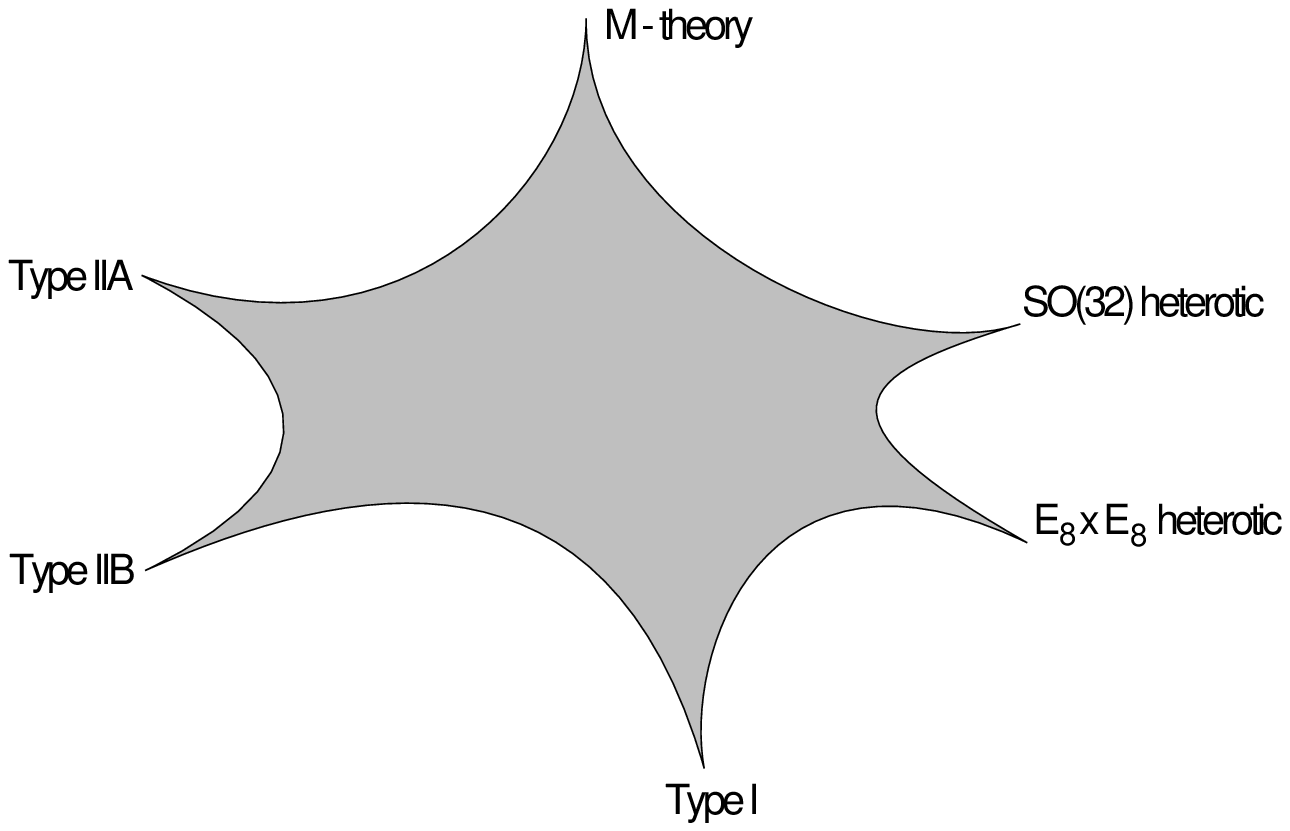}

\centerline{{\bf Figure 2:} A Cartoon of the Moduli Space of \mth.}
\vskip.1in

\noindent
There is a single theory, which we now call
\mth\footnote{or at least some of us do.  Some 
people reserve the name \mth\ for the region of moduli
space where 11D SUGRA is a good approximation.  I
consider that a waste of a good name since we can 
call this the 11D SUGRA region.} which has a large
moduli space.  All of the known perturbative
string theories, and 11D SUGRA are limits of this
theory in certain extreme regions, or boundaries,
of moduli space (the cusps in the picture).
There is another class of limits, called F-theory,
which are not amenable to complete
analysis, but about which many nontrivial statements
can be made.   An example of an F-theory region
is the strongly coupled heterotic string compactified
on a two torus.  

One of the lessons of duality is that no one of
these regions is {\it a priori better} than any
other.  Each of them tells a partial story about
\mth, and we learn a lot by trying to patch these
stories together.  However, the 11D SUGRA limit
has a distinct advantage when one is trying to
explain \mth\ to non-string theorists, particularly
if they have a good background in GR.  In this
limit, most of the arguments are completely 
geometrical and can be understood on the basis of
classical field theory and the classical Lagrangians
for various extended objects.   For this reason,
I will begin in the next section with a discussion
of the 11D SUGRA Lagrangian\footnote{at least its
purely bosonic part.  Fermions, like virtue in the
world of politics, are entities often talked
about, but rarely seen, in discussions of SUSY
theories.}.

\section{Eleven Dimensional Supergravity}

In eleven dimensions, the graviton has $44$
spin states transforming in the symmetric
traceless tensor representation of the transverse
(transverse to the graviton's lightlike momentum)
$SO(9)$ rotation group.  The gravitino is a tensor-
spinor of this group, satisfying the constraint
$\gamma^i_{ab} \psi^i_b = 0$ which leaves $128$
components.  The remaining $84 = {9\times 8 \times 
7 \over 3\times 2\times 1}$ bosonic states in the SUGRA
multiplet transform as a third rank antisymmetric
tensor. 

The covariant Lagrangian for the bosonic fields
in the multiplet is
\eqn{suglag}{M_P^9 \{ \sqrt{-g} [- {1\over 2} R + 
 - {1\over 48} g^{\mu_1 \nu_1} \ldots
g^{\mu_4 \nu_4} G_{\mu_1 \ldots \mu_4} 
G_{\nu_1 \ldots \nu_4} ] - {\sqrt{2}\over 3456}\epsilon^{\mu_1 \ldots
 \mu_{11}} C_{\mu_1 \ldots \mu_3} G_{\mu_4 \ldots 
\mu_7} G_{\mu_8 \ldots \mu_{11}} \}}
The supersymmetry transformation of the gravitino is
\eqn{sutrans}{\delta \psi^{\mu}_{\alpha} = 
D_{\mu} \epsilon_{\alpha} + {\sqrt{2}\over
288}(\Gamma_{\mu}^{\nu\lambda\kappa\sigma} -
\delta_{\mu}^{\nu}\Gamma^{\lambda\kappa\sigma})G_{\nu\lambda\kappa\sigma}
\epsilon_{\alpha}.}

The existence of a three form gauge potential, suggests that
the theory may contain a membrane, which couples to the
three form via
\eqn{membcoup}{Q_2 \int C_{\mu\nu\lambda}{\partial x^{\mu} 
\over \partial \xi^a}{\partial x^{\nu} 
\over \partial \xi^b}{\partial x^{\lambda} 
\over \partial \xi^c} d\xi^a d\xi^b d\xi^c}
where the $\xi^a$ are coordinates on the membrane world volume.
The dual of the four form field strength $G_{\mu\nu\lambda\kappa}$
is a seven form $G_7$, whose source, defined by $d*G_7 (\equiv * 
dG_4) = J_6$\footnote{The large number of spacetime dimensions
lead me to resort to differential form notation even for an 
audience of cosmologists.}, is a six form current.  This is a current of
five dimensional objects, which we will call M5 branes.

The low energy SUGRA approximation to \mth\ gives us evidence
that both M2 and M5 branes exist, since there are soliton 
solutions of the SUGRA equations of motion with the requisite
properties.  This would not be a terribly convincing argument,
since the scale of variation of the fields of these objects is
(what else?) the eleven dimensional Planck scale, and SUGRA is 
only an effective field theory.  However, these solitons have
the BPS property.  That is, we can find them by insisting that
half of the gravitino SUSY variations vanish.  This leads to first 
order equations, which are much easier to solve than the full second
order equations, but give a subclass of solutions of the latter. 
Since these solutions are constructed so that half of the SUSY variations
vanish, their Poisson brackets with half the SUSY generators (in a canonical
formulation of 11D SUGRA) vanish.  This is the classical approximation to
the statement that the quantum states represented by these soliton solutions
are annihilated by half of the SUSY generators.  

The solutions are
\eqn{M2}{ds^2 = (1 + {k\over r^6})^{-2/3} (-dt^2 + d\sigma^2 + d\rho^2)
+ (1 + {k\over r^6})^{1/3} d{\bf x}_8^2 }
\eqn{M22}{A_{\mu\nu\lambda} = \epsilon_{\mu\nu\lambda} (1 + {k\over
r^6})}
\eqn{M23}{k \equiv {L_P^9 \ T_2 \over 3\ \Omega_7}}
\eqn{M5}{ds^2 = (1 + {T_5\over r^3})^{-1/3} (-dt^2 + d{\bf x}_5^2)
+ (1 + {T_5\over r^3})^{2/3} d{\bf y}_5^2 }
\eqn{M52}{G_{\mu_1 \ldots \mu_4} = 3\ T_5\ \epsilon_{\mu_1 \ldots \mu_4
\nu}{y^{\nu} \over r^5}}
for the two brane and five brane respectively. 

In each of these equations, $r$ denotes the transverse distance from the
brane.
These solutions contain arbitrary parameters $T_2$ and $T_5$
which control the strength of the coupling of these objects
to the three form gauge potential and to gravity.  However, an elementary 
argument leads to a determination of these parameters.
Compactify the theory on a seven torus and wrap the M2 brane 
around two of the dimensions of this torus and the M5 brane 
around the other five.  The integral of the three form over
the two torus on which the membrane is wrapped give us an 
ordinary Maxwell (1 form) potential.  It is easy to see that the
wrapped membrane is a charged particle with charge $Q_2$
with respect to this Maxwell field (the membrane coupling of
(\ref{membcoup}) dimensionally reduces to the standard Maxwelll coupling
to a charged particle).  It is a little harder to
see that the wrapped M5 brane is a magnetic monopole for
this field.  Thus, the Dirac quantization condition implies
\eqn{dquan}{2\pi Q_2 Q_5 \in {\bf Z}}
This is a quick and dirty proof of the Nepomechie-Teitelboim
\cite{claudio} generalization of the Dirac quantization condition
to $p$-form gauge fields.  This condition determines the tension of the
minimally charged M2 and M5 branes to be:
\eqn{T1}{T_5 = {1\over 2\pi} T_2^2}
\eqn{T2}{T_2 = L_P^{-3}}

Given the existence of these infinite flat branes, we can also
study small fluctuations of them which describe (slightly)
curved branes moving in spacetime.  The most useful way to do this
is to introduce a world volume field theory which contains
the relevant fluctuations.  Among the variables of such a theory
should be a set of scalar fields which describe small fluctuations
of the brane in directions transverse to itself.
It turns out that these world volume theories are, in the present
case, completely determined by SUSY.  

Let us begin with the M2 brane.  
The SUSYs that preserve the brane satisfy 
$\gamma^3 \ldots \gamma^{10}\ Q = Q $\footnote{To see this,
note that the condition specifying which charges annihilate the membrane 
state must be linear (the sum of two such charges is another)
and invariant under both the transverse rotation and world volume
Lorentz groups.  This is the only such condition since the product of 
all the 11D Dirac matrices is 1.}.  There are $16$ solutions
to this equation, which transform as $8$ spinors under the 
$SO(2,1)$ Lorentz group of the brane world volume. 
Each two component world volume spinor transforms in the
eight dimensional spinor representation of the transverse 
$SO(8)$ rotation group. From the point of view of the world
volume field theory, the latter is an internal symmetry.
We expect
the world volume theory to contain $8$ scalar fields,
 representing the transverse fluctuations of the membrane.
A SUSY Lagrangian containing these fields is given by
\eqn{M2wv}{{\cal L}_{M2} = \partial_a x^i \partial_a x^i + 
\bar{\theta^J} \Gamma^a \partial_a \theta^J}
where $\Gamma^a$ are three world volume Dirac matrices.
The SUSY generators are:
\eqn{M2sugen}{Q_{\alpha}^J = \int d^2 \xi \gamma^i_{JK} 
[\partial_0 x^i  \theta_{\alpha}^K + \Gamma^a_{\alpha\beta} 
\theta_{\beta}^K \partial_a x^i ] }
Using the canonical commutation relations for the world volume fields
it is straightforward to verify that these satisfy the SUSY algebra.  
Here we have used $\gamma^i$ to represent the eight dimensional
Dirac matrices, despite the possibility of confusion with the
eleven dimensional matrices of the paragraphs above and below.\footnote{We 
have also passed over in silence the two different
types of
eight dimensional spinor which appear in these equations.
Experts will understand and amateurs would only be confused by
this detail.}

The world volume theory of the M5 brane is more interesting.
The SUSYs preserved satisfy $\gamma^6 \ldots \gamma^{10}\ Q = Q$
(using the same argument as in the footnote above).
The world volume Lorentz group $SO(5,1)$ has two different 
chiralities of spinor representation, and (ultimately because
the product of all eleven dimensional Dirac matrices is 1) this
condition says that all the SUSY generators have the same
chirality.  There are sixteen real solutions of these constraints
which can be arranged as two complex ${\bf 4}$ representations
of the world volume Lorentz group.  Under the transverse $SO(5)$
rotation group, they transform as four copies of the fundamental
pseudoreal spinor.  This kind of SUSY is called $(2,0)$ SUSY
in six dimensions.

The coordinates of transverse fluctuations are five scalar world
volume fields, which transform as the vector of $SO(5)$.
There is a unique SUSY representation which includes these fields.
Their superpartners are two fermions in the $\bar{\bf 4}$ 
representation of the Lorentz group, and a second rank 
antisymmetric tensor gauge field, $B_{\mu\nu}$ whose field
strength satisfies the self duality condition
$H_{\mu\nu\lambda} = \epsilon_{\mu\nu\lambda\kappa\sigma\rho}
H_{\kappa\sigma\rho}$.  Indeed, in a physical light cone
gauge, $B_{AB}$ (the A,B indices indicate the four 
transverse dimensions in the
lightcone frame inside the world volume) has $6$ components and
the self duality cuts this down to $3$.  Combined with the five
scalars this makes eight bosonic degrees of freedom which balance
the eight degrees of freedom of a Weyl fermion.
The self dual antisymmetric tensor field is chiral and its
field equations cannot be derived from a covariant Lagrangian
(without a lot of extra complications and gauge symmetries).
As we will see, this is the origin of the world sheet chirality
of the heterotic string.

\section{Forms, Branes and BPS states}

\subsection{Differential forms and topologically nontrivial cycles}

Before proceeding with our discussion of compactification
of 11D SUGRA, and its relation to string theory,
I want to insert a short remedial course on the mathematics
of differential forms.  We have already used this above, and
will use it extensively in the sequel.
Differential $n$ forms or 
totally antisymmetric covariant tensor
fields, were invented by mathematicians
as objects which can be integrated
over $n$ dimensional submanifolds of
a manifold of dimension $d$.  The basic
idea is that at each point, such a form
picks out $n$ linearly independent 
tangent vectors to the manifold and 
assigns a volume to the corresponding
region on the submanifold.
If $t^{\mu}_i$ are the tangent vectors,
then $\omega_{\mu_1 \ldots \mu_n} t^{\mu_1}_1
\ldots t^{\mu_n}_n$ is the volume element.

Mathematicians introduced Grassmann
variables $dx^{\mu}$ as placeholders
for the $n$ independent tangent vectors
Thus, an $n$ form becomes 
$\omega_{\mu_1 \ldots \mu_n} dx^{\mu_1}
\ldots dx^{\mu_n}$, which is a commuting
or anticommuting element of the Grassman
algebra according to whether $n$ is even
or odd.  In this way, the set of all
forms of rank $0 \rightarrow d$ is turned
into an algebra.   A derivative operator
$d$ is defined on this algebra by
$d = dx^{\mu}{\partial\over \partial 
x^{\mu}}$.  This definition is independent
of the metric or affine connection on
the manifold.  Note that $d^2 = 0$. 
Forms satisfying $d\omega = 0$ are called
closed.  Trivial solutions of the form
$\omega = d\lambda$ are called exact, 
and the set of equivalence classes of 
non exact closed forms (modulo addition
of an exact form) is called the 
{\it cohomology} of the manifold.

If a submanifold is parametrized as
a mapping $x^{\mu} (\xi)$ from some
$n$ dimensional parameter space 
into the
manifold, then the integral of a form
over the submanifold is given by
\eqn{intform}{\int_M \omega =
\int d\xi^{a_1} \ldots d\xi^{a_n}
{\partial x^{\mu_1} \over \partial 
\xi^{a_1}}\ldots {\partial x^{\mu_n} \over \partial 
\xi^{a_n}} \omega_{\mu_1 \ldots \mu_n}
(x(\xi))}

If $\omega$ is an $n-1$ form and $S$ 
an $n$ dimensional submanifold with 
boundary $\partial S$, then the 
generalized Stokes theorem says that
$\int_S d\omega = \int_{\partial S} 
\omega $.  In particular, the integral
of an exact form over a submanifold
without boundary, vanishes.

Most $n$ dimensional 
submanifolds without boundary, 
are themselves boundaries
of $n+1$ dimensional submanifolds.  
However, in 
topologically nontrivial situations there
can be exceptions, called nontrivial $n$-
cycles.  You can see this in the example
of the $a$ or $b$ cycle in Fig.  1.  Generally
there are many such nontrivial cycles
if there are any, but they often differ
by trivial cycles (think of two different
circles which go around the circumference
of the torus).  Again, the $n$ cycle
is considered to be the equivalence 
class of nontrivial submanifolds modulo
trivial ones.

One of the most important theorems in
mathematics is the de Rham theorem, which
states that there is a one to one
correspondence between the cohomology of
a manifold and the independent nontrivial
closed cycles. 
That is, one can choose a basis 
$\omega_i  $ in the
space of closed modulo exact $n$
forms such that $\int_{C_j} \omega_i = 
\delta_{ij}$. Here $C_j$ are the 
independent nontrivial $n$ cycles.

So much for pure mathematics.  The reason
that all of this math is interesting in
M-theory is that the theory contains
dynamical extended objects called $p$-branes,
and the theory of differential forms
allows us to understand the most important
low energy dynamical properties of these
objects as a beautiful generalization
of Maxwell's electrodynamics.  In addition
this leads us to a new and deep
mechanism for generating nonabelian
gauge groups, which is connected to the
theory of singularities of smooth 
manifolds.  This in turn allows us to
obtain an understanding of certain
spacetime singularities in terms of
Wilson's ideas about singularities of
the free energy at second order phase
transitions.  It was long known that the
free energy of statistical systems at second order
phase transitions had singularities as 
a function of the temperature and other
thermodynamic variables.  Wilson realized
that these singularities could be understood
using the equivalence between statistical mechanics
and Euclidean quantum field theory.  At values
of the parameters corresponding to phase transitions,
massless particles appear in the field theory 
and the singularities of the free energy are attributed
to infrared divergences coming from integrating over the
fluctuations of these particles.  

In classical geometry, singularities of manifolds can be classified
by asking which nontrivial cycles shrink to zero as parameters are varied
in such a way that a smooth manifold becomes singular. In \mth\  
there are states described by BPS branes wrapped around these nontrivial
cycles, which become massless when the cycles shrink to zero.
The singularities in classical geometry are then understood to be a 
reflection of the quantum fluctuations of these massless particles.
That is, singular quantities in classical geometry can be calculated
in terms of Feynman diagrams with loops of the massless states that
\mth\ predicts at these special points in moduli space 
(only these states contribute to the infrared divergence).
The quantum theory itself is nonsingular at these points, but its
description in terms of classical geometry breaks down because there
are light degrees of freedom (the wrapped branes) other than the
gravitational field.  Branes and singularities
are at the heart of string duality.

Let us begin the discussion of branes
by recalling the Lagrangian for the
coupling of the electromagnetic field
to a charged particle.  It is
\eqn{partlag}{\int dt A_{\mu} (x(t)) {d x^{\mu} \over dt} = \int_C A_1,}
where in the second equality we have used
the notation of forms.  If the particle
path is closed, this action is invariant
under gauge transformations 
$ A_1 \rightarrow A_1 + d\Lambda_0 $.
If we add to the action the simplest
gauge invariant functional of the $A_1$ 
field $\int dA_1 *dA_1 $\footnote{Here the
* denotes the Hodge dual, but for the
purposes of this lecture we can just
think of this as a shorthand for 
Maxwell's action.}, we obtain Maxwell's
theory of the coupling of charged 
particles to electromagnetism.

There is an obvious generalization of
all of this to the coupling of a $p+1$
form potential to a $p$-brane.  The
interaction is given by
\eqn{plag}{\int_{C_{p+1}} A_{p+1}} 
where the integral is over the world
volume of the $p$-brane.  By the 
generalized Stokes theorem, this enjoys
a gauge invariance $A_{p+1} \rightarrow
 A_{p+1} + d\Lambda_p$.  By virtue of
the fundamental equation $d^2 = 0$,
$dA_{p+1}$ is a gauge invariant object
and we can write an immediate 
generalization of Maxwell's action,
\eqn{genmax}{\int d^n x\ * F \wedge F.}

Once we have normalized $A$ by writing
the free Maxwell Lagrangian, we are left
with a free coefficient in the coupling
of the brane to the gauge field.  In
the electromagnetic case, we know that 
this coefficient, the electric charge,
is in fact quantized if we introduce 
magnetic monopoles and quantum mechanics, 
an observation first made by Dirac.  

The analogous observation for general $p$-branes 
was made by Nepomechie and 
Teitelboim \cite{claudio}.  A $p$-brane
couples to a rank $p+2$ field strength.
In $d$ spacetime dimensions we can 
introduce, given a metric, a dual
field strength
\eqn{dualtwo}{*F_{\mu_1 \ldots \mu_{d-p-2}} = 
\epsilon_{\mu_1 \ldots \mu_d} F^{\mu_{d-p-1} \ldots \mu_d}}
where we have raised indices with the
metric tensor.  One thus sees that the
natural dual object to a $p$-brane is a
$d-p-4$ brane.  

The easiest way to see the 
Dirac-Nepomechie-Teitelboim condition is
to compactify the system on a torus
of dimension $d-4$.  We wrap the $p$-brane
around $p$-cycles of this torus and 
its dual around the remaining $d-p-4$.
The integral, $\int_{T^p} A_{p+1}$ defines
a one form or Maxwell field in the 
uncompactified four dimensional spacetime,
and the $p$-brane is an electrically charged
particle.  It is easy to convince oneself
that the wrapped dual brane is
a monopole.   Thus we obtain a 
quantization condition relating the
couplings of the two dual branes to
the $p+1$ form gauge potential.  
Nepomechie and Teitelboim show that there
are no further consistency conditions.

\subsection{SUSY algebras and BPS states}

Now let us recall what we learned in the
previous section about BPS states.  I will repeat
that material briefly here, but readers who
feel they have absorbed it adequately
can skip the first few paragraphs. 
We pointed out above that many
SUSY theories have classes of massive
states whose masses are protected from
renormalization in the same way that those
of massless particles of spin greater than
or equal to one half are.   These are
called Bogolmony-Prasad-Sommerfield
or BPS states.
The easiest to understand are the Kaluza
Klein states of toroidal compactification,
but there is a vast generalization of
this idea.  The theorem of Haag Lopuzanski
and Sohnius (a generalization of the
Coleman-Mandula theorem) \cite{hlscm}
shows that the ordinary SUSY algebra
is the most general algebra compatible 
with an S matrix for particle states.
Purely algebraically though, the right
hand side of the SUSY algebra could
have contained higher rank antisymmetric
tensor charges (the general representation
appearing in the product of two spinors).

Our discussion of branes and gauge fields
provides us with a natural source of
such charges, as well as showing us the
loophole in the HLS theorem.  
Indeed, following our analogy with 
Maxwell electrodynamics, it is easy to
see that an infinite, flat, static
$p$-brane carries a conserved 
rank $p$-antisymmetric tensor charge
as a consequence of the equations of
motion of the $p+1$ form gauge field
it couples to.  The fact that these
branes are infinite extended objects and
carry infinite energy is the loophole
in the theorem.  It referred only to
finite energy particle states.  All
of the tensor charges vanish on finite
energy states.

However, when we compactify a theory,
we can imagine wrapping one of these
$p$-branes around a nontrivial $p$-cycle
in the compact manifold.  The resulting
state propagates as a particle in
the noncompact dimensions.  It has
finite energy, proportional to the
volume of the cycle it was wrapped
around.  Its tensor charge becomes a
scalar in the noncompact dimensions
and is called a central charge of
an extended ({\it i.e.} larger than
the minimal algebra in the noncompact
dimensions) SUSY algebra.  Thus the central
charges in extended SUSY algebras in low
dimensions may come from wrapped brane charges
as well as KK momenta.

Perhaps the most remarkable fact about
this statement is that as the volume
shrinks to zero, the mass of the wrapped
BPS state does as well.  If the volume
of the relevant $p$-cycle parametrizes
a continuous set of supersymmetric
vacuum states, then this conclusion is
exact and can be believed in all regimes
of coupling even though it was derived
by crude semiclassical reasoning.
Indeed, even if we don't know the theory
we are trying to construct, we can still
believe in the existence of massless
wrapped brane states as long as we
posit that the SUSY algebra is a symmetry.
We will make extensive use of this
argument in the sequel.

\section{Branes and Compactification}

\subsection{A tale of two tori}

We are now in a position to study many of the important dualities
of \mth, at least at a cursory level.  We will not have time
to delve here into the many computations and cross checks which
have convinced most string theorists that all of these dualities
are exact.  Many of the duality statements remain conjectures supported
by a lot of circumstantial evidence.  Obviously, they cannot be proven
until a full nonperturbative form of \mth\ is discovered.  However,
an important subclass of the dualities can actually be proven in
a Discrete Light Cone Quantization (DLCQ) of \mth\ known as Matrix Theory
\cite{bfss}.
It applies to Compactifications of \mth\ with at least 16 unbroken
SUSYs and at least $6$ noncompact Minkowski spacetime dimensions.
In DLCQ, one gives up Lorentz invariance by compactifying a lightlike
direction on a circle.  One then gets an exact 
description of \mth\ in terms of an auxiliary quantum field theory
living on a fictitious internal space.  
All of the duality symmetries
relevant to this class of compactifications (the only ones we will talk
about in these lectures) can be {\it derived} as properties 
of the auxiliary field theory.  This includes statements (such as
rotation invariance of the Type IIB theory constructed by compactifying
\mth\ on a two torus) for which there was no other evidence
prior to the advent of Matrix Theory. 

 We begin by compactifying \mth\ on a circle of radius
$R_{10}$.   When $R_{10}$ is much larger than the eleven 
dimensional Planck length $L_P$ there is a good description of the
low energy physics of the system in terms of 11D SUGRA
compactified on a circle.  The SUGRA Lagrangian incorporates
all of the low energy states of the system and gives a good 
approximation to their low energy scattering amplitudes.

As $R_{10}$ drops below $L_P$, this description breaks down.
An 11D low energy physicist might guess that the low energy
states of the system are just the zero modes (on the circle) 
of the SUGRA fields.  This gives 10D Type IIA SUGRA, which has
the following fields: $g_{\mu\nu}, \phi, B_{\mu\nu},
C_{\mu\nu\lambda}^{RR}, C_{\mu}^{RR}$.  These can be identified
as the ten dimensional metric (in string conformal frame), the 
dilaton field which describes local variations of the radius of
the eleventh dimensional circle, a two form gauge potential
which is given by the integral of the 11D three form around the
 circle, the three form itself, and a Kaluza-Klein one form
gauge field.  The effective Planck mass, $M_P^{10}$,of 
this ten dimensional theory is given by 
\eqn{planck}{(M_P^{10})^8 \sim R_{10} (M_P^{11})^9}
Thus, when $R_{10}$ is small,
the effective 10D SUGRA description breaks down at 
a much lower energy scale than the 11D Planck mass.

In fact, the existence of BPS M2 brane states tells us that
there is an even lower energy scale in the problem.
Consider a configuration of an M2 brane ``wrapped on
the circle'':
\eqn{wrap}{x^{\mu} (t,\sigma,\xi) = x^{\mu} (t, \sigma);\ 
\mu = 0 \ldots 9}
\eqn{wraptoo}{x^{10} (t, \sigma,\xi) = R_{10} \xi}
The main part of the action of an M2 brane is the volume of
the world surface swept out by the brane, multiplied by the
brane tension, which is of order $L_P^{-3}$.  For wrapped
configurations, this reduces to the ten dimensional area
swept out by the string $x^{\mu} (t, \sigma)$ in units of
the string tension $L_S^{-2} \sim R_{10} L_P^{-3}$.  This
gives an energy scale for string oscillations
$m_S \sim \sqrt{R_{10}} M_P^{11}$ which is much smaller than
the ten dimensional Planck mass.  

Thus, we are led to expect that \mth\ on a small circle is
dominated at energies below the eleven dimensional Planck scale
by low tension string states.  At the energy scale set by
the string length gravitational couplings are weak.  This can 
be seen by rewriting the dimensionally reduced action in terms
of the string length.  The coefficient of the Einstein action
becomes $(L_P/R_{10})^3 L_S^8$, indicating that at the energy scale
defined by the string tension, gravitational couplings are determined by
a small dimensionless parameter, $g_S^2 = e^{2\phi (\infty)} =
(R_{10}/L_P)^3$.   In fact, using the technology of Matrix Theory,
\cite{bfssmbsdvv} one can show that in the small $g_S $ limit, \mth\
becomes a theory of free strings.

There is in fact a unique consistent ten dimensional theory of
free strings with the supersymmetry algebra of 11D SUGRA 
compactified on a circle (the so called IIA algebra).  It
is the Type IIA superstring.  In fact, one can directly 
derive the full Green-Schwartz action for the superstring by
considering the supermembrane action of \cite{sezgin} restricted
to the wrapped M2 brane configurations above.  However,
this derivation is entirely classical, while the existence of
the string and its action actually follow purely from SUSY and
are therefore exact quantum mechanical results.

This, the first of many dualities, exhibits the general
strategy of the duality program.  Starting from a limiting
version of \mth\ valid only in a certain domain of moduli 
space and/or energy scale we exhibit some heavy BPS state
whose mass can be extrapolated into regimes where the original
version of the theory breaks down, and goes to zero there.
We then find that the effective theory of these new light
states is another version of the theory.  For the most part,
we find only weakly coupled string limits and limits where
11D SUGRA is a valid approximation.  This is exactly true
if we restrict attention to vacuum states with three or
more noncompact space dimensions and 32 supercharges.  With less SUSY
there are limiting regimes (many of which are called by the
generic name F-theory) where we do not have a systematic
expansion parameter, though many exact results can be derived.

If we try to repeat this exercise on a two torus something 
really interesting happens.  The new regime corresponds to
taking the area of the torus to zero with the rest of its
geometry fixed.   As is well known, up to an overall scale,
the geometry of a two torus is determined by a parallelogram
in the complex plane with one side going from zero to one
along the real axis.  This parallelogram describes the periodic
boundary conditions which define the torus.  It is completely
fixed by its other side, which is a complex number $\tau$ in
the upper half plane. $\tau$ is called the complex 
structure of the torus.
In fact, different $\tau$s can describe
the same torus. The $SL(2,Z)$ group generated by $\tau 
\rightarrow \tau + 1$ and $\tau \rightarrow - 1/\tau$ maps
all complex numbers which define the same torus onto each other.

In the zero area limit, we can define a whole set of low tension
strings, by choosing a closed path of nontrivial topology
on the torus, and studying M2 branes wrapped on this path.
The inequivalent nontrivial paths on the torus are characterized
by two fundamental cycles, called $a$ and $b$ in Figure 1.
A general path consists of going $p$-times around $a$ and $q$ 
times around $b$.  It turns out that the $(p,q)$ strings with
relatively prime integers are stable and can be viewed as bound
states of the $(1,0)$ and $(0,1)$ strings.  When the integers
are not relatively prime the state is not bound.  This picture
is derived from the BPS formula \cite{aspsch} for the 
string tension, which 
follows from a classical calculation in 11D SUGRA and
is promoted to an exact theorem by invoking SUSY.  The proof that
the states with integers having a common divisor are not bound
is more complicated \cite{witbound}.

Something even more interesting occurs when we consider M2 branes
which wrap the whole torus.  A state with an $m$ times wrapped
brane has energy 
\eqn{wrp2}{\sim m A L_P^{-3} \equiv {m\over R_B}}
 in the limit that the area 
goes to zero.  It turns out \cite{witbound} that the $m$ wrapped
states are stable against the energetically allowed decay into
$m$ singly wrapped states.  So in the area goes to zero limit
we get a new continuum.  Any other state in the theory can bind
with these wrapped membranes at little cost in energy. 
The result is that the states are labelled by a new continuous
quantum number in addition to their momenta in the 
eight noncompact dimensions.  Even more remarkable 
(the only extant proof of this requires Matrix Theory \cite{bsss})
is that the new continuum is related to the old one by an
$SO(9,1)$ Lorentz symmetry \cite{bsss}.  Thus, in \mth\
$11 - 2 = 10$.  

Since the origin of the new Lorentz group is obscure, we have
to resort to Matrix Theory again to find out which
kind of ten dimensional SUSY the theory has\footnote{
Actually, a consideration of the field content of the low
energy theory is enough.  In particular the fact that 
variations of the complex structure $\tau$ 
over the noncompact dimensions should appear as a complex
scalar field, is enough to tell us that we are in the 
IIB theory.  The ten dimensional Type IIA theory has only 
a single real massless scalar. Matrix Theory is only necessary to prove
that the statement is consistent at all energies.}.  It turns out 
that both of the ten dimensional Weyl spinors 
have the same chirality, and we are in the IIB theory.

There is of course a weakly coupled string theory with this
SUSY algebra; the Type IIB Green Schwarz superstring.  
In fact, the SUGRA limit of this theory has
an $SL(2,R)$ symmetry which one can argue is broken to
$SL(2,Z)$ by instanton effects.  It acts in the expected
way on $\tau$.   Furthermore there are actually two different
two form gauge potentials, which form an $SL(2,Z)$ doublet.
Thus we expect to find two different kind of strings, 
the F(undamental) string and the D(irichlet) string.  
The latter is a soliton, whose tension goes to infinity
in the weak coupling limit.  Consulting the eleven 
dimensional picture we realize that the weak coupling limit
should be identified with the ${\rm Im} 
\tau \rightarrow\infty$ limit in
which one of the cycles of the torus is much smaller than
the other.  The F(D) string is then identified with the
M2 brane wrapped around the shorter (longer) cycle.

This trick of dimensional reduction by $2-1$ dimensions
is interesting because it gets around old theorems which
stated that Kaluza Klein reduction cannot produce chirality.
It can be generalized in the following interesting way.
Certain higher dimensional manifolds can be viewed as
``elliptic fibrations''.  That is, they consist
of an $m$ dimensional base manifold with coordinates $z$
and a family of two tori $\tau (z)$ (the area also varies
with $z$), making altogether an $m+2$ dimensional manifold.
Now one varies parameters in such a way that the area of the
two tori all shrink to zero.  Naively this would give 
a dimensional reduction by two dimensions.  However, given
enough SUSY one can again verify that an extra noncompact
dimension appears in the limit so that the result is 
dimensional reduction by one.  The name given to this
general procedure is F-theory \cite{vafaf}.  It is very
useful for describing strong coupling limits of the 
heterotic string.

If we try to pull the shrinking torus trick in 3 dimensions
we run into a disappointment.  The new low tension state
which appears is a membrane obtained by wrapping the
M5 brane around the torus.  The effective low energy theory
is then \mth\ again with a new Planck length defined in terms
of the light membrane tension.  Indeed it can be shown
\cite{witvar} \cite{bfm} that for three or more noncompact dimensions
the only limiting theories one can obtain without breaking
any SUSY are the Type II string theories and 11D SUGRA.
We will actually prove this theorem below in our discussion of extreme 
limits of the moduli space.

For my last example of a duality I will study the moduli
space of \mth\ compactifications which break half of the
11D SUSY.  This is achieved by compactifying on four 
dimensional spaces called K3 manifolds.  We will have to 
understand a little bit about the geometry of such manifolds,
but I promise to keep it simple.  The equation for the
SUSY variation of the gravitino is 
\eqn{suvar}{\delta \psi_{\mu} = D_{\mu} \epsilon}
This must vanish for certain values of the SUSY parameters
$\epsilon$ in order to leave some SUSY unbroken.
A consistency condition for this vanishing is
\eqn{curv}{R_{\mu\nu}^{ab}\sigma_{ab} \epsilon = 0}
where $R_{\mu\nu}^{ab}$ is the curvature tensor in an
orthonormal frame and $\sigma_{ab}$ are the spin matrices
in the Dirac spinor representation.  We will always be
dealing with strictly Euclidean $n$ dimensional manifolds
so these are generators of $O(n)$.

In two and three dimensions, the spinor has only two
components and the generators are the Pauli matrices.
The only solution of (\ref{curv}) is to set the curvature
equal to zero, but then we do not break any SUSY.
We can do better with four compact dimensions.
The group $SO(4) = SU(2) \times SU(2)$ has two different
two dimensional spinors (familiar to particle physicists
after an analytic continuation as left and right handed
Weyl spinors), transforming as $(1,2)$ and $(2,1)$ under
the two $SU(2)$ subgroups.  Thus, if the curvature lies
in one of these two subgroups and we choose $\epsilon$ to be
a singlet of that subgroup, then the consistency condition
is satisfied.  

The stated condition on the curvature tensor is
\eqn{selfdual}{R_{\mu\nu}^{ab} = \epsilon^{abcd} 
R_{\mu\nu}^{cd} }
It is easy to see, using one of the standard identities
for the Riemann tensor, that this implies that the
Ricci tensor, $R_{\mu}^a = R_{\mu\nu}^{ab}e_{\nu b}$, 
vanishes.  Thus, insisting that half the SUSY is preserved
implies that the manifold satisfies the vacuum Einstein
equations (Euclidean) or is, as we say, Ricci flat.

There is one more immediate consequence of the SUSY equations
which I want to note.  Just like the spinor representation,
the second rank antisymmetric tensor representation of
$SO(4)$ breaks up into a direct sum of $(1,3)$ and $(3,1)$.
Thus, there will be, on a manifold which preserves 
half the SUSY, three independent covariantly constant
(and therefore closed and nowhere vanishing) 
two forms, $\omega_{\mu\nu}^a$.  Modulo some technical 
questions which we will ignore, this implies that the 
manifold is {\it hyperk\"ahler}.  Compact, four dimensional
hyperk\"ahler manifolds are called K3 manifolds (this is part of
an elaborate joke having to do with the famous Himalayan peak K2).

Noone has ever seen a K3 metric, but mathematicians are
adept at dealing with objects they can't write down 
explicitly.  We will only need a tiny bit of the vast
mathematical literature on these spaces.  In particular,
I want to remind you of the famous de Rham theorem, which
relates topologically nontrivial submanifolds in a space
to the cohomology of differential forms.  Remember that
a differential form is just a totally antisymmetric tensor
multiplied by Grassmann variables that mathematicians
call differentials
\eqn{diff}{\omega = \omega_{\mu_1 \ldots \mu_p} dx^{\mu_1} \ldots dx^{\mu_p}}
The operator
\eqn{d}{d = {\partial\over \partial x^{\mu}} dx^{\mu}}
maps $p$-forms into $p+1$-forms and satisfies $d^2 = 0$.
This defines what mathematicians call a cohomology problem.
Namely, one wants to characterize all solutions of $d\omega =0$, modulo
trivial solutions of the type $\omega = d \psi$ (such trivial
solutions are called exact forms) where
$\psi$ is a well defined $p-1$-form.   
 This is 
a generalization of finding
things with zero curl which cannot be written as gradients.
A well known physics example is a constant magnetic field
on the surface of a sphere or a torus.  The set of closed
but not exact $p$-forms is called the cohomology at dimension $p$.

The importance of $p$-forms stems from the fact that their
integrals over $p$-dimensional submanifolds are 
completely defined by the differential topology of the
manifold.  No metrical concepts are needed to define these
integrals.

Another important concept is that of a nontrivial $p$-cycle
on a manifold.  Basically this is a $p$-dimensional submanifold
which cannot be shrunk to a point because of the topology of
the manifold.  The simplest examples are the $a$ and $b$ 
1-cycles on the torus of Figure 1.  Actually, it is an equivalence
class of submanifolds because any curve which circles around
the $a$ cycle and then does any kind of topologically trivial 
thing on the rest of the torus is equivalent to the a cycle.
de Rham's theorem tells us that there is a one to one 
correspondence between $p$-cycles and $p$-forms, as we have mentioned
above.

After that brief reminder,
we can turn to the question of what the cohomology of
K3 manifolds is.  Since it is a topological question we can answer it
by examining an example.  Every 4-manifold has cohomology 
at dimension 0 (the constant function) and dimension 4 (the volume
form, $\epsilon_{abcd}e_{\mu_1}^a e_{\mu_2}^b  
e_{\mu_3}^c  e_{\mu_4}^d $).  The simplest K3 manifold is
the ``physicists K3'', the singular orbifold $T^4 /Z_2$.
This is defined by taking a rectilinear torus with axes $2\pi R_i$ 
and identifying points related by $x^i \rightarrow \pm
x^i + 2n_i\pi R_i$.  This has 16 fixed points in the fundamental
domain : $x^i = R_i (1 \pm 1) \pi/4 $.  The space is flat except at the
fixed points but has curvature singularities there.  It can be
verified that the holonomies around the fixed points are in
a single $SU(2)$ subgroup of $O(4)$ so the space is a K3.

It is easy to see that the nontrivial one cycles on the torus
all become trivial on the orbifold (the corresponding one forms
are odd under the orbifold transformation and are projected out).
The torus has six obvious 2 cycles, which are the six different $T^2$s
in the $T^4$.  In addition, when one studies this singular manifold
as a limit of smooth K3's by the methods of algebraic geometry
(realizing the manifold as the solution set of polynomial equations)
one finds that each of the fixed points is actually a two sphere of
zero area.  Thus there are twenty two non-trivial two cycles on 
a K3 manifold.  By the de Rham theorem, there are twenty two linearly
independent elements of the cohomology at dimension two of K3.

One can introduce a bilinear form on two forms in a four manifold.
The product of two two cycles is a four form, which can be integrated
over the manifold.  Define:
\eqn{intpair}{I_{ij} = \int \omega_i \omega_j}
Remember that $\int *\omega \omega$ is the usual Euclidean 
Maxwell action for a two form field strength is thus positive
definite.  The form $I$ is thus negative on antiselfdual tensors and
positive on self dual ones.  We have already established that there are 
three independent antiselfdual covariantly constant (and therefore 
closed but not exact) two forms.  It can be shown that 
the rest of the cohomology consists of self dual two forms, so that $I$
has signature $(19,3)$.  A basis can be chosen in which it has the form
$I = \sigma_1 \oplus \sigma_1 \oplus \sigma_1 \oplus E_8 \oplus
E_8$, where $\sigma_1$ is the familiar Pauli matrix and $E_8$ is the 
Cartan matrix of the Lie Group $E_8$ (the matrix of scalar products of
simple roots).  

This is very suggestive.  The heterotic string
compactified on a three torus, has nineteen left moving and three right
moving currents (the sixteen $E_8 \times E_8$ gauge currents and linear
combinations of the momentum and winding number currents on the torus).
Indeed, Narain \cite{narain} introduced the same scalar product, where 
left movers have opposite signature to right movers, in his study of
heterotic compactifications on tori.  At this point, readers who are not
familiar with the heterotic string will undoubtedly benefit from $\ldots$.

\subsection{An heterotic interlude}

The bosonic string ``lives'' in 26 bosonic dimensions, while the
superstring
lives in 10.  This discrepancy his two sources, both of which have to do with the
difference between the world sheet gauge groups of the two theories.  The bosonic string
has only worldsheet diffeomorphism invariance and the 26 is required to cancel the
anomaly in this symmetry against a corresponding anomaly coming from Fadeev-Popov
ghosts.  Type II superstrings have worldsheet supergravity\footnote{Not 
to be confused with spacetime supergravity, which is another beast
entirely.}.  On the one hand, this
requires the embedding coordinates $X^{\mu}$ to have superpartners $\psi^{\mu}$ which
also contribute to the anomaly.  On the other hand, since the world sheet gauge group is
larger, there are more ghosts.  The net result of these two effects is to reduce to 10 the 
maximal number of Minkowski dimensions in which the Type II strings can propagate.
Smaller numbers of $X^{\mu}$ can be achieved by compactification.

In two dimensions, the smallest SUSY algebra is called $(1,0)$ and has a single
right moving spinor supercharge. There is a corresponding chiral worldsheet supergravity.
Type II strings have the vector like completion
of this, $(1,1)$ SUSY, which consists of one left moving and one right moving
supercharge.  The heterotic string is defined as a perturbative string theory
with only $(1,0)$ worldsheet SUSY.  Its maximal number of Minkowski dimensions
is 10.

In ten Minkowski dimensional target space, the world sheet field theory of any string theory
is a collection of free massless fields each of which can be separated into its left and
right moving components.  The ten dimensional heterotic string has 10 right moving
$X^{\mu}$s and their superpartners, and 26 left moving $X^{\mu}$s.  In order to eliminate
an extra continuum from the 16 extra bosonic dimensions, we can compactify them on a torus.
This simply means that we eliminate all states which are not periodic functions
in these 16 coordinates.  

The restriction to toroidal compactification in fact follows from a deeper principle.
The construction outlined so far was a consistent gauge fixed quantum theory
with infinitesimal $(1,0)$ superdiffeomorphism invariance in two dimensions.
We have seen above that perturbative string theory requires us to evaluate the
world sheet path integral on Riemann surfaces of arbitrary genus.  For genus
one and higher, there are disconnected pieces of the diffeomorphism group
and we must require invariance under those as well.  This is called the
constraint of modular invariance.
It turns out that this
restriction is satisfied iff we choose the sixteen dimensional torus to be the
Cartan torus of one of the groups $E_8 \times E_8$ or $SO(32)$.  
The operators $(\partial_{\tau} - \partial_{\sigma}) X^i $\footnote{$\tau$ and 
$\sigma$ are worldsheet  coordinates and the $X^i$ satisfy $(\partial_{\tau} +
 \partial_{\sigma}) X^i = 0$.}, with $i = 1 \ldots 16$,
are then the current algebra for the $U(1)^{16}$ Cartan subgroup.  The currents
corresponding to raising and lowering operators of the group have the form
of exponentials $e^{i r_i X^i}$ where $r_i$ are the roots of the algebra.

If we further compactify the heterotic string on a $d$-torus, we get $d$ pairs
of $U(1)$ currents (which, for generic radii of the toru are {\it not} completed
to a nonabelian group) from $(\partial_{\tau} \pm \partial_{\sigma}) X^a$.
Half of these are purely left moving and the other half purely right moving.
The vectorlike combinations are simply the Kaluza Klein symmetries expected
when compactifying GR on a torus. The axial combinations couple to the winding
number of strings around the $d$ torus.  Perturbative string theory always
has a two form gauge field $B$, which couples to the string world sheet as
$\int_{worldsheet} B$.  When integrated around the $d$ 1-cycles of the torus,
it gives rise to $d$ one forms which couple to string winding number.
In addition to this gain in the rank of the symmetry group, a generic toroidal
compactification will lose the nonabelian parts of the group.  This is because
we can have Wilson lines around the cycles of the torus.  Thus, 
at a generic point on the moduli space of toroidal compactifications of the
heterotic string, the gauge group is $U(1)^{16 +d} \times U(1)^d$, where we have
separated the contributions coming from left and right moving currents.

Thus, one way of viewing toroidally compactified heterotic string theory
is to say that it consists of the modes of $16 + d$ left moving and $d$
right moving scalar fields, where the zero modes of these fields live on
independent tori (there are also fermionic partners and fields
representing the noncompact dimensions, but we do not need to discuss
them here).  A given compactification can then be specified by talking
about the allowed values of the dimensionless momenta around the torus,
a discrete set of numbers $(l^R_m, l^L_n)$.  One must insist that the
vertex operators with any allowed momenta are all relatively local
on the world sheet in order that the expressions for tree level string
amplitudes make sense\footnote{Left moving or right moving fields are
not local operators.  The vertex operators are exponentials of these
fields and are generally not local either.  But certain discrete subsets
of these vertex operators are relatively local.}.  Furthermore one must
impose a condition called {\t modular invariance} to guarantee that one
loop amplitudes make sense.   These conditions turn out to be equivalent
\cite{narain} to the restriction
\eqn{naraincomp}{({\bf l^L})^2 - ({\bf l^R})^2 \in 2{\bf Z}}
combined with the requirement that the lattice of all possible momenta
be self dual\footnote{The dual of a lattice with a scalar product
defined on it is the set of all vectors which have integer scalar
product with the vectors of the original lattice.}.  Such lattices turn
out to be unique up to an $O(16+d,d)$ rotation.  It can be shown that
the parameters of these rotations are equivalent to choices of
background Wilson lines, constant andtisymmetric tensor fields on the
$d$ torus and the choice of the flat metric on the torus.

Thus, heterotic string theory compactified on a $d$-torus with generic Wilson lines.
has a natural $O(16+d, d, Z)$ invariant scalar product defined on 
the space of worldsheet currents.  The fact that the same sort of scalar
product arises as the intersection matrix of cohomology classes of K3
manifolds is the first hint that the two systems are related.

\subsection{Enhanced gauge symmetries}

One of the reasons Type II strings did not receive much attention after
the discovery of the heterotic string was that they did not appear to
have the capability of producing gauge groups and representations like
those of the standard model.  The same was true of 11D SUGRA.  
However, the suggestive connection with heterotic strings leads one to
suspect that a mechanism for producing nonabelian gauge symmetries has
been overlooked.  The theory of singularities of K3 manifolds was worked
out by Kodaira and others in the 1950's.  It turns out that one can 
characterize singular K3 manifolds in terms of topologically nontrivial
cycles which shrink to zero size.  The singularity is determined by the
intersection matrix $I$ restricted to the shrinking cycles.  It turns
out that in almost all cases, the resulting matrix was the Cartan matrix
of some nonabelian Lie group.  In the purely mathematical study of
four manifolds, there is no way to understand where the Lie group is.

However, viewed from the point of view of \mth\ compactification on K3,
a nonabelian group jumps into view.  Indeed, imagine BPS M2 branes
wrapped around the shrinking cycles of the singularity.  These will be
massless particles in the uncompactified seven dimensional spacetime.
Since we have 16 SUSYs in the effective seven dimensional theory these
must include massless vector fields, since the smallest representation
of this SUSY algebra is the vector multiplet.   Furthermore, even away
{}from the singularity, we have 22 $U(1)$ vector multiplets.  Indeed
one can write three form potentials in 11D SUGRA of the form 
$A_{\mu\nu\lambda} dx^{\mu} dx^{\nu} dx^{\lambda} = a_{\mu}^i (X) dX^{\mu}  
\omega_i$, where $\omega_i$ are the 22 independent harmonic two forms
on K3, and $X$ are the seven noncompact coordinates.  The $a_{\mu}^i$ are 
gauge potentials in a product of $U(1)$ algebras which will
be the Cartan subalgebra of the nonabelian group that appears at the
singularity.
Since membranes are charged under the three form, we see, using the de
Rham connection between forms and cycles, that the new massless vector
bosons are charged under the Cartan subalgebra, {\it i.e.} we have
a nonabelian gauge theory.

One further point of general interest.  As is obvious from the $T^4
/Z_2$ orbifold example, the singularities that give rise to nonabelian
gauge groups live on manifolds of finite codimension (or branes) in the
compact space.  If the volume of the compact space is large, this will
lead to a large ratio between the gauge and gravitational couplings in
the noncompact effective field theory.  We will discuss the
phenomenological implications of this observation in the context of
the Ho\v rava Witten scenario below.

The emergence of nonabelian gauge theory from singularities is one of
the most beautiful results of the \mth\ revolution.  It combines Wilson's 
observation that singularities in the free energy functional at second
order phase transitions could be correlated with the appearance of
massless states, with the mysterious occurrence of Lie groups in 
singularity theory, uniting physics and mathematics in a most satisfying
fashion.   One can go much further along these lines.  When studying
singularities of Calabi-Yau manifolds of dimension three or four one
encounters cases which cannot be explained in terms of gauge theory, but
which do have an explanation in terms of nontrivial fixed points of the
renormalization group.  The interplay between SUSY, singularity theory,
and the theory of the renormalization group in these examples, is a
stunning illustration of the power of \mth\ \cite{ms}.

So far, we have seen how enhanced gauge symmetries arise from \mth\ on K3
but have not yet delivered on our promise to make a connection with the
heterotic string.  We have seen in toroidal examples that the key to
string duality is the existence of light BPS states when cycles of a
manifold shrink to zero.  The limit of \mth\ on K3 which gives rise to
weakly coupled heterotic string theory (on a torus, $T^3$)
is the limit where the K3 volume shrinks to zero in Planck units. 
The M5 brane wrapped around the K3 gives rise to a low tension string 
in this limit \cite{hsschw}.  Recall that the world volume of the
fivebrane carries an antisymmetric tensor gauge field with self dual
3 form field strength, $H = * H$, which satisfies $d H = 0$.  
For configurations of the five brane wrapped on K3 one can study
configurations of $H$ of the form $H = j_i \omega_i$, where $j_i$ is a world
volume one form which depends only on the two coordinates of the world
volume which are not wrapped on K3, and $\omega_i$ is one of the 22
harmonic forms on K3.  In order to satisfy $H = * H$, $j_i$ must satisfy
$j_i = \epsilon_i * j_i$, where $\omega_i = \epsilon_i * \omega_i$
(recall that 19 of the forms on K3 have $\epsilon_i =1$, while for the other
three it is negative).  $d H = 0$ implies $d j_i = 0$.  In more familiar
notation, the string formed from the M5 brane wrapped on K3 will have
19 left moving ($j^a = \epsilon^{ab}$) and 3 right moving conserved
currents.  This is precisely the bosonic field content of the (bosonic
form of) the heterotic string on a three torus.  The evident SUSY of the
wrapped brane configuration guarantees the existence of the appropriate
world sheet fermions.   

The heterotic string was discovered as a solution to the consistency
conditions of perturbative string theory.  Though it was obviously the
perturbative string most closely connected to real physics, no one ever
claimed that it was beautiful.  The derivation of its properties from
the interplay between the K3 manifold and the M5 brane of 11D SUGRA can
make such an aesthetic claim.  It is another triumph of string duality.

This construction automatically gives rise to the heterotic string
compactified on a three torus.  Note that again the geometry of K3
disappears from the ken of low energy observers and is replaced by a
space of a different dimension and topology.  Following Aspinwall
\cite{psa} we can try to recover the ten dimensional heterotic string
{}from the K3 picture.  The mathematics is somewhat complex but in the end
one recovers the picture of Ho\v rava and Witten \cite{horwita} 
(if one is careful to
keep the full $E_8 \times E_8$ gauge symmetry manifest at all times).
That is, one finds 11D SUGRA compactified on an interval, with 
$E_8$ gauge groups living on each of two 10
dimensional boundaries\footnote{As one takes the limit corresponding to
infinite three torus, one is forced to K3 manifolds with two $E_8$
singularities.}.   The heterotic string coupling is related to
the size of the interval, $L$, by $g_S = (L/L_P)^{3/2}$.

The Ho\v rava-Witten description of the strongly coupled heterotic string
in ten dimensions was originally motivated by considerations of anomaly
cancellation and matching onto various weakly coupled string limits.
It is somewhat more satisfying to realize it as a singular limit of
compactification of \mth\ on a K3 manifold.

Witten \cite{horwitb} has pointed out that the strong coupling limit 
of this picture can resolve one of the phenomenological problems of
weakly coupled heterotic string theory.  Among the few firm predictions
of heterotic perturbation theory is the equality
between the gauge coupling unification scale $M$ and the four dimensional
Planck scale $m_P$.  In reality these differ by a factor of $100$.
Careful consideration of threshold corrections brings this discrepancy to a
factor of $20$, but one may still find it disturbing.  Witten points
out that in the picture of 11D SUGRA on an interval it is easy to remove
this discrepancy.  Indeed, if we compactify the system to four
dimensions on a Calabi-Yau 3-fold of volume $V_6$\footnote{Actually, due
to details which we will not enter into, the Calabi-Yau volume varies
along the interval.  The parameter $V_6$ is its value at the end of the 
interval where the standard model gauge couplings live.}, 
then the four dimensional gauge
couplings are given approximately by $1/g_G^2 \sim (V_6 /L_P^6)$, while
the four dimensional Planck mass is given by $m_P^2 \sim (L V_6 /L_P^8)$.
Tuning $L$ and the volume to the experimental numbers
(and taking into account various numerical factors, gives a linear
size for the 3-fold of order $2 L_P$ and  $L \sim 70 L_P$.  The
unification scale $M$ is of order $L_P^{-1}$.   

I want to emphasize three features of this proposal.  First, the four
dimensional Planck mass is not a fundamental scale of the theory.
Rather, it is the unification scale, identified with the eleven
dimensional Planck scale, which plays this role.  Secondly, since we
have seen that gauge groups generically arise on branes in \mth,
Witten's proposal may be only one out of many possibilities for
resolving the discrepancy between the unification and Planck scales.
A possible advantage of a more flexible scenario might be the
elimination of all large dimensionless numbers from fundamental physics,
in particular the factor of $70$ in Witten's scenario.  If the
codimension of the space on which the standard model lives is large,
then the factor of order $100$ which is attributed to the volume of this
space in Planck units, might just be $2^6$.
Finally, let us note that in this brane scenario, the bulk physics
enjoys a larger degree of SUSY (twice as much) than the branes.  This
will be useful in our discussion of inflationary cosmology below, and
may also help to solve the SUSY flavor problem \cite{horwitc} 
\cite{rsetal}.

\subsection{Conclusions}

In this brief summary of \mth\ and its duality symmetries, we have seen
that classical geometry can undergo monumental contortions while the
theory itself remains smooth.  When there are enough noncompact dimensions
and enough supersymmetry, there are exact moduli spaces of degenerate vacua
which interpolate between regions which have very different classical
geometric interpretations by passing through regions where no geometrical
interpretation is possible (for the compact part of the space).  The
most striking example is perhaps the K3 compactification, where the 80 
geometrical parameters describing K3 manifolds are interpreted in an 
appropriate region of the parameter space as the geometry, {\it and
background gauge and antisymmetric tensor fields}, of a three torus with
heterotic strings living on it.  The clear moral of the story is that
``geometry is in the eye of the (low energy) beholder'', and must
actually be a low energy approximation to some other concept, which we
do not as yet understand.  

Equally important for our further discussion is that the modular parameters
interpolate smoothly between different geometrical regions and exist
even in regions which can not be described by geometrical concepts.
In different regimes of moduli space, the moduli can be viewed as zero modes
of different low energy fields living on different background geometries.
But, although their interpretation can change, the moduli remain intact,
and (with enough SUSY), their low energy dynamics is completely determined.  
In subsequent sections we argue that they are the appropriate 
variables for discussing the evolution of the universe.

\section{Quantum Cosmology}

\subsection{Semiclassical cosmology and quantum gravity}

In today's lecture we will leave behind our brief survey of \mth\ and
duality and proceed to cosmological questions.  We will begin by
discussing some ``fundamental'' issues in quantum cosmology and proceed
to a somewhat more practical application of \mth\ to inflationary models.
None of this work will lead to the kind of detailed model building
and comparison with observation that is the bread and butter of most
astroparticle physics.  In my opinion, the current theoretical
understanding of \mth\ does not warrant the construction of such detailed
models.  Detailled inflationary model building requires, among other
things, knowledge of the inflaton potential.  In an \mth\ context this
means that we have to have control over SUSY breaking terms in the
low energy effective action.  Even the advances of the last few years
have not helped us to make significant progress in understanding SUSY
breaking.

My aim in these lectures will be to address general questions like what
the inflaton is likely to be, the relation between the energy scales of 
inflation and SUSY
breaking, the connection between various scales and pure numbers
encountered in cosmology with the fundamental parameters, and so on.
We will see that a rather amusing picture can be built up on this basis,
which is quite different from most conventional cosmological models.
I will concentrate here primarily on my own work (and that of my
collaborators) rather than trying to give a survey of all possible
approaches to cosmology within \mth.  Prof. Veneziano will be giving a
detailed exposition of one of the other major approaches, so between the
two of us you will get some idea of what is possible.  

The discussion will be divided into two parts, one more ``fundamental''
and the other more ``practical''.  The aim of the first part will be
to pose the problem of how the conventional equations of cosmology
may eventually be derived from a fully quantum mechanical system.
We will also begin to address the question of why \mth\ does not choose
one of its highly supersymmetric vacua for the description of the world
around us.  We end this exposition by introducing a heterodox 
{\it antiinflationary} cosmology.  
The ``practical'' section will concentrate on issues related to moduli
and SUSY breaking.  We will see that cosmological considerations suggest 
a vacuum structure similar to that proposed by Ho\v rava and Witten, and
put further constraints on the form of SUSY breaking. One also obtains
an explanation of the size of the fluctuations in the microwave background
in terms of the fundamental ratio between the unification and Planck 
scales.
We will conclude
with an inflationary cosmology very different from most of those in the
literature.  Among its virtues is the possibility of supporting a QCD
axion with decay constant as large as the fundamental scale.  Indeed,
the assumption that such an axion exists gives an explanation of the
temperature of matter radiation equality in terms of the fundamental 
parameters of the theory.
 
We will begin our discussion of ``fundamental'' cosmology by recalling
the treatment of quantum cosmology in GR.
One of the more bizarre consequences of an attempt to marry GR to QM 
is the infamous
{\it Problem of Time}.  A generally covariant theory is constructed for the
precise purpose of not having a distinguished global notion of time.  
In classical
mechanics this is very nice, but quantum mechanically it turns the 
conventional
Hamiltonian framework on its head.  The problem can be seen in simple systems
with time reparametrization invariance, that is, actions of the form
\eqn{timereplag}{\int dt e L(q,\dot{q}/e)}
where $q$ represents a collection of variables which transform as
scalars 
under
time reparametrization, and $e$ is an einbein ({\it i.e. } $e dt$ is
time 
reparametrization
invariant).  We can use the symmetry to set $e$ equal to a constant 
(gauge fixing), but the
$e$ equation of motion then says that the canonical Hamiltonian of the 
$q$ vanishes.
\eqn{wdsimp}{H = \dot{q} {\partial L \over \partial\dot{q}} - L = 0.}

In simple covariant systems like Chern-Simons gauge theory, one can 
solve this Hamiltonian
constraint and quantize the system in the sense that the classical 
observables
are realized as operators in a Hilbert space.  However, the notion of 
time evolution
is still somewhat elusive.  More generally, in realistic systems where the
constraints are not explicitly soluble, one recovers time evolution by 
finding
classical variables.  For example, if spacetime has a boundary, with 
asymptotically 
flat or asymptotically Anti deSitter boundary conditions, then one can 
use one of
the symmetry generators of the classical geometry at infinity as a time 
evolution
operator.  

In cosmology one generally does not have the luxury of a set of
variables whose quantum fluctuations are frozen by the boundary 
conditions. The notion 
of time evolution 
is tied to a semiclassical approximation for a particular set of 
variables.  Different cosmological evolutions may not be described by
the same semiclassical variables.
One of
the challenges of this framework is to find a generic justification for 
the semiclassical
approximation. To see how the idea works, one ``quantizes'' the
$g_{00}$ Einstein equation by writing it in Hamiltonian form and naively 
turning the
canonical momenta into differential operators (at the level of 
sophistication of this
analysis, it does not make sense to worry about ordering ambiguities).  
This gives
the Wheeler-DeWitt equation, a second order PDE which is supposed to 
pick out
the physical states of the system inside a space of functionals of the 
fields on
a fixed time slice.  The challenge is to put a positive metric Hilbert 
inner product
on the space of physical states and identify a one parameter group of 
unitary operators
that can be called time evolution.

It is well known that, viewed as a conventional field theory, 
the conformal factor
of the gravitational field has negative kinetic energy.  In 
quantization of GR in perturbation theory around any classical
solution of the field equations, the negative modes are seen to be 
gauge artifacts and
a positive definite Hamiltonian is found for gravitons. 

In general 
closed cosmologies, the analogous
statement is the following: 
the Wheeler DeWitt constraint completely eliminates all negative modes
{}from the physical Hilbert space.  It is convenient to think of GR
in synchronous gauge, where the $g_{0i}$ components of the metric
vanish and $g_{00} = 1$.  Such a gauge is built by choosing a spacelike
hypersurface and following timelike geodesics orthogonal to this
hypersurface to define the evolution into the future.  It can then be shown
that all of the negative modes represent the freedom to change the
choice of the initial hypersurface (the many fingered time of GR).  
The Wheeler-DeWitt equation is then the constraint which says that
physics must be independent of this choice.  It is often convenient
to solve the contraint in stages.  Namely, among all spacelike surfaces
in a given spacetime geometry, there are one parameter 
families related to each other by propagation along orthogonal timelike
geodesics.  The choice of such a family eliminates all but one of the
negative modes, the last one being related to the choice of which surface 
in the family is called the initial surface.  That is, it is related to
the {\it time} as measured by observers following the timelike geodesics
which define the family\footnote{This discussion is purely classical, but
mirrors the less intuitive mathematical operations which one carries out
in semiclassical quantization of the WD equation.}.  It can be chosen
to be any monotonic function along these trajectories, and it is often
convenient to choose the volume of the spatial metric.  

The upshot of all this, is that once a family of hypersurfaces is chosen,
one still has a single component of the Wheeler-DeWitt constraint which 
has not yet
been imposed.  Classically this is the familiar Friedmann equation 
relating the expansion
rate to the energy density.  A naive quantization of this equation gives a
hyperbolic PDE on a space with signature $(1,n)$.   A form 
of this equation sufficiently general for our purposes is
\eqn{wd}{[h G^{ab} (m) \partial_a \partial_b + g^{AB}(q,m) \partial_A 
\partial_B + {1\over h}
V(m) + U(q,m)] \Psi = 0}
We have separated the variables into classical ($m^a$) and quantum
($q^A$) and introduced
a formal parameter $h$ to organize the WKB like approximation for 
the classical variables.
The metric $G$ is hyperbolic with one negative direction, while 
the metric $g$ has 
Euclidean signature.  The analysis we are presenting goes back to 
\cite{rbbfs}.
Up to terms of order $h$, it is easy to see that the solution of this 
equation has the
form
\eqn{wdsoln}{\Psi (m,q) = e^{i S(m)/h} A(m) \psi (m,q)}
where
\eqn{wkba}{- G^{ab}\nabla_a S \nabla_b S + V = 0}
\eqn{wkbb}{G^{ab}(\nabla_a S \nabla_b A + A \nabla_a \nabla_b S  =0)}
\eqn{wkbc}{i G^{ab}\nabla_a S \nabla_b \psi + H\psi = 0}
\eqn{wkbd}{H \equiv g^{AB}\nabla_A \nabla_B + U}
The first of these equations has a Hamiltonian-Jacobi form.
It can be solved by finding classical motions $\dot{m}^a (t) =
G^{ab}\nabla_a S(m)$.  $S$ is then the action of the classical solution,
and (\ref{wkba}) is satisfied if the solution has zero ``energy''.  The
existence of real zero energy solutions (and thus real $S$) depends on
the fact that $G^{ab} $ has nonpositive signature.

Using the classical solution $m^a (t)$, we recognize that (\ref{wkbc})
can be written as a conventional Schr\"odinger equation:
\eqn{schrod}{i {\partial \psi \over \partial t} = H \psi}
Positivity of the Hamiltonian (\ref{wkbd}) requires that $g_{AB}$ have
Euclidean signature.  Note that since $H$ depends on the $m$'s, the
Hamiltonian will in general be time dependent.
Furthermore, the quantum fluctuations of the
classical variables $m$ will have a sensible Hamiltonian only if
the metric $G_{ab}$ has only a single negative eigenvalue.  
Thus, we see that within the naive approach to quantization of
Einstein's equations, the existence of a Hilbert space
interpretation of the physical states, with a positive definite scalar
product and a unitary time evolution with a sensible Hamiltonian
operator, is closely tied to the fact that Einstein's equations coupled
to matter with positive kinetic energy have a hyperbolic metric with
exactly one negative eigenvalue (after gauge fixing).

One may wonder whether these observations will survive in a more
realistic theory of quantum gravity.  We know that Einstein's action is
only a low energy effective description of \mth.  Even those heretics
who refuse to admit that \mth\ is the unique sensible theory of quantum
gravity\footnote{One hopes that the world has not come to a state in
which one has to emphasize that a sentence like this is a joke, but let
me record that fact in this footnote just to be on the safe side.}
are unlikely to insist that quantization of this famously
nonrenormalizable field theory by the crude procedure described above is
the final word on the subject of quantum gravity.  I would like to
present some evidence that in \mth\ the $(1,n)$ signature of the metric
on the space of classical variables is indeed guaranteed by rather
robust properties of the theory.   

Before doing so I want to point out how \mth\ addresses the question of
the existence of semiclassical variables $m^a$.  There are actually two
desiderata for the choice of such variables: we want the semiclassical
approximation for these variables to be valid during most of cosmic
history\footnote{As Borges pointed out long ago \cite{borges} it is
almost impossible to avoid self referential paradoxes when trying to
conceptualize a system in which the notion of time is an illusion or an
approximation.  According to the paragraphs above, cosmic history and
its implied notion of time {\it only exist} because of the classical
nature of the $m^a$.  Rather than attempting the impossible task of
being logically and linguistically precise, I will make the common
assumption that ``any sensible physicist who has followed my discussion
understands exactly what I mean by these imprecise phrases''.}.
Secondly, given the notion of energy implied by the classical solution
for the $m^a$, one often wants to be able to make a Born Oppenheimer
approximation in which the $m^a$ are slow variables or collective
coordinates. Note that the classical nature of the $m^a$ is crucial, while
the Born-Oppenheimer approximation is not.  Without classical variables
we would have no notion of time evolution.  The Born-Oppenheimer 
approximation allows us to discuss the evolution of the classical variables
in terms of an effective action in which other degrees of freedom are
ignored.  This is particularly useful below the Planck energy, since we
have no idea how to describe the full set of degrees of freedom of the
theory in the Planck regime, but are comfortable with a quantum field
theory description below that.  
Nonetheless, we will argue below that the classical 
variables might still provide a useful notion of  time evolution during
the Planck era, as long as the variable which we identify as the
spatial volume of classical geometry below the Planck scale, is large.  
It is important for any such pre-Planckian endeavor that \mth\ gives
an unambiguous meaning to (at least highly supersymmetric) moduli spaces
even in regimes not describable by low energy Einstein equations.
In a regime of super-Planckian energy and large volume,
 one would have to know something about the
dynamics and the state of all the degrees of freedom in the system
to understand how they effect the evolution of the classical variables.  

As suggested in the last paragraph, both classicality and slow evolution
 can be understood in \mth\ if we identify the $m^a$
as {\it moduli}, though with a slightly different definition of that
word than the usual ``parameters describing continuous families of 
supersymmetric vacua
with $d \geq 4$ asymptotically flat dimensions''.  In a theory of
quantum gravity, SUSY can only be defined nonperturbatively if we insist
on studying states with certain {\it a priori} boundary conditions.
The SUSY charges, just like the Hamiltonian, are defined as generators of
certain asymptotic symmetries of the whole class of metrics satisfying
the boundary conditions.  However, if we restrict attention to the
classical SUGRA equations, then we can define what we mean by
solutions which preserve a certain amount of supersymmetry.  
Since the Hamiltonian appears in the SUSY algebra, they will all be
static solutions. To find them, we simply
require that certain SUSY variations of all the fields vanish at the
solution.  Typically, we find a {\it moduli space} of continuously
connected solutions preserving a particular SUSY subalgebra.  The
parameters $m^a$ are coordinates on this space. In particular, for 11D
SUGRA, each solution will be a static, compact ten geometry, and the
volume of the compact space, $V$ will be one of the moduli.

Now consider classical motions in which the $m^a$ become functions of
time.  The effective action derived by plugging such time dependent
moduli into the SUGRA action has the form
\eqn{modact}{S = G_{ab}(m) \dot{m^a}\dot{m^b} e^{-1} (t)}
where $e$ is an einbein which imposes time reparametrization invariance.
$G$ is a hyperbolic metric with signature $(1,n)$.  In fact, it is
easy to see that the only modulus with negative kinetic energy is the
volume $V$.  This is because our choice of parametrization of the
spacetime metric has implicitly chosen a family of spacelike hypersurfaces
in these spacetimes (those of constant $t$).
The constraint equation coming from varying $e$ can be
written
\eqn{friedmann}{({\dot V \over V})^2 = 
\hat{G_{ab}}\dot{\hat{m}^a}\dot{\hat{m}^b}}
and is just the Friedman equation for a Robertson Walker cosmology.
The hatted quantities stand for the moduli space of SUSY solutions with
volume $1$.

It is easy to prove (and well known to those who have studied 
cosmologies with minimally coupled massless scalar fields) that the
field equations of this system give, for the Volume variable, exactly
the equations of a Robertson-Walker universe with equation of state
$p = \rho$.  The ``energy density'' $\rho$ then scales like $1/V^2$.
The $\hat{m}$ variables satisfy the equations of geodesic motion in
the metric $\hat{G}$, under the influence of cosmological friction.
This is equivalent to free geodesic motion in the reparametrized time
$s$ defined by $ds/dt = V^{-{1\over 2}}$.  The volume is always
monotonically decreasing or increasing in these solutions.
The derivation of these facts
is an enjoyable exercise in classical mechanics which I urge the
students to do.  

Finally I want to note that under the transformation
$V \rightarrow c V$, the action scales as $S \rightarrow c S$.  Thus,
Planck's constant $h$ can be absorbed in $V$, and the system is
classical at large $V$.  I want to emphasize that the actual spacetime
geometries described by these evolution equations can be quite complex.
That is, the $m^a$ might parametrize a set of Calabi-Yau manifolds.  However
the simple properties of the evolution on this moduli space described
above are unaffected by the complexity of the underlying manifolds.

The point of all of these classical SUGRA manipulations is that, given
enough SUSY, there are nonrenormalization theorems which protect this
structure in regimes where the classical SUGRA approximation is invalid.
For example, if there are 16 or more SUSYs preserved, then one can prove
that there is no renormalization of the terms with $\leq 2$ derivatives
in the effective Lagrangian for the moduli, to all orders in the
expansion around classical SUGRA.  Furthermore, these are the cases
where SUGRA is dual to Type II (32 SUSYs) or Heterotic (16 SUSYs) string
theories, compactified on tori.  The weak coupling string expansions are
in some sense expanding around the opposite limit from the SUGRA
expansion (extremely small volumes, in $L_P$ units, 
of compact submanifolds rather than extremely large ones).  To all
orders in the weak coupling string expansions one can establish that the
moduli space exists ({\it i.e.} that no potential term is generated in
the effective Lagrangian for the moduli) and that its 
topological and metrical structure is the same as that given 
by 11D SUGRA\footnote{If one is willing to decompactify three of the toroidal
directions and view the remaining moduli as zero modes of fields in
$3+1$ dimensional Minkowski space, then one can prove these statements
{}from SUSY without recourse to any expansion.  It is likely that these
proofs can be adapted to the completely compactified situation, but this
has not yet been done.}. 

There is thus ample evidence that there is some exact sense in which 
the configuration space of \mth\ contains regions which map precisely on
to the classical moduli spaces of SUGRA solutions preserving at least
16 SUSYs.  For 8 SUSYs, the situation is a bit more complicated.  
The well understood regions of moduli space here correspond to 11D SUGRA 
(or Type II strings) compactified on a manifold which is the 
product of a Calabi-Yau 3-fold and a torus, or heterotic strings
compactified on K3 manifolds times a torus.  Calabi-Yau 3-folds come
in different topological classes, but there is a conjecture that all
of these regions are on one continuously connected moduli space once
quantum mechanics is taken into account.  This statement depends on the
fact that the strong nonrenormalization theorems described above are not
valid.  The metric on moduli space is modified by higher order
corrections.  However, one can still prove a nonrenormalization theorem
for the potential on moduli space (namely that it is identically zero)
so that the moduli space still exists as an exact concept.

This is all that is needed to establish that the moduli are good
candidates to be the semiclassical, Born-Oppenheimer variables that are
necessary for the derivation of a cosmology from a generally covariant
quantum system.   Indeed, the absence of a potential on moduli space
means that the classical moduli can execute arbitrarily slow motions
with arbitrarily low energy.  Thus, in regimes where the classical motion 
has energy density small compared to the fundamental scale of \mth,
they are good Born-Oppenheimer
variables.  Furthermore, the $V$ rescaling symmetry of the action shows
that whenever $V$ is large the moduli will behave classically.  Indeed
this will even be true in regions where the Born-Oppenheimer
approximation breaks down, that is regions where the energy density is
of the order of or larger than the Planck scale, but the volume is
large.
In such a regime, a description of the evolution in terms of classical
moduli coupled to a stochastic bath of high energy degrees of freedom
might be appropriate.  The mystery will be to understand the equation of
state of the stochastic bath.   

The necessity of coupling the moduli to another, stochastic set of
degrees of freedom appears also very late in the history of the
universe.  The modular energy density scales to zero much faster than
either matter or radiation.  Thus if there is any mechanism which
generates matter or radiation, they will quickly dominate the energy
density of the universe.  In \cite{preva} it was shown that when the
moduli can be treated as the homogeneous modes of quantum fields, there
is an efficient mechanism for converting modular energy into radiation.
Thus, at late times, one must study the motion of the moduli coupled to
a stochastic bath of radiation and/or matter.

To summarize, the existence of a set of approximately classical, low
energy collective coordinates which take values in a space of hyperbolic
signature $(1,n)$ seems to be a very robust property of \mth.  
These would seem to be just what we need for a derivation of cosmology
{}from the theory.
{}From the point of view of someone who is deeply attached to ``the real
world'', the only problem with this analysis is that the universes it
describes become highly supersymmetric in the large volume limit.
We will defer the discussion of moduli in the context of broken SUSY 
to section 7.

\subsection{Extreme moduli}

In this subsection I will present some results about the beginning and end
of cosmic evolution in the highly supersymmetric situations I have just
described.  One motivation for this is to provide a controlled model for
more realistic cosmologies.  Another is to try to address the question 
with which we began these lectures, of why the universe as described by
\mth\ does not end up in a highly supersymmetric state.  Finally, we will
discover some very interesting results about duality and singularities
which are closely related to the hyperbolic structure of moduli space
and the question of the arrow of time.

We will discuss only the case of maximally SUSY moduli spaces, which are
obtained by compactifying \mth\ on a ten torus.  The parameters are a
flat metric on the torus, and the expectation value of the three form
potential, $A_{\mu\nu\lambda}$ on three cycles of the torus.  Most of
these are compact angle variables.  Among the metric variables, only
the radii $R_i$ of a rectilinear torus are noncompact, while the three
form expectation values are all angle variables because of the
Dirac-Nepomechie-Teitelboim quantization condition \cite{claudio}
(their conjugate momenta are quantized).  Thus, intuitively, we 
can restrict our
discussion of the possible extreme regions of moduli space to the radii
of a rectilinear torus.  This argument can be made mathematically
precise using the description of the moduli space as a homogeneous
space.  We will call the restricted rectilinear moduli space, the Kasner
moduli space.

The metrics which describe motion on the Kasner moduli space have the
form
\eqn{metric}{ds^2 = - dt^2 + R_i^2 (t) (dx^i)^2}
where the $x^i$ have period $2\pi$.
Inserting this ansatz into the action, 
we find that the solution of the equations
of 11D SUGRA for 
individual radii
are
\eqn{kassoln}{R_i (t) = \lp (t/t_0)^{p_i}}
where
\eqn{kascond}{\sum p_i^2 = \sum p_i = 1}
Note that the equation (\ref{kascond}) implies that at least one of the
$p_i$ is 
negative. 
We have restricted attention to the case where the volume expands
as time goes to infinity.  We will see below that, although the
equations are time reversal invariant, all of these solutions visit
two very different regions of moduli space at the two endpoints of
their evolution.  One of the regions has a simple semiclassical
description, while the other does not.  This introduces a natural
arrow of time into the system -- the future is identified as the
regime where the semiclassical approximation becomes better and better.    

It is well known that all of these solutions are singular 
at both infinite and zero time.  Some of the radii shrink to zero at
both ends of the evolution.
Note that if we add a matter or radiation energy density to
the system
then it dominates the system in the infinite volume limit and changes
the 
solutions for the
geometry there.  However, near the singularity at vanishing volume both 
matter and radiation 
become negligible (despite the fact that their densities are 
becoming infinite) 
and the solutions retain their Kasner form.

All of this is true in 11D SUGRA.  In M-theory we know that many regions
of moduli space which are apparently singular in 11D SUGRA can be
reinterpreted as living in large spaces described by weakly coupled Type
II string theory or a dual version of 11D SUGRA. The vacuum Einstein
equations are of course invariant under these U-duality transformations.
So one is lead to believe that many apparent singularities of the Kasner
universes are perfectly innocuous.

Note however that phenomenological matter and radiation densities which
one might add to the equations are not invariant under duality.  The
energy density truly becomes singular as the volume goes to zero.  How
then are we to understand the meaning of the duality symmetry?  The
resolution is as follows.  We know that when radii go to zero, the
effective field theory description of the universe in 11D SUGRA becomes
singular due to the appearance of new low frequency states.  We also know
that the singularity in the energy densities of matter and radiation
implies that scattering cross sections are becoming large.  Thus, it seems
inevitable that phase space considerations will favor the rapid
annihilation of the existing energy densities into the new light degrees
of freedom.  This would be enhanced for Kaluza-Klein like modes, whose
individual energies are becoming large near the singularity.

Thus, near a singularity with a dual interpretation, the contents of the
universe will be rapidly converted into new light modes, which have a
completely different view of what the geometry of space is. The most
effective description of the new situation is in terms of the transformed
moduli and the new light degrees of freedom.  The latter can be described
in terms of fields in the reinterpreted geometry.  We want to emphasize
strongly the fact that the moduli do not change in this transformation,
but are merely reinterpreted.  This squares with our notion that they are
exact concepts in M-theory.  By contrast, the fields whose zero modes they
appear to be in a particular semiclassical regime, do not always make
sense.  The momentum modes of one interpretation are brane winding modes
in another and there is no approximate way in which we can consider both
sets of local fields at the same time.  Fortunately, there is also no
regime in which both kinds of modes are at low energy simultaneously, so
in every regime where the time dependence is slow enough to make a low
energy approximation, we can use local field theory.

This mechanism for resolving cosmological singularities leads naturally to
the question of precisely which noncompact regions of moduli space can be
mapped into what we will call the {\it safe domain} in which the theory
can be interpreted as either 11D SUGRA or Type II string theory with radii
large in the appropriate units.

\subsection{The moduli space of  M-Theory on rectangular tori}

In this section, we will study the structure of the moduli space
of M-theory compactified on various tori $T^k$ with $k\leq 10$.  We
are  especially interested in noncompact regions of this space which
might represent either singularities or large universes. As explained above, 
the three-form potential $A_{MNP}$ will be
set to zero and the circumferences of the cycles of the torus
will be expressed as the exponentials
\eqn{radiiexp}{ {R_i \over \lp} = s^{p_i},\qquad
i=1,2, \dots, k.}

The remaining coordinates $x^0$ (time) and $x^{k+1}\dots x^{10}$ are
considered to be infinite and we never dualize them.  It is important
to distinguish the variable $s$ here from the time in the Kasner
solution.  Here we are just parametrizing possible asymptotic domains
in the moduli space, whereas the Kasner solution is to be used as
a metric valid for all values of the parameter $t$.  We will see that it
interpolates between two very different asymptotic domains.

The radii are encoded in the logarithms $p_i$. We will study limits of
the moduli space in various directions which correspond to keeping
$p_i$ fixed and sending $s\to\infty$ (the change to $s\to
0$ is equivalent to $p_i\to -p_i$ so we do not need to study
it separately).  In terms of this parametrization of the extreme 
regions of moduli space, we can see that a Kasner solution 
with parameters $p_i$ will visit the regime of moduli space characterized
by $p_i$ as $t\rightarrow\infty$ and the regime $- p_i$ as $t\rightarrow 0$.

\subsection{The \rut}

M-theory has dualities which allow us to identify the vacua with
different $p_i$'s.  A subgroup of this duality group is the $S_k$ which
permutes the $p_i$'s.
Without  loss of generality, we can assume that
$p_1\leq p_2\leq \dots \leq p_{10}$. We will assume this in most of
the text.
The full group that leaves invariant rectilinear tori with
vanishing three form is the Weyl group of the noncompact $E_k$ group
of SUGRA. We will denote it by $\Rut_k$.  We will give an elementary
derivation of the properties of this group for the convenience of
the reader.  
$\Rut_k$ is generated
by the permutations of the cycles on the torus, 
and one other transformation which acts as follows:
\eqn{rutdef}{(p_1,p_2,\dots, p_k)
\mapsto
(p_1-{2s\over 3},
p_2-{2s\over 3},
p_3-{2s\over 3},
p_4+{s\over 3},
\dots,
p_k+{s\over 3}).}
where $s=(p_1+p_2+p_3)$.  
Before explaining why this transformation is a
symmetry of M-theory, let us point out several of its properties
(\ref{rutdef}).

\begin{itemize}
 \item The total sum $S=\sum_{i=1}^k p_i$ changes to $S\mapsto
S+(k-9)s/3$. So for $k<9$, the sum increases if $s<0$, for $k=9$
the total sum is an invariant and for $k>9$ the sum decreases for
$s<0$.

 \item If we consider all $p_i$'s to be integers which are
equal modulo 3, this property will hold also after
the \rut. The reason is that, due to the assumptions, $s$ is a multiple
of three and the coefficients $-2/3$ and $+1/3$ differ by an integer.

 \item As a result, from any initial integer $p_i$'s we get $p_i$'s
which are multiples of $1/3$ which means that all the matrix elements
of matrices in the \rut{} are integer multiples of $1/3$.

 \item The order of $p_1,p_2,p_3$ is not changed (the difference
$p_1-p_2$ remains constant, for instance). Similarly,
the order of $p_4,p_5,\dots, p_k$ is unchanged. However the
ordering between $p_{1...3}$ and $p_{4...k}$ can change in general.
By convention, we will follow each \rut{} by a permutation which places
the $p_i$'s in ascending order.

\item The bilinear quantity $I= (9-k) \sum (p_i^2) + (\sum p_i)^2 = (10
- k) \sum(p_i^2) + 
 2 \sum_{i < j} p_i p_j$ is left invariant by $\Rut_k$.
\end{itemize}

The fact that \rut{} is a symmetry of M-theory can be proved as follows.
Let us interpret $L_1$ as the M-theoretical circle of a type IIA string
theory. Then the simplest duality which gives us a theory of the same kind
(IIA) is the double T-duality. Let us perform it on the circles $L_2$
and $L_3$. The claim is that if we combine this double T-duality
with a permutation of $L_2$ and $L_3$ and interpret the new $L_1$ as the
M-theoretical circle again, we get precisely (\ref{rutdef}).

Another illuminating way to view the transformation \rut{} is to
compactify M-theory on a three torus.  The original M2-brane and the
M5-brane wrapped on the three torus are both BPS membranes in eight
dimensions.  The tension of the original M2-brane is of order $\lp^{-3}$,
while that of the membrane which comes from the wrapped M5 is $V \lp^{-6}$
where $V$ is the volume of the three torus.  When the three torus is
large and the 11D SUGRA approximation is valid, the wrapped M5-brane is
much heavier than the M2-brane, while in the small volume limit, the
opposite is true.  We have seen previously that in limits where classical
geometrical descriptions are breaking down, one can find a new classical
description by following those BPS states which become lightest in the
limit.  This suggests that we try to define $\bar{l}_P^{-3} = V \lp^{-6}$
and $\bar{V} = \lp^{-3} \bar{l}_P^6$ and try to imagine a duality
transformation in \mth\ which takes a compactification on a small three 
torus to a compactification on a large one, with corresponding redefinition
of the Planck scale.  Aharony \cite{ofer}
has given arguments that such a duality transformation exists, and it 
can be demonstrated rigorously in Matrix Theory. 
In the limit in which one of the cycles of
the $T^3$ is small, so that a type II string description becomes
appropriate, it is just the double T-duality of the previous paragraph.  
The fact that this transformation plus permutations generates $\Rut_k$ was
proven by the authors of \cite{elitzur} for $k \leq 9$.
I leave it to the reader to verify that the effect of this transformation
on the variables $p_i$ is precisely that described above.

In the following subsection we will use this group of duality 
transformations to prove that extreme regions of the moduli space fall
into a number of distinct categories.  One is such that some kind of
semiclassical description of the physics is valid, and breaks up into regions that are described by 11D SUGRA or weakly coupled Type IIA or IIB
string theory.  The other is completely mysterious and has no known
semiclassical description.  Each Kasner solution visits both of these
regions at the extreme ends of its trajectory.  It is thus reasonable to
identify the past with the unknown region and the future with the
semiclassical regime.   

The derivations below are based primarily on elementary algebra and the
definition of the duality transformations given above.  
However, many cosmologists may want to skip the technical details.
   
\subsection{The boundaries of moduli space}

There are three types of boundaries of the toroidal moduli space which
are amenable to detailed analysis.  The first is the limit in which
eleven-dimensional supergravity becomes valid. We will
denote this limit as 11D. The other two limits are weakly coupled
type IIA and type IIB theories in 10 dimensions. We will call the domain
of asymptotic moduli space which can be mapped into one of these limits,
the safe domain.

\begin{itemize}

 \item For the limit 11D, all the radii must be greater than $\lp$.
Note that for $t\to\infty$ it means that all the radii are much greater
than $\lp$. In terms of the $p_i$'s, this is the inequality $p_i>0$.

 \item For type IIA, the dimensionless coupling constant
$g_s^{IIA}$ must be smaller than 1 (much smaller for $t\to\infty$)
and all the remaining radii must be greater than $\ls$ (much
greater for $t\to\infty$).

 \item For type IIB, the dimensionless coupling constant
$g_s^{IIB}$ must be smaller than 1 (much smaller for $t\to\infty$)
and all the remaining radii must be greater than $\ls$ (much
greater for $t\to\infty$), including the extra radius whose momentum
arises as the number of wrapped M2-branes on the small $T^2$ in the
dual 11D SUGRA picture.

\end{itemize}

If we assume the canonical ordering of the radii, i.e. $p_1\leq p_2\leq
p_3\leq \dots \leq p_k$, we can simplify these requirements as follows:

\begin{itemize}
 \item 11D:    $0<p_1$
 \item IIA:    $p_1<0<p_1+2p_2$
 \item IIB:    $p_1+2p_2<0<p_1+2p_3$
\end{itemize}
To derive this, we have used the familiar relations:
\eqn{fama}{  {L_1\over \lp}=(g_s^{IIA})^{2/3}=
\left({\lp\over\ls}\right)^2=
\left({L_1\over \ls}\right)^{2/3}}
for the 11D/IIA duality ($L_1$ is the M-theoretical circle) and similar
relations for the 11D/IIB case ($L_1<L_2$ are the parameters of the
$T^2$ and $L_{IIB}$ is the circumference of the extra circle):
\begin{eqnarray}
{L_1\over L_2}=g_s^{IIB},\quad
1={L_1\ls^2\over\ls^3}={g_s^{IIB}L_2\ls^2\over\ls^3}=
{L_{IIB}L_1L_2\over\ls^3},\\
\frac{1}{g_s^{IIB}}\left(\frac{\lp}{\ls}
\right)^4=\frac{L_1L_2}{\lp^2}=\frac{\lp}{L_{IIB}}=
(g_s^{IIB})^{1/3}\left(\ls\over L_{IIB}\right)^{4/3}\label{famb}
\end{eqnarray}

Note that the regions defined by the inequalities above cannot overlap,
since the regions are defined by $M,M^c\cap A,A^c\cap B$ where
$A^c$ means the complement of a set.
Furthermore, assuming $p_i<p_{i+1}$ it is easy to show that 
$p_1+2p_3<0$ implies $p_1+2p_2<0$ and $p_1+2p_2<0$ implies
$3p_1<0$ or $p_1<0$. 

This means that (neglecting the boundaries where
the inequalities are saturated) the region outside 
$\mbox{11D}\cup\mbox{IIA}\cup\mbox{IIB}$ is defined simply by
$p_1+2p_3<0$.  The latter characterization of the safe domain 
of moduli space will simplify our discussion considerably.

The invariance of the bilinear form defined above gives an important
constraint on the action of $\Rut_k$ on the moduli space.  For $k=10$ it
is easy to see that, considering the $p_i$ to be the coordinates of a ten
vector, it defines a Lorentzian metric on this ten dimensional space.  
Thus the group $\Rut_{10}$ is a discrete subgroup of $O(1,9)$.  The
direction in this space corresponding to the sum of the $p_i$ is timelike,
while the hyperplane on which this sum vanishes is spacelike. We can
obtain the group $\Rut_9$ from the group $\Rut_{10}$ by taking $p_{10}$ to
infinity and considering only transformations which leave it invariant.  
Obviously then, $\Rut_9$ is a discrete subgroup of the transverse Galilean
group of the infinite momentum frame. For $k \leq 8$ on the other hand,
the bilinear form is positive definite and $\Rut_k$ is contained in
$O(k)$.  Since the latter group is compact, and there is a basis in which
the $\Rut_k$ matrices are all integers divided by $3$, we conclude that in
these cases $\Rut_k$ is a finite group. In a moment we will show that
$\Rut_9$ and {\it a fortiori} $\Rut_{10} $ are infinite. Finally we note
that the \rut{} is a spatial reflection in $O(1,9)$.  Indeed it squares to
$1$ so its determinant is $\pm 1$.  On the other hand, if we take all but
three coordinates very large, then the \rut{} of those coordinates is very
close to the spatial reflection through the plane $p_1 + p_2 + p_3 = 0$,
so it is a reflection of a single spatial coordinate.

\vspace{0mm}

\epsfig{file=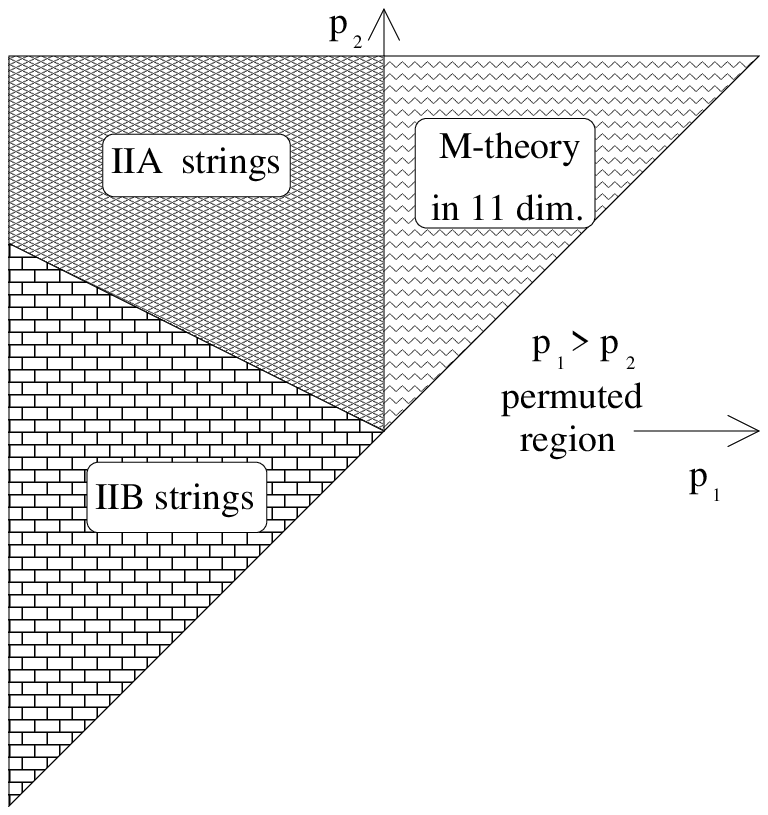}
\nopagebreak
\nopagebreak\newline
\ \ \ \ {\bf Figure 3:} The structure of the moduli space for $T^2$.
\vskip.1in

We now prove that $\Rut_9$ is infinite.
Start with the first vector of $p_i$'s given below and iterate
(\ref{rutdef}) on the three smallest radii (a strategy which we will use
all the time)  and sort $p_i$'s after each step, so that their index
reflects their order on the real line. We get
\eqn{ninefinite}{
\begin{array}{lcr}
(-1,-1,-1,&-1,-1,-1,&-1,-1,-1)\\
(-2,-2,-2,&-2,-2,-2,&+1,+1,+1)\\
(-4,-4,-4,&-1,-1,-1,&+2,+2,+2)\\
(-5,-5,-5,&-2,-2,-2,&+4,+4,+4)\\
{ }&\vdots&{ }\\
(3\times (2-3n),&3\times (-1),&3\times (3n-4))\\
(3\times (1-3n),&3\times (-2),&3\times (3n-2))
\end{array}
}
so the entries grow (linearly) to infinity.
\vskip.2in
\subsection{Covering the moduli space}

We will show that there is a useful strategy which can be used to
transform any point $\{p_i\}$ into the safe domain in the case of $T^k$, $k<9$.
The strategy
is to perform iteratively \rut{s} on the three smallest radii.

Assuming that $\{p_i\}$ is outside the safe domain, i.e.
$p_1+2p_3<0$ ($p_i$'s are sorted so that $p_i\leq p_{i+1}$),
it is easy to see that $p_1+p_2+p_3<0$ (because $p_2\leq p_3$).
As we said below the equation (\ref{rutdef}), 
the \rut{} on $p_1,p_2,p_3$ always increases the total
sum $\sum p_i$ for $p_1+p_2+p_3<0$. But this sum cannot increase
indefinitely because the group $\Rut_k$ is finite for
$k<9$. Therefore the iteration proccess must terminate at some
point. The only way this can happen is that
the assumption $p_1+2p_3<0$ no longer holds, which means that
we are in the safe domain. This completes the proof for $k<9$.

\vspace{3mm}

For $k=9$ the proof is more difficult. The group $\Rut_9$ is infinite
and furthermore, the sum of all $p_i$'s does not change. In fact 
the conservation of $\sum p_i$ is the reason that only points with
$\sum p_i>0$ can be dualized to the safe domain.
The reason is that if $p_1+2p_3\geq 0$, also $3p_1+6p_3\geq 0$
and consequently
\eqn{ninep}{p_1+p_2+p_3+p_4+p_5+p_6+p_7+p_8+p_9 \geq
p_1 +p_1+p_1 + p_3+p_3+p_3+p_3+p_3+p_3\geq 0.}
This inequality is saturated only if all $p_i$'s are equal
to each other. If their sum vanishes, each $p_i$ must then vanish.
But we cannot obtain a zero vector from a nonzero vector
by \rut{s} because they are nonsingular. If the sum $\sum p_i$ is
negative, it is also clear that we cannot reach the safe domain.

However, if $\sum_{i=1}^9 p_i>0$, then we can map the region of moduli space
with $t \rightarrow \infty $ to the safe domain. 
We will prove it for rational $p_i$'s only. This assumption compensates
for the fact that the order of
$\Rut_9$ is infinite.
Assuming $p_i$'s rational is however sufficient
because we will see that a finite product of \rut{s} brings us
to the safe domain. But a composition of a finite number
of \rut{s} is a continuous map from $\IR^9$ to $\IR^9$ so there must be 
at least a
``ray'' part of a neighborhood which can be also dualized to the
safe domain. Because $\IQ^9$ is dense in $\IR^9$, our argument proves the 
result
for general values of $p_i$.

{}From now on we assume that the 
$p_i$'s are rational numbers. Everything is scale invariant so we
may multiply them by a common denominator to make integers. In fact, we choose
them to be integer multiples of three since in that case we will have integer
$p_i$'s even after \rut{s}. The numbers $p_i$ are now integers equal
modulo 3 and their sum is positive. We will define a critical quantity
\eqn{cq}{C=\sum_{i<j}^{1...9} (p_i-p_j)^2.}
This is {\it a priori} an integer greater than or equal to zero 
which is invariant
under permutations. What happens to $C$ if we make
a \rut{} on the radii $p_1,p_2,p_3$? The differences
$p_1-p_2$, $p_1-p_3$, $p_2-p_3$ do not change and this holds for
$p_4-p_5$, \dots $p_8-p_9$, too. The only contributions to
(\ref{cq}) which are changed are from $3\cdot 6=18$ ``mixed'' terms like
$(p_1-p_4)^2$. Using (\ref{rutdef}),
\eqn{rutcq}{(p_1-p_4) \mapsto (p_1-\frac{2s}3) -(p_4+\frac{s}3)=
(p_1-p_4)-s}
so its square 
\eqn{rs}{(p_1-p_4)^2\mapsto [(p_1-p_4)-s]^2=(p_1-p_4)^2 - 2s(p_1-p_4)
+s^2}
changes by $- 2s(p_1-p_4) +s^2$. Summing over all 18 terms we get
($s=p_1+p_2+p_3$)
\eqn{delta}{\Delta C= -2s[6(p_1+p_2+p_3)-3(p_4+\dots+p_9)]+18s^2
=6s^2 + 6\left((\sum_{i=1}^9 p_i)-s\right)=6s\sum_{i=1}^9 p_i.}
But this quantity is strictly negative because $\sum p_i$ is positive
and $s<0$ (we define the safe domain with boundaries, $p_1+2p_3\geq 0$).

This means that $C$ defined in (\ref{cq}) decreases after each
\rut{} on the three smallest radii. Since it is a non-negative integer,
it cannot decrease indefinitely. 
Thus the assumption $p_1+2p_3<0$ becomes invalid after a finite number
of steps and we reach the safe domain.

Now let us turn to the fully compactified case.  As we pointed out, the
bilinear form $I \equiv 2\sum_{i < j} p_i p_j$ defines a Lorentzian
signature metric on the vector space whose components are the $p_i$.  The
\rut{} is a spatial reflection and therefore the group
$\Rut_{10}$ consists of orthochronous Lorentz transformations.  
Now consider a vector in the safe domain.  We can write it as
\eqn{safevec}{(-2, -2 + a_1, 1 + a_2, \ldots,  1+a_9)S,
\qquad S\in\IR^+}
where the $a_i$ are positive.  It is easy to see that $I$ is positive
on this configuration.  This means that only the inside of the light
cone can be mapped into the safe domain.  Furthermore, since $\sum p_i$
is positive in the safe domain and the transformations are
orthochronous, only the interior of the 
future light cone in moduli space can be mapped
into the safe domain.  

We would now like to show that the entire interior of the forward light
cone can be so mapped.  We use the same strategy of rational coordinates
dense in $\IR^{10}$.  If we start outside the safe domain, the sum of the
first three $p_i$ is negative.  We again pursue the strategy of doing a
\rut{} on the first three coordinates and then reordering
and iterating.  For the case of $\Rut_9$ the sum of the coordinates was
an invariant, but here it decreases under the \rut{} of the three
smallest coordinates, if their sum is negative.  
But $\sum p_i$ is (starting from rational values
and rescaling to get integers congruent modulo three as before) a
positive integer and must remain so after $\Rut_{10}$ operations.  
Thus, after a finite number of iterations, the
assumption that the sum of the three smallest coordinates is negative
must fail, and we are in the safe domain.  In fact, we generically enter
the safe domain before this point.  The complement of the safe domain
always has negative sum of the first three coordinates, but there are
elements in the safe domain where this sum is negative.

It is quite remarkable that the bilinear form $I$ is proportional to 
the Wheeler-De~Witt Hamiltonian for the Kasner solutions:
\eqn{wdI}{\frac{I}{t^2}=\left(\sum_i \frac{dL_i/dt}{L_i}\right)^2
- \sum_i\left(\frac{dL_i/dt}{L_i}\right)^2=\frac{2}{t^2}\sum_{i<j}p_ip_j.}
The solutions themselves thus lie precisely on the future light cone in
moduli space. Each solution has two asymptotic regions ($t \rightarrow
0,\infty$ in (\ref{metric})), one of which is in the past light cone and
the other in the future light cone of moduli space.  The structure of the
modular group thus suggests a natural arrow of time for cosmological
evolution.  The future may be defined as the direction in which the
solution approaches the safe domain of moduli space.  All of the Kasner
solutions then, have a true singularity in their past, which cannot be
removed by duality transformations.

Actually, since the Kasner solutions are on the light cone, which is
the boundary of the safe domain,  we must add
a small homogeneous energy density to the system in order to make this
statement correct.  The condition that we can map into the safe domain
is then the statement that this additional energy density is positive.
Note that in the safe domain, and if the equation of state of this matter
satisfies (but does not saturate) 
the holographic bound of \cite{lenwilly}, this energy density
dominates the late time 
evolution of the universe, while near the singularity, it
becomes negligible compared to the Kasner degrees of freedom.  The assumption
of a homogeneous negative energy density is manifestly incompatible with
Einstein's equations in a compact flat universe so we see that the
spacelike domain of moduli space corresponds to a physical situation
which cannot occur in the safe domain.

The backward lightcone of the asymptotic moduli space is, as we have
said, visited by all of the classical solutions of the theory.

To summarize: the U-duality group $\Rut_{10}$ divides the asymptotic
domains of moduli space into three regions, corresponding to the
spacelike and future and past timelike regimes of a Lorentzian manifold.
Only the future lightcone can be understood in terms of weakly coupled SUGRA or
string theory.  The group theory provides an exact M-theoretic meaning for the
Wheeler-De~Witt Hamiltonian for moduli.  Classical solutions of the low
energy effective equations of motion with positive energy density for
matter distributions lie in the timelike region of moduli space and 
interpolate between the past and future light cones.
We find it remarkable that the purely group theoretical considerations
of this section seem to capture so much of the physics of toroidal
cosmologies.

\subsection{Moduli spaces with less SUSY}

We would like to generalize the above considerations to situations 
which preserve less 
SUSY.  This enterprise immediately raises some questions, the first of which is
what we mean by SUSY.  Cosmologies with compact spatial sections have no
global symmetries in the standard sense
since there is no asymptotic region in which one can define the generators.
We will define a cosmology with a certain amount 
of SUSY by first looking for Euclidean
ten manifolds and three form field configurations which are solutions of the
equations of 11D SUGRA and have a certain number of Killing spinors.
The first approximation to cosmology will 
be to study motion on a moduli space of
such solutions.
The motivation for this is that at least 
in the semiclassical approximation we are guaranteed
to find arbitrarily slow motions of the moduli.  
In fact, in many cases, SUSY
nonrenormalization theorems guarantee that the semiclassical 
approximation becomes valid for
slow motions because the low energy effective Lagrangian of the
moduli is to a large
extent determined by SUSY.  There are however a number of 
pitfalls inherent in our approach.
We know that for some SUSY algebras, the moduli space of 
compactifications to four or six 
dimensions is not a manifold.  New moduli can appear at 
singular points in moduli space
and a new branch of the space, attached to the old one at the 
singular point, must
be added.  There may be cosmologies which traverse from one branch to 
the other in the
course of their evolution.  If that occurs, there will be a point at
which the moduli space approximation breaks down.
Furthermore, there are many examples of SUSY vacua of M-theory which 
have not yet been 
continuously connected on to the 11D limit, even through 
a series of ``conifold'' 
transitions such as those described above \cite{islands}.   
In particular, it has been suggested that there
might be a completely isolated vacuum state of M-theory \cite{evadine}.
Thus it might not be possible to imagine that all cosmological solutions
which preserve a given amount of SUSY are continuously connected to the
11D SUGRA regime.

Despite these potential problems, we think it is worthwhile to 
begin a study of compact, 
SUSY preserving, ten manifolds.  Here we will only study 
examples where the
three form field vanishes.  The well known local condition for a 
Killing spinor, 
$D_{\mu} \epsilon = 0$, has as a condition for local integrability 
the vanishing
curvature condition
\eqn{killspin}{R_{\mu\nu}^{ab} \gamma_{ab} \epsilon = 0}
Thus, locally the curvature must lie in a subalgebra of the Lie 
algebra of $Spin (10)$ which
annihilates a spinor.  The global condition is that the holonomy 
around any 
closed path must lie in a subgroup which preserves a spinor.  
Since we are dealing with
11D SUGRA, we always have both the $16$  and $\bar{16}$ representations
of 
$Spin (10)$ so SUSYs 
come in pairs.

For maximal SUSY the curvature must vanish identically and the space 
must be a torus.
The next possibility is to preserve half the spinors and this is
achieved 
by manifolds 
of the form $K3 \times T^7$ or orbifolds of them by freely acting
discrete 
symmetries. 

We now jump to the case of 4 SUSYs.   To find examples, it is convenient to
consider the decompositions $Spin (10) \supseteq
Spin (k) \times Spin (10-k) $.

The $16$ is then a tensor product of two lower dimensional spinors.  For
$k=2$, the holonomy must be contained in $SU(4) \subseteq Spin (8)$ in
order to preserve a spinor, and it then preserves two (four once the
complex conjugate representation is taken into account). The corresponding
manifolds are products of Calabi-Yau fourfolds with two tori, perhaps
identified by the action of a freely acting discrete group.  This moduli
space is closely related to that of F-theory compactifications to four
dimensions with minimal four dimensional SUSY. The three spatial
dimensions are then compactified on a torus. For $k=3$ the holonomy must
be in $G_2 \subseteq Spin (7)$.  The manifolds are, up to discrete
identifications, products of Joyce manifolds and three tori.  For $k=4$
the holonomy is in $SU(2) \times SU(3)$.  The manifolds are free orbifolds
of products of Calabi-Yau threefolds and K3 manifolds.  This moduli space
is that of the heterotic string compactified on a three torus and
Calabi-Yau three-fold. The case $k=5$ does not lead to any more examples
with precisely 4 SUSYs.

It is possible that M-theory contains U-duality transformations which map
us between these classes.  For example, there are at least some examples
of F-theory compactifications to four dimensional Minkowski space which
are dual to heterotic compactifications on threefolds.  After further
compactification on three tori we expect to find a map between the $k=2$
and $k=4$ moduli spaces.

It is clear that the metric on the full moduli space still has Lorentzian
signature in the SUGRA approximation.  In some of these cases of lower
SUSY, we expect the metric to be corrected in the quantum theory.  
However, we do not expect these corrections to alter the signature of the
metric.  To see this note that each of the cases we have described has a
two torus factor.  If we decompactify the two torus, we expect a low
energy field theoretic description as three dimensional gravity coupled to
scalar fields and we can perform a Weyl transformation so that the
coefficient of the Einstein action is constant.  The scalar fields must
have positive kinetic energy and the Einstein term must have its
conventional sign if the theory is to be unitary.  Thus, the
decompactified moduli space has a positive metric.  In further
compactifying on the two torus, the only new moduli are those contained in
gravity, and the metric on the full moduli space has Lorentzian signature.

Note that as in the case of maximal SUSY, the region of the moduli space
with large ten volume and all other moduli held fixed, is in the future
light cone of any finite point in the moduli space.  Thus we suspect that
much of the general structure that we uncovered in the toroidal moduli
space, will survive in these less supersymmetric settings.

The most serious obstacle to this generalization appears in the case
of 4 (or fewer) supercharges.  In that case, general arguments do not
forbid the appearance of a potential in the Lagrangian for the moduli.
Furthermore, at generic points in the moduli space one would expect
the energy density associated with that potential to be of order the
fundamental scales in the theory.  In such a situation, it is difficult
to justify the Born-Oppenheimer separation between moduli and high
energy degrees of freedom.  Typical motions of the moduli on their
potential have frequencies of the same order as those of 
the ultraviolet degrees of freedom.  In section 7 we will try
to present a solution to this conundrum.

\subsection{Chaotically avoiding SUSY}

The considerations of this section also allow us to achieve some insight
into the problem of why \mth\ has not chosen to set in one of its stable
highly supersymmetric vacua in the world we observe.  The discussion
which follows is completely rigorous on the branches of moduli space
with 16 or more SUSYs.  It is probably valid for 8 SUSYs as well, for
in that case the moduli space exists although its topology and metric
are not determined by classical considerations.  Nonetheless, all known
extreme regions of the moduli space have the properties we will use
below. 

The key point is that our analysis of extreme regions of moduli space
showed a monotonic flow from the unsafe to the safe regions.  We have
neglected extreme regimes corresponding to partial decompactification,
and also the motion of the other moduli, and of the non modular degrees
of freedom which surely dominate the energy density in regimes where the
universe has expanded a lot.  In fact, inclusion of these other degrees
of freedom reinforces the conclusion that the universe will always end
up in the safe domain.

Horne and Moore \cite{greg} have shown that motion on the full moduli
space (as opposed to its Kasner subspace) is chaotic.  Furthermore, 
the Euclidean metric on the subspace of moduli with unit 
spatial volume has finite volume in the metric on moduli space, 
which means that the extreme regions
of this space (which correspond to partial decompactifications) have
vanishingly small measure.  The chaotic nature of the motion, as well
as the fact that the moduli are, at least at late times, coupled to a
stochastic radiation bath, imply that the generic cosmological solution
will in fact sample regions of the moduli space in proportion to the
measure defined by the kinetic energy of the moduli.  
In particular, partial decompactifications, which are of of measure
zero on the moduli space, will not be generic final states of the
cosmological evolution.

We conclude that the generic cosmological solution in these
supersymmetric regions of the moduli space will asymptote to a ten or
eleven dimensional universe filled with radiation.  All of low energy
physics is weakly coupled, there are no finite energy scales apart from
the Planck or string scales, and there are no apparent candidates for
long lived nonrelativistic particles\footnote{I do not see any source
for a population of large and therefore long lived black holes.}.  
It seems safe to conclude that
none of these model universes could ever contain galaxies.  Thus, if we
are willing to entertain the very weak form of the anthropic principle
which claims that galaxies are necessary for intelligent life, we can
find an explanation of why we do not live in a universe with 8 or more
SUSYs. 

I do not claim to find this a completely satisfactory resolution of the
question.  On the one hand, I maintain that this sort of use of
anthropic reasoning is scientifically valid.  That is, we appear, in
\mth, to be
faced with a model of physics which predicts the possibility of
alternate universes which do not resemble what we observe.  I have tried
to give an honest account of what happens to a generic universe of this
sort (within the class with maximal SUSY) and found that it lacks what
would appear to be a very weak requirement for the existence of life.  I
did not have to speculate about unknown results in extra universal
biology to come to this conclusion.  
On the other hand, one might wish for a sharper distinction between
our own universe and these unobservable ones.  Wouldn't it be nicer if
they all suffered some sort of satisfyingly final cosmic catastrophe and
sank back into the ultraviolet muck of creation?\footnote{We will see
something of the sort happening to another class of undesirable
universes in the next section.}.  Or perhaps one could, with a more
comprehensive knowledge of \mth, argue that generic cosmological
solutions of the whole theory do not end up in the maximally SUSY
regions. 

One direction in which to search for such an argument has to do with
inflation.  I have purposely avoided mentioning that cosmologies which
remain on the moduli spaces with 8 or more SUSYs cannot inflate.  The 
obvious retort to such a remark is that inflation could have occurred
somewhere else in configuration space, and the system could then have
rolled down to the moduli space.  One cannot investigate the probability
of such a motion without a much more thorough understanding of \mth\ than
we now possess.   So the {\it galactothropic} explanation of the
absence of SUSY ground states is the best we can do at the moment.  
Perhaps it will be the best we can ever do.

\subsection{Against inflation}

To an audience of astroparticle physicists the suggestion that inflation
might not be a necessary feature of our explanation of the universe is akin
to heresy.  I therefore thought it would be amusing to insert some
speculations here about alternative ways to solve the cosmological conundra
which led to the invention of inflation.  Those of my readers who actually
attended these lectures will not that I did not actually present this 
material.  Let me assure you that it was only for lack of time, and not
because I was afraid of being mauled by an angry crowd of true believers.

To begin our trek down the path of heterodoxy let me attack the common
wisdom about the horizon problem.  This is the observation that in
conventional Big Bang cosmology, the horizon at early times is much smaller
than the backward extrapolation of our current horizon.  Thus, regions
of the universe that we can observe today were out of causal contact.
How one then asks can their contents be in thermal equilibrium at a
uniform temperature?  I would like to contend that the M theorist's answer
might be ``very easily''.  Local field theory is only an approximation
to \mth.  At sufficiently high energies it is clear that locality
breaks down in some way.  The typical high energy state in perturbative
string theory is an extremely long single string.  Beyond the perturbative
approximation, large branes of other dimensions may be relevant.
Although brane interactions are local on the brane ({\it e.g.} strings
split and join at points in spacetime) this does not seem to be an
argument which forces one to conclude that the correct state of 
the string is unlikely to be a typical member of the ensemble of strings
with given energy (as one argues for a quantum field theory in a Big Bang
cosmology when one says that the fields in causally disconnected regions
have not had a chance to thermalize).  If the system is in thermal
equilibrium at very high energy, and if the expansion is slow enough, then
it will remain at equilibrium at lower energies.

Another argument against naive locality at the fundamental scale (which
might be much lower than $10^{19} $~GeV) has to do with black holes.
Once the typical energy and impact parameter in particle collisions
are such that black hole formation is common, the spacetime geometry
is distorted in a way which modifies the naive causality arguments.  If
we believe that black hole evaporation is a unitary process, then standard
causality arguments are only valid outside black hole horizons (I am assuming that if the universe is closed, then its radius is much larger than
the relevant black hole horizons).  All states associated with a given
black hole are in thermal equilibrium with each other, and black holes
will tend to coalesce, bringing more and more of the system into equilibrium.   

The claim then is that the horizon problem is not a problem (I am
being deliberately provocative here -- I don't know whether I believe
these arguments).  Rather, the principle of thermodynamic equilibrium
{\it  i.e.} that systems tend to be in typical states consistent with
their energy content is more fundamental than the causality principle
applied to a simple averaged classical geometry and a model of its matter
content as localized particles interacting via local field theory.

Similar remarks apply to the monopole problem, at least in those regions
of moduli space where there is no grand unified group below the fundamental
scale.  Monopoles then belong to the high energy theory, and the 
conventional field theoretic estimates (again based on causality) of their
abundance are incorrect.

One can make an even more convincing attack on
 the arguments for the flatness problem.
This puzzle is based on the model of a homogeneous isotropic universe.
This should properly be regarded as a phenomenological model rather than
a fundamental starting point for cosmology.  Indeed although descriptions
of inflationary cosmology usually start from standard Robertson-Walker
ideology, they in fact reject that ideology.  Homogeneity and
isotropy arise as late time fixed point behavior.  However, if one is going
to start from more general initial conditions, one can get rid of the
flatness problem in a simpler way.  Indeed, I have argued above that a
more fundamentally motivated approach to cosmology might start from
geometries (and configurations of other fields) on a moduli space of 
static classical solutions of the SUGRA equations.  It is a generic
feature of such models, that unless the energy density is allowed to be
negative, the universe evolves monotonically toward large volume.  Thus
spatial curvatures (and many of the Calabi-Yau manifolds on these moduli
spaces are curved) are generally evolving towards zero without any fine
tuning.  The general cosmological solution for motion on a moduli space
of geometries, coupled to positive energy density matter evolves toward
zero spatial curvature if we are only willing to wait long enough.
The only real issue left among the conventional cosmological
puzzles is the Entropy Problem.

To explain this in more detail let us consider a simple example of
the kind of model we are discussing. Consider the moduli space of
solutions of weakly coupled heterotic string theory compactified on
a three torus large compared to the string scale, times a Calabi-Yau 
threefold.   Let us agree to ignore the phenomenological problems with the
dilaton which make this regime problematic as a model of the real
world.  The Friedmann equation for this model has the form
\eqn{torfried}{m_P^2 (\dot{a} /a)^2 =  m_P^4 [b/a^6 + d /a^4  + 
e / m_P a^3 + \Lambda].}
$a$ is the scale factor of the three torus, and $b$, $d$, $e$ and $\Lambda$
represent the contributions to the energy density of the moduli,
radiation, nonrelativistic matter, and a cosmological constant, all measured
in Planck units.
We choose conventions such that $a=1$ is the present scale factor.  
The volume of the torus is $a^3 V_0$, where $V_0$ is the volume today.
Observation tells us that the periods of the three torus are 
of the same order as, or larger than our horizon volume, whose size is
$10^{60}$ Planck units.  We neglect processes which convert one form of
energy density into another and do not attempt to explain why all of
the constants $d,e$ and $\Lambda$ 
are within an order of magnitude or so of each other.  

The moduli of the torus are the ratios of its periods, the angles
between the different toroidal directions, Wilson lines for the
heterotic gauge fields and ``Wilson two surfaces'' for the antisymmetric
tensor potential of heterotic string theory.  These evolve as a
nonlinear sigma model of Goldstone type.  The analysis of \cite{preva}
implies that motion on this space stops early in the history of the
universe, its kinetic energy being converted into a gas of momentum
modes of the corresponding fields, which contributes to the constant
$d$.   The torus then expands indefinitely with fixed shape.  
Thus, if we wait long enough, all remnant of the finiteness of space is
wiped out, without fine tuning of initial conditions.  In cases where
the moduli space in question is a family of curved Calabi-Yau
spaces, the same analysis applies and the spatial curvature is erased
without any fine tuning.  

The real difficulty for this solution of the flatness problem is simply
that if we wait long enough for spatial finiteness and curvature to be
stretched away, there may not be enough matter and radiation in our
model to account for the universe we observe.  This is what is commonly
referred to as the Entropy problem in the literature of inflationary
cosmology.   The models we are discussing show that it is logically
separate from the flatness problem, which is rather specific to
homogeneous isotropic models, where the spatial geometry at each instant
is not a static solution of the Einstein equations.  In these models,
generic initial values for curvature would have long ago led to a
curvature dominated regime of expansion and substantially modified much
of cosmic history.  In models based on moduli, generic initial
conditions would not have changed the expansion rate very much at late times
and would probably not show up in local physics.  Their discrepancy with
observation would simply come from the absence of evidence for global
structure or anisotropy in the background geometry.

Another way to phrase the Entropy Problem
is the discrepancy between the universe's
energy content and its size at the ``moment of the Big Bang''.  
If one follows the conventional Robertson-Walker cosmology back to
the Planck energy density, then the linear 
size of our horizon volume at that time is $10^{27}$ Planck units.
The size of any closed universe would have to be larger than this.
I do not have any explanation of this large pure number in the present
context.  In inflationary cosmology it is solved by creating the matter
and radiation {\it after} a period of inflationary expansion.

At the level of the semiclassical analysis we have done, there does not
seem to be any strong objection to such initial conditions.  We have
emphasized that the semiclassical treatment of the moduli requires only
that the volume of the universe be large.  At energies above the Planck
scale, there will be new terms in the equations of motion of the moduli
representing their interaction with the full set of high energy degrees 
of freedom of \mth.  But in principle one could imagine following
the evolution back to a Planck size for the whole universe before the
semiclassical approximation breaks down.  The statement of the Entropy
Problem at that time would be that the energy density was many orders
of magnitude higher than the Planck scale.  Is there some principle
which prevents this?

It would be nice to find one, because one would like to have a clean
reason for rejecting alternatives to inflation.  Alternatively, it would
be interesting to find an explanation of this large number, and to take
the anti-inflationary cosmology more seriously.  In the latter event
one would be required to come up with an explanation for the fluctuations
in the cosmic microwave background at least as convincing as that provided
by inflationary models.  

What should the serious cosmologist take away from this discussion?
I hardly hope or wish to convince anyone to abandon the inflationary
paradigm.  However, I think it is salutary to recognize that many of
the theoretical arguments which one thinks of as the basic {\it raison
d'etre} of inflationary cosmology, are on rather shaky ground in the light 
of current theory. The clear cut triumphs of inflation are reduced to
two: the explanations of the entropy of the current universe and of the
fluctuations in the microwave background.  

In the next section, we will abandon this heresy and pursue a more orthodox
path.

\subsection{Conclusions}

We argued that the supersymmetric moduli of \mth\ were the natural 
semiclassical variables which provide the clock for cosmology.  Our
argument was based on the naive Wheeler-Dewitt quantization of gravity
but we presented some evidence that the general structures assumed
in that quantization were more robust than their derivation from a
low energy effective theory would have led us to believe.  We showed that
duality transformations resolve some but not all cosmological 
singularities, and provided a first draft of an argument for the
absence of highly supersymmetric vacuum states of \mth\ in the
list of Natural Phenomena in the Real World.
We also briefly explored a heterodox, noninflationary, approach to
cosmology which resolves some but not all of the problems that inflation
was invented to solve.

\section{Moduli and Inflation}

\subsection{Introduction}

In this lecture we will finally start to discuss more realistic sectors
of M theoretic
cosmology.  As I have warned you several times, this area is still under 
development and 
there is no justification for trying to build detailed models which can
be compared to
observation.  Indeed, towards the end of my presentation I will describe 
my own favorite
scenario for cosmology in \mth.  It turns out that its viability
depends heavily on
numerical factors of order one which cannot be reliably calculated at 
present.  Such factors 
in fundamental quantities have a tendency to get raised to high powers
in a cosmological context
({\it e.g.} the widths of unstable states depend on the cube of their 
masses and the square
of their couplings.  These in turn might be estimated by formulae which 
depend on high powers of 
some fundamental scale.  Mistakes of order one can thus be amplified.).  
Also, 
experience with weakly coupled string theory shows that order of
magnitude estimates can
miss factors like $16 \pi^2$.  Our fundamental contention about \mth\ is 
that neither 
the true vacuum
state nor the point where inflation takes place are likely to sit in one
of the weakly
coupled or large radius regimes where systematic calculations can be 
done.  Thus, we are
unlikely to be able to extract detailed numbers from \mth\ until we 
learn a lot more about
the nonperturbative formulation of the theory.  In this situation it 
seems wisest to try
to investigate very general problems, and that is what we will try to
do.  
I will deviate
{}from this formula only towards the end of my lectures, in order to
present the amusing
scenario that I favor.

\subsection{Moduli as inflatons?}

In view of our discussion in the previous section, one might have
thought that the appropriate 
title for this section was ``Cosmology on the Moduli space with 4 
SUSYs''.
At first sight, 
the phrase in quotes does not appear to 
make any sense.  \mth\ has no global internal
symmetries -- all of its
symmetries are residual gauge symmetries which leave some class of 
configurations invariant\footnote{As usual, there are two arguments
for this, one based on SUGRA, the other on perturbative string theory.
Their agreement is taken as evidence that the statement is exact.
The SUGRA argument is simply that all symmetries of SUGRA are
diffeomorphisms, thus gauge symmetries.  Global symmetries arise only
as diffeomorphisms which leave invariant the asymptotic behavior of
the noncompact portion of space. All other symmetries are gauged.
In perturbative string theory an internal symmetry would arise as
a symmetry of the superconformal field theory describing the internal
space.  One can show, \cite{lance}, that a continuous global symmetry
implies the existence of a Kac-Moody current algebra in the superconformal
field theory (basically just Noether's theorem plus conformal
invariance -- up to technicalities).  
The Kac-Moody currents can be used to construct vertex operators for
massless gauge bosons.  }.
With only 4 SUSYs, supersymmetry alone permits a superpotential on the 
space of chiral 
superfields.  The full effective potential is the sum of a term coming 
{}from the
so-called D-terms of continuous gauge groups, and a term coming from the 
superpotential\footnote{See Keith Olive's lectures at this school for a 
concise introduction to four dimensional SUSY, chiral superfields, 
superpotentials, D terms, {\it etc.}.}.
The D-terms are positive, and the moduli space of fields on which they 
vanish can be parametrized
in terms of gauge invariant composite fields.  The superpotential can be 
viewed as
a function on this space.  The only symmetries which act on the
composites 
are discrete gauge
symmetries\footnote{The only difference between gauged and nongauged 
discrete symmetries
{}from a practical point of view is the absence of stable domain walls for 
gauged 
discrete
symmetries.  }.  In most cases, a 
discrete symmetry cannot imply the 
vanishing of
a function on an entire submanifold 
(we will explore the exception below).  

The apparent implication of this is that
the phrase ``moduli space of \mth\
compactifications with 4 SUSYs''
has no apparent meaning.
There is no moduli space
in the true sense of the word (with the
exception noted in the last parenthesis).
Nonetheless, the authors of 
\cite{binetry} proposed and \cite{preva}
and others explored the idea, that moduli
of such compactifications were the 
natural inflaton candidates in 
string/\mth.  Note that inflatons, by
their nature, must have a potential
so the idea of moduli as inflatons
is truly oxymoronic.

However, I hope to demonstrate for you
that this idea is not at all idiotic,
and that it has many attractive features.
The original proposals were based on
string perturbation theory.  Here the
idea of a moduli space of 
{\it quadrisusic\footnote{A recently
rediscovered ancient Latin word meaning:
having four supersymmetries.}} 
compactifications makes perfect 
mathematical sense.  At string tree
level, a vacuum state is characterized
as a conformal field theory with
certain extra properties.  There is
an exact theorem which guarantees the
existence of continuous families of
solutions to this constraint.  The most
famous among them are those which
correspond to compactification of
the heterotic string on a CY 3-fold
with the standard embedding of the 
spin connection of the manifold in the
gauge group.  Here the theorem follows
{}from the fact that the same conformal
field theories can be used to compactify
Type II string theories to four 
dimensions, preserving 8 spacetime
SUSYs.  The extra spacetime SUSY 
guarantees the existence of moduli.
The heterotic and Type II theories
compactified on these backgrounds differ
at the one loop level and beyond, and
the heterotic theory has only 4 SUSYs.
Nonetheless, to all orders in the loop
expansion, no superpotential is generated
on the tree level moduli space 
in the heterotic theory.  Indeed, the
heterotic coupling, like a generic 
gauge coupling, can be viewed as the
real part of a chiral superfield $
S = {8\pi \over g_S^2} + i \theta$,
whose imaginary part is an axionlike
field called the model independent
string axion.  This field arises by a 
duality transformation on a second rank
antisymmetric tensor gauge field.
As a consequence, to all orders in 
perturbation theory there is a 
continuous  shift symmetry $S \rightarrow
S + i a$.  This symmetry, combined with
holomorphy, forbids any perturbative
correction to the superpotential.

The idea behind most previous work on the
subject was that the real world 
corresponds to a point in moduli space
where the perturbative estimates of the
superpotential were correct.  
The string coupling was supposed to 
correspond more or less to the 
perturbative gauge couplings we see in
nature, or to be related to them by
simple group theoretical factors. The
superpotential on the perturbative moduli
space was then much smaller than the
fundamental scales of the theory, and
it made sense to think about an 
approximate moduli space.  

This set of ideas had a number of related
difficulties.  The first was the Dine
Seiberg problem \cite{ds}.  These authors
made the simple observation that for
most functions, the leading asymptotic
formula in some extreme region (here
the weak coupling region) is monotonic
and does not have minima\footnote{
Exceptions to this are somewhat 
pathological.  The leading asymptotic
behavior could contain a factor
$\sin (1/g^2)$ which has an infinite
number of more and more closely spaced
minima as one approaches the weak 
coupling regime.  }.  
There have been two mechanisms proposed for stabilizing
\mth\ in the weak string coupling regime, which go under the names 
of K\"ahler stabilization \cite{bdcoping} and racetrack models
\cite{krasnikov}.  Both imply that, although the couplings are weak,
many quantities cannot be calculated in a systematic expansion.

A related cosmological problem with the weak coupling regime was
pointed out by Brustein and Steinhardt \cite{brustein}.  
There is a distinct possibility that the universe would ``overshoot''
a weak coupling minimum and evolve into the regime of extreme weak
coupling where \mth\ is in violent disagreement with observation.

When combined with Witten's analysis \cite{horwitb} of the possible
resolution of the discrepancy in the weak coupling prediction 
of the ratio between the unification and Planck scales, these
observations compel one to consider the possibility that weakly coupled
string theory is not a good description of nature.  A somewhat better
starting point is the 11D SUGRA analysis begun in \cite{horwita}
The analyses of
\cite{horwitb} and \cite{horwitc} indicate that
\begin{itemize}
\item In the regime of moduli determined by the fit to the unified
coupling strength and the four dimensional Planck mass, the volume of
the Calabi-Yau manifold on the brane where the standard model lives
is not really in the regime where the SUGRA expansion can be trusted.
However, the small size of the four dimensional effective coupling,
combined with holomorphy, is enough to guarantee the usual tree level 
unification relations between standard model couplings.  This gives rise
to a situation similar to that hypothesized in the K\"ahler stabilization
mechanism, where holomorphic quantities can be calculated reliably but
the K\"ahler potentials of chiral fields are unknown.  
\item Witten's hypothesis that the coupling of the gauge fields on the
second brane is strong, and gives rise to a gaugino condensate whose
magnitude is of order the unification scale (which is also the
fundamental 11D Planck scale) induces too high a scale of SUSY breaking
on the standard model brane.  We will discuss a resolution of this
problem below.
\item In the analysis of \cite{horwitc} 
the SUSY breaking F term comes from the modulus which 
parametrizes the radius of the single large
dimension transverse to the Ho\v rava-Witten ninebranes.  To leading order
in the SUGRA expansion this leads to no-scale SUSY breaking with
vanishing cosmological constant, and also gives rise to degenerate
squarks\footnote{The degeneracy in mass of the squarks is a desirable
phenomenological feature.  To the extent that it is valid it eliminates unwanted flavor changing neutral currents which threaten the viability of
generic SUSY models.  This success of the scenario is mitigated by the
failure to stabilize the radial mode.  The terms necessary to stabilize
the radius come from corrections to its K\"ahler potential.  Similar 
corrections could ruin the degeneracy of squarks.}.  
\item The radial mode is not stabilized in this approximation and we
have a sort of Dine-Seiberg problem within the SUGRA approximation.
It is unclear how many of the good features of the model will survive
the resolution of this problem.  It is clear that the vanishing
cosmological constant will not.

\end{itemize}

In short, this scenario is better than perturbative string theory, but
not without its own flaws.  
On the other hand, the observation that gauge theories
arise on branes of finite codimension is generic in \mth\ and leads one
to expect that Witten's explanation of the Planck and unification scales
is a correct one.  

At first sight, the above conclusions would seem to rule out the idea of
modular inflation.  If we are in the strong coupling regime and
there is no reason for the superpotential to be small then what is our
excuse for separating the moduli out from all the other variables of
\mth?  What does the word moduli mean in the strong coupling regime
with only four SUSYs?
Worse, one of the points of \cite{preva} was that within the
context of modular inflation, the energy scale during inflation is
predicted to be near the unification scale.  In Witten's scenario, this
scale is identified with the fundamental scale of quantum gravity and it
seems unreasonable to use any sort of effective field theory
description to describe this situation.

In fact, I claim that the Ho\v rava-Witten scenario and Witten's use of it
to explain the ratio between $m_P$ (the four dimensional Planck
scale $\sim 2 \times 10^{18}$~GeV) and $M$ (the unification scale $\sim
2 \times 10^{16}$~GeV) may resolve all of
these problems.  The key is that the higher dimensional theory has more
SUSY than the effective theory below the KK scale.  The higher
dimensional SUSY is broken by the branes, but if the bulk volume is
large then this
breaking can be ignored for some purposes.  In particular, we can
identify the moduli space as that of the higher dimensional theory.
Thus, in such scenarios, a clearcut notion of approximate 
moduli survives at all energy
scales, as long as we remain in a regime where the compact volume is large.
We will call these approximate moduli the {\it inflamoduli} to
distinguish them from certain fields we will discuss below, which get
their potential only from lower energy physics.

Note that this is all compatible with the existence of a superpotential of
order $M^3$ for the inflamoduli, and indeed this order of magnitude is
reasonable for fields which parametrize properties of the bulk higher
dimensional theory {\it only if there is enhanced SUSY in the bulk}.
Otherwise we would have expected the effective superpotential of the
moduli to contain a factor of the volume of the internal space.
On the other hand, if the superpotential comes only from the vicinity of
the branes, it has, by dimensional
analysis, the form
\eqn{supot}{W = M^3 w(\theta_a)} 
where $\theta_a$ are dimensionless parameters characterizing 
the internal geometry.
On the other hand, the kinetic term for these zero modes, just 
like the Einstein term
for the zero modes of the gravitational field, is proportional to 
the volume $V_7$ of the 
internal manifold, and has the form
\eqn{kin}{M^9 V_7 \sqrt{-g} G_{ab} (\theta) \nabla \theta_a \nabla \theta_b.}
Note that $M^9 V_7 = m_P^2 = {1\over 8\pi G_N}$ is, 
as the notation indicates, the same
coefficient which multiplies the Einstein action.  Furthermore, although 
the volume
$V_7$ is itself a modulus, when we pass to the Einstein conformal frame 
in which $V_7$ is replaced 
by its vacuum value, the kinetic term of the moduli is rescaled 
in precisely the same manner
as the gravitational action.   It is then natural to define canonical
scalar fields 
by $\phi_a = m_P \theta_a$.  Their action has the form
\eqn{action}{\int \sqrt{-g} [G_{ab} (\phi /m_P) 
\nabla \phi^a \nabla \phi^b - {M^6 \over m_P^2} v(\phi /m_P)].}

Now let us examine the implications of a Lagrangian of this form for
inflationary cosmology.
The slow roll equations of motion derived from this action are
\eqn{slowroll}{3H d \phi^a /dt = - {M^6\over m_P^2} G^{ab} 
{\partial v \over \partial\phi^b}.}
and lead to the equation
\eqn{vdot}{d v /dt = {M^3 \over 3 m_P^2 \sqrt{v}} 
\partial_a v G^{ab} \partial_b v.}
where $\partial_a$ refers to the derivative with respect to the 
dimensionless variable $\theta^a$.
We have also used the slow roll expression for $H$ in terms of the potential.
{}From \ref{vdot} we immediately derive an expression for the number 
of $e$-foldings
\eqn{efold}{N_e = 3 \int {v\over \partial_a v 
G^{ab} \partial_b v} \partial_c v d\theta^c.}
where the integral is over the trajectory in moduli space that 
the system follows
during the time interval when the slow roll approximation is valid.  
We see that in order to
obtain a large number of $e$-foldings we need a potential which is 
flat in the sense that
$|\partial v|/v \sim 1/N_e$.  The phenomenologically 
necessary $N_e \sim 60$ can be achieved with
only a mild fine tuning of dimensionless coefficients.  
Correspondingly, the conditions on the
potential which ensure the validity of the slow roll approximation 
are order one conditions on
the derivatives of the potential and do not contain any exponentially
small dimensionless numbers.

An additional feature of modular dynamics, which provides extra
frictional damping of the motion of the moduli, was discovered in
\cite{preva}.  If we completely ignore the potential on moduli space, it
is still an interacting nonlinear system.  In \cite{preva} the equations
for small fluctuations of the modular field theory around a solution of
the equations of motion (without potential) for the zero modes, was
studied, and an unstable mode was found.  This was interpreted as an
efficient mechanism for converting kinetic energy of the zero modes into
energy of a gas of nonzero modes.  It was estimated that the zero modes
were effectively brought to a halt by this mechanism in less than a
Lyapunoff time of the chaotic motion on moduli space.
In the inflationary context, this
mechanism will act as a source of friction which should make inflation
much more probable.  In particular, it is an avenue in which the large
dimension of the moduli space (which can be a number of order $10^2$)
could effect inflation, by providing a large number of degrees of
freedom for efficient frictional damping of the zero mode motion.  This
is a topic which has not been investigated and deserves much more
thorough study.

The fact that actions of the form (\ref{action}) give rise to 
inflation with minimal fine tuning, 
and that such actions naturally arise for moduli in string theory 
was pointed out in \cite{preva}.
The general point that moduli might provide the flat potentialled, 
weakly coupled fields 
necessary to inflation was first made in \cite{binetry}.  Here we note 
that in brane scenarios,
it is the {\it bulk inflamoduli} which play this role.   There may also be
moduli associated with branes, but they will have a natural scale
$M$ and have a quite different role to play.

Another pleasant surprise awaits us when we plug the potential 
{}from (\ref{action}) into the
standard formula for the amplitude of the primordial energy density 
fluctuations generated
by inflation.  Up to numbers of order one we find
\eqn{deltarho}{{\delta\rho \over \rho} \sim N_{\lambda} (M/m_P)^3 \sim 10^{-5}}
where the numerical value comes from the measured cosmic microwave
 background fluctuations, and $N_{\lambda} \sim 50$.
This gives $M \sim (2/10)^{1/3} \times 2 \times 10^{16}$~GeV, which, 
given the crudeness of the
calculation, is the unification scale.   To put this in the most 
dramatic manner possible, we
can say that a brane scenario of the Ho\v rava-Witten type, given 
the unification scale as
input, {\it predicts the correct amplitude for inflationary 
density fluctuations}.  Furthermore,
the whole scenario only makes sense because of the same large volume 
factor that underlies
Witten's explanation of the ratio between the Planck and unification 
scales.  This is necessary 
at a conceptual level to understand why it is sensible to think 
about a modulus with a super
potential of order the fundamental scale, and at a phenomenological 
level to understand the
magnitude of the density fluctuations.  

A detailed calculation of the fluctuation {\it spectrum} as opposed to
its absolute normalization requires more knowledge of the potential $v$
than we possess.  A crucial question (posed during my lecture by
Andre Linde) is how natural the phenomenologically
necessary flat spectrum is in this context.  I leave it as an exercise
for the enterprising student.

Although it has no connection with our discussion here I cannot 
resist pointing out the other
piece of evidence for a scale of the same order as $M$.  Any theory of 
the type we are discussing
would be expected to contain corrections to the standard model Lagrangian 
of the form (in superfield notation) ${1\over M} L L H^2 $, which gives 
rise to neutrino masses.
It is a matter of public record \cite{superK} now that such masses 
exist, with an estimated value
for $M$ between $.6$ and $1.8 \times 10^{15}$~GeV.  Although this 
is an order of magnitude shy
of the unification scale I believe the uncertainties in 
coefficients of order one in dimensional
analysis could easily make up the difference.  If not, we will have 
the interesting problem
of explaining the existence of two close but not identical energy 
scales in fundamental physics.
\cite{wilczeketal}.  

We also want to note that this scenario for inflation does not 
suffer from the runaway problem pointed out by Brustein and Steinhardt 
\cite{brustein}.  These authors noted that the inflationary vacuum energy
is much larger than the SUSY breaking scale.  Furthermore, the minimum of
the effective potential was assumed close to the region of weak string
coupling.  There was then a distinct possibility that the inflaton field would
overshoot the small barrier separating it from the extreme weak coupling
regime where string theory is incompatible with experiment.  In the present
scenario, the coupling is not assumed to be weak (nor the volume extremely
large).  Furthermore the inflationary potential has nothing to do with
SUSY breaking.   There is no runaway problem at all.

The authors of the papers in \cite{preva} agonized over 
the discrepancy between the
unification scale and the scale of SUSY breaking.  In fact, they 
discussed and discarded what
I now believe is the obvious solution of this problem, because 
of problems specific to weakly
coupled string theory\footnote{Namely the fact that superpotentials
are exponentials of exponentials of the canonically normalized dilaton
field.}.  The obvious way to avoid SUSY breaking 
at the scale $M$, is to
insist that the superpotential (\ref{supot}) has a SUSY minimum.  
In fact, the existence of
such minima is  generic
, requiring only the solution of $n$ complex equations for $n$ unknowns.  
However, in general, the superpotential will not vanish at such 
a minimum but instead 
will give rise to a negative cosmological constant. 

It was pointed out in \cite{prevb} that in postinflationary cosmology,
the universe's attempt to access such a SUSY minimum of the effective
potential leads to a very welcome cosmological disaster.  The key point
is that inflation has completely eliminated the spatial curvature terms
from the cosmological equations, so that the Friedmann equation reads
\eqn{fried}{m_P^2 (\dot{a} /a)^2 = G_{AB}\dot{m^A} \dot{m^B} + V}
This does not have static solutions with $m^A$ resting at a minimum of V
with negative value.  What happens instead is that a generic solution of
the cosmological equations\footnote{There are very special solutions in
which the universe is static and the scalar fields oscillate in the
potential with exactly zero energy, and I once thought that these were
relevant to the cosmological constant problem.  However, they are
unstable to small perturbations.} reaches a point where $\dot{a} = 0$
and then begins to recollapse to infinite energy density.  This happens
on a microscopic time scale.  Thus inflationary cosmology eliminates
generic SUSY preserving minima of the effective potential from 
the list of late time attractors of the cosmological equations.

The stable postinflationary attractors of a supersymmetric cosmology 
are points in inflamoduli space
with vanishing superpotential and SUSY order parameters.  These can 
be characterized in terms
of a symmetry.  Namely, any complex R symmetry forces the 
superpotential to vanish, and if there
are no fields of R charge 2 then the SUSY order parameter 
vanishes as well.  The R symmetry must
of course be discrete, since we are discussing M-theory\footnote{This is
an example of the nonexistence of continuous global symmetries.}.  If in 
addition, there do exist
fields of R charge 0, then there will be an entire submanifold on 
which the superpotential
vanishes and SUSY is preserved.  Our future considerations will 
concentrate on this submanifold, 
which from now on we call the 
true moduli space, since it is the oft advertised exception to our
statement that quadrisusic backgrounds had no moduli space.
It is the locus of restoration of a discrete R symmetry with the 
above properties.  We should expect the true moduli space to have more
than one connected component, each characterized by a different R symmetry.

\subsection{Radius stabilization}

Every silver lining has its cloud.  The discussion above treated the
four dimensional Planck scale as a fixed parameter.  In fact, in the
Ho\v rava-Witten scenario, it is determined by the radius of the fifth
dimension, which is one of the moduli.  In fact it is one of the bulk
moduli and might be expected to vary during inflation.

At first glance, the situation appears to be
much worse than that.  In the limit of large
$R$, the Lagrangian of the field $R$ is highly constrained by extended
SUSY.  In this limit the K\"ahler potential of the superfield $T$ which
contains $R$ is fixed to be $ - 3 m_P^2 \ {\rm ln} (T + T^*)$.
In the analysis of \cite{horwitb} \cite{horwitc}, the superpotential was
supposed to be generated only by gaugino condensation on the hidden
brane, separated by a distance $R$ from the brane where the standard
model lives.  This is a function only of a particular linear combination 
$S$, where $S$ is the superfield
which controls the 
coupling of the hidden sector gauge group.  The superpotential
can also depend on the other moduli, {\it e.g.} the complex structure 
moduli of the Calabi-Yau threefold, as well as the vector bundle moduli 
in the hidden gauge group.  Although this superpotential is not
explicitly calculable, it will generically have a supersymmetric point
with $S$ fixed to be small (the hidden gauge theory is strongly coupled)
and the complex structure and hidden sector gauge bundle moduli fixed.
Unless there are points of enhanced discrete R symmetry, as described
above, the superpotential will be nonvanishing at the SUSY point and of 
order $M^3$.  

The fact that the superpotential is of order $M^3$ means that it cannot
really be considered to have originated in some ``low energy effective
theory'', but comes from physics at the fundamental \mth\ scale.  The
possibility of superpotentials generated at short distance was not
appreciated in \cite{horwitb} and \cite{horwitc}, nor as far as I can
tell in any of the papers on \mth\ phenomenology which have appeared
since that time.  I do not see any good argument for omitting such terms
in the low energy Lagrangian.  However, there is a symmetry argument 
that such a superpotential will be of the form $ \sum_{n > 0} w_n (S,C)
e^{-n k m_P T / M^2}$, where $k$ is a number of order one.
The factors in the exponent will be explained below.
Here $C$ is a collection of superfields representing the 
complex structure moduli, as well as vector bundle moduli for the gauge
configurations on each wall\footnote{Here and henceforth we restrict
attention to CY threefolds with only a single K\"ahler modulus and
disregard the possibility of inserting M5 branes in the bulk between the
two walls.}.   The imaginary part of $T$ comes from a pure gauge mode of
the bulk graviphoton, which is chosen to vanish on the hidden sector
wall.  The gauge symmetry becomes a shift symmetry for ${\rm Im} T$.
One may expect this symmetry to be broken by effects involving membrane
instantons stretched between the walls, and by fivebranes (which, in the
walls, are gauge theory instantons).  As a consequence, a discrete
remnant of the shift symmetry remains, and this is what constrains the
superpotential in the manner described above.

Thus, in the large $R$ limit, one expects the K\"ahler potential of the
field $T$ to be given by its asymptotic form, and the superpotential to
be independent of $T$.  As a consequence, even if we assume the
inflamoduli are slowly rolling at some point away from the minimum of
their potential, the dynamics of the universe will be strongly
influenced by the motion of $T$.  It is easy to see that the real part
of $T$ is, in Einstein frame, related to a canonically normalized scalar
field with an exponential potential.  The slope in the exponent is
outside the range in which (power law) inflationary solutions of the
equations of motion exist.  Other sources of friction for $T$ must be
found if inflation is to take place.

There are several obvious sources for such extra friction.  The first is
the imaginary part of $T$, which, in the large $R$ approximation,
behaves like a Goldstone field.  Unfortunately, this means that the
energy density associated with this field, and the extra friction
associated with it,  scales away like $1/ a^6$.   While I have not done
a proper numerical study of this system it seems unlikely that it will
have long periods of inflation for generic initial
conditions\footnote{Remember that unlike the case of the other moduli fields,
there are no unknown parameters in the asymptotic Lagrangian for $T$.}.

Two other sources of extra friction are the excitation of nonconstant
modes of the $T$ field, and Kaluza-Klein particle production.  In
\cite{preva} it was argued that the first of these mechanisms is very
efficient at stopping the chaotic motion on moduli space with no
potential.   As noted above,
there is an instability which converts modular zero mode 
kinetic energy into a gas of nonzero modes within less than a Lyapunoff
time of the chaotic motion on moduli space.  It seems quite plausible
that in the presence of an exponential potential one would then have
inflationary solutions.   Kaluza-Klein particle production is also to be
expected in the presence of a rapidly moving $T$ field, because the real
part of $T$ directly influences the masses of these particles.  

Obviously, more work is needed to see whether these mechanisms can
really salvage the inflationary scenario of the previous section.
Even if they do, one mystery still remains.  Although some combination
of these effects can explain why $T$ is slowly varying during inflation,
there is no explanation of why it is close to its vacuum
value.    Since the four dimensional Planck mass (and through it our
successful prediction of the magnitude of energy density fluctuations)
depends exponentially on the canonically normalized field constructed
{}from the real part of $T$, it is extremely important to explain this
coincidence.

Another possibility for rescuing inflation comes from the recognition
that the radial modulus has a Dine Seiberg instability.  That is to say,
although we would like to be doing a systematic asymptotic expansion 
in $R$, we know that we will never find a stable minimum for $T$ in this
approximation.  Thus we should admit that near the vacuum value for $T$, 
the large radius expansion for (at least) the effective potential of
this field has broken down.  Let us recall that we defined $T$ in terms
of the deviation of the radius from its vacuum value \cite{horwitc}.
Thus, $RM \sim (m_P T / M^2)$.  On physical grounds, we expect
corrections to the asymptotic form of the Lagrangian to be functions of
$RM$.  In the case of the $T$ dependence of the
superpotential discussed above, this guess can be verified by
analytic continuation from the region of weakly coupled string theory
\cite{horwitc}.  

The potential for $T$ during inflation has two terms.  The first, coming
{}from the F terms of the other moduli was discussed above, and is all
that exists in the extreme asymptotic limit.  In that limit, it gives an
exponential potential with slope of order $1/m_P$ for the canonically
normalized field $\sim m_P {\rm ln\ Re\ }(T/m_P)$. 
The second term has the
form:
\eqn{tpot}{V \sim e^{K/m_P^2}
[K^{TT^*} |K_T /m_P |^2 -3]|W/m_P|^2,}
where $K$ is the K\"ahler potential of $T$.  
The implication of the previous paragraph is that there is a region of
$T/m_P$ of order one, where $K$ is very different from its asymptotic
form, and varying rather rapidly as a function of this variable.
Now consider initial conditions where $RM$ starts out close to one
and growing.  The $T$ field will then have to cross a regime in which
the rapidly varying piece of the potential is significant before it can
access the asymptotic regime.  If the other moduli are slowly rolling,
it is clear that it will instead be rapidly 
driven very close to the minimum of its potential.  Unfortunately, I
have no argument that this is the same as its VEV.

Indeed, we will see in the next section that the minimum
of the low energy potential for $T$ is the same as that of (\ref{tpot}).  
There is no obvious reason to expect the first term of the potential
(proportional to the F terms of other chiral fields) to be negligible
compared to (\ref{tpot}).  Thus, although this mechanism saves inflation,
it is not clear that it preserves our explanation of the size of primordial 
fluctuations.

Our discussion of the end of inflation is also modified.  Once the
contribution of $T$ is taken into account, the cosmological constant 
(at the end of inflation, but neglecting low energy gauge dynamics) is
given by the value of (\ref{tpot}) at its minimum, with the other moduli
set at SUSY preserving values.  Points with nonvanishing superpotential
will now have SUSY spontaneously broken by the F term of the $T$ field.
If we insist that the low energy cosmological constant vanishes exactly
(in the scenario with discrete R symmetry broken by low energy dynamics),
then these points will also have vanishing cosmological constant and
will be attractors of the postinflationary cosmological equations.  
This is unfortunate, because these points have gravitino masses of order
$M^3/m_P^2$ and are ruled out by phenomenonology.  It would have been
pleasant to find that they were also disfavored by cosmological evolution.
In the next section we will see that we can still 
recover acceptable phenomenology at points of enhanced R
symmetry (broken only by low energy dynamics). 

There {\it is} a (weakly anthropic) way of understanding why 
points in moduli space with R symmetry broken at high energies
could be ruled out by cosmology as well as
phenomenology, if we accept that there is a nonvanishing cosmological
constant in the world we observe.  Then the ratio of cosmological
constants between the R asymmetric worlds and our own is 
$\sim (M/\mu)^6$, where $\mu$ is the scale of low energy R symmetry
breaking. If one insists on a low energy SUSY breaking scale of order a TeV 
$\mu$ is fixed at about $10^{13}$~GeV (see below).
This gives the R asymmetric worlds a De Sitter horizon size of about a
light year.  There is certainly no galaxy formation in such a universe,
and it does not take a degree in exobiology to conclude that no life is
possible there.  There is no plausible initial (post primary inflation) matter
distribution which leads to any appreciable late time matter inside a 
horizon volume, unless it is collapsed into black holes.

Finally, one should note that at small values of $T$ (values of $RM$ of
order 1) there might be a SUSY minimum of the potential for $T$.
This regime is hard to discuss because effective field theory does not
apply to it and the notion of effective potentials, approximate moduli,
and classical spacetime are all suspect.  However, even if one assumed
that such a minimum existed, one would find that one could not access it
after inflation and it would be irrelevant to macroscopic physics.

I have divided the discussion of inflation on moduli space into two
parts, initially ignoring the problem of the radial modulus, because 
I suspect that it may be possible to find other scenarios in which this
problem is completely absent.  I will make a similar division of the
discussion of SUSY breaking below.

\subsection{SUSY breaking}

Before proceeding to the discussion of SUSY breaking on the true 
moduli space, we should
introduce the final characters in our story, the boundary or 
brane moduli.   In Calabi-Yau compactification of weakly coupled string
theory, there are moduli which correspond to the parameters of the $E_8
\times E_8$ gauge field configuration on the manifold (these are
called vector bundle moduli in the string compactification literature).  
In a brane scenario these
moduli should be thought of as living on the branes where the gauge
fields live.  In the strong coupling regime, these fields will have a
superpotential of the form $M^3 W(b/M)$ and it is not clear that they
should be called moduli at all.  Some of them may be invariant under the
discrete complex R symmetry, and thus belong to the true moduli space.
In perturbative string theory, some vector bundle moduli have components
$\theta_b $ which couple to gauge fields like axions : 
$\theta_b F \tilde{F}$.  The decay constants of these axions are of order 
$M$ because, since they live on the brane, no other scale can enter their
kinetic terms.

In our later considerations, we will have need of a field with a decay
constant of order $M$ and a very small potential energy.  The vector
bundle moduli on the standard model wall have the first of these properties. 
In perturbative string theory these fields have Peccei-Quinn symmetries
which are broken only by world sheet instantons.  It is then plausible
that in the Ho\v rava-Witten regime the dominant breaking of these
symmetries
comes from nonperturbative QCD.
The potential energy of one of the gauge bundle axions 
would be much smaller than any
fundamental scale, and would have the form $\Lambda_{QCD}^4 u(a/M)$.  
We will consider the possibility that there are
other moduli of this type, with a variety of scales replacing $\Lambda_{QCD}$.

In addition to these moduli fields, any brane scenario will 
contain a variety of gauge fields and
matter fields in nontrivial representations of the gauge group.  
The moduli will interact with
these fields via the moduli dependence of bare gauge and yukawa 
coupling parameters in the effective theory as well as thru a variety of 
irrelevant operators.  If the gauge couplings are asymptotically 
free and do not run to
infrared fixed points at low energy, this description of the
physics only makes sense if the bare gauge couplings are sufficiently 
small that the scale 
at which the effective 
coupling becomes large is substantially below the scale $M$.  
Otherwise it is not
consistent to include the gauge degrees of freedom in the low energy 
effective theory.
The weakness of bare couplings in these scenarios is not evident a 
priori, as it would be 
in a purely perturbative approach.  The underlying physics is assumed to 
be strongly coupled.
Witten \cite{horwitb} 
has shown how the small unified coupling of the standard model 
can be explained in terms
of a product of a large number of factors of order one in a geometry 
of large dimensions.
We will assume that similar numerical factors explain the strength of
the gauge interactions that lead to SUSY breaking.  

The main role of the gauge interactions is not to break SUSY, but 
rather the discrete R
symmetry.  If we fix the moduli and treat the gauge theory as a 
flat space quantum field
theory, then SUSY remains unbroken even though a nonperturbative 
superpotential is generated.
The scale of this superpotential is determined via a standard 
renormalization group
analysis in terms of the bare gauge coupling function 
$f(\phi /m_P, \chi / M)$, where we 
have indicated dependence on both bulk and boundary moduli.  
For simplicity we assume that
$f$ is a large constant $f_0$ plus a smaller, moduli dependent, 
term.  The conclusions are 
not affected by this assumption.  The scale $\mu$ of the 
nonperturbative superpotential
is then determined by $f_0$.  It takes the form
\eqn{supottwo}{W_1 = \mu^3 w_1 (\phi /m_P, \chi /M)}
We have eliminated all (composite) superfields related to the 
gauge interactions from this
expression by solving their F and D flatness conditions for fixed 
values of the moduli.
The possibility of doing this is equivalent to the statement that 
the gauge theory does not itself break SUSY.
We assume that $W_1$ does not vanish at any minimum of the effective potential.
This is the statement of spontaneous R symmetry breaking.  
As a consequence, SUSY minima of the potential have negative 
cosmological constant of
order at least $\mu^6 / m_P^2$ and are not attractors of the 
cosmological equations.
Thus, cosmologically, R symmetry breaking forces the moduli to choose 
a minimum with 
spontaneously broken SUSY\footnote{The tunneling amplitudes of such 
nonsupersymmetric
vacua into supersymmetric AdS vacua are incredibly tiny and might be 
identically zero, 
as discussed in \cite{prevb}.}.  

Phenomenology puts an upper bound on the value of $\mu$  because it
contributes directly to squark 
masses.  The nonvanishing
F terms are of order ${\mu^3 \over m_P}$.  A standard argument shows 
that squark masses
will be of order $ {\mu^3 \over m_P^2}$, about the same as the gravitino.  
Assuming this is about a TeV we find $\mu \sim 10^{13}$~GeV.   An
attractive feature of this scenario is that the positive and 
negative terms
in the SUGRA potential are naturally of the same order of magnitude.  
Although we have no
real understanding of why the cosmological constant is so small, this 
fact of nature is an
indication of a relation between the scales of R symmetry breaking 
and of SUSY breaking.
In models in which the SUSY breaking F term originates as a bulk modulus the
correct order of magnitude relation between these scales arises 
automatically.  

As we now recall, a deficiency of this scenario for SUSY breaking is
that it leads to the
cosmological moduli problem.  The scalar fields in the bulk moduli 
multiplets acquire masses from 
the SUSY violating potential of order $m_M \sim \mu^3 / m_P^2$ which is 
the same order of magnitude
as the gravitino and squark masses, {\it i.e.} a TeV.  They have only 
nonrenormalizable
couplings to ordinary matter, scaled by $m_P$.  Thus, their nominal 
reheat temperature, 
$\sqrt{m_M^3 /m_P}$ is of order $\sim 3 \times 10^{-2}$~MeV, 
and the universe is matter
dominated at the time that nucleosynthesis is supposed to be taking
place.  The thermal 
inflation scenario \cite{therminf} can solve this problem, 
and we will now review another solution
\cite{bdaxion}.

Suppose that
the coefficient in the order of magnitude relation between the moduli 
mass and the fundamental
parameters is $m_M = 5 \times \mu^3 / m_P^2$, while the squark mass is 
actually $m_{\tilde{q}}
= \mu^3 /4m_P^2 = 1$~TeV.  Then the reheat temperature for the bulk
moduli is multiplied by
a factor of $20^{3/2} \sim 10^2$ and is just above $1$~MeV.  
Thus, an innocent looking insertion of factors of order one can cause
the moduli to decay just in time to light the furnace in which the
primordial elements are forged.

One still has to account for baryogenesis.  Adopting a mechanism 
suggested long ago by Holman, Ramond and Ross
\cite{hrr} we aver that 
this can come from the decay of the moduli themselves.  
All of their interactions
are of order the fundamental scale of M-theory, so there is no 
reason for them to preserve
accidental symmetries like baryon and lepton number.  It is quite 
reasonable that they also
violate CP, though the status of CP in M-theory is somewhat more 
obscure.  The decay itself
is an out of equilibrium process, so all of the Sakharov criteria 
for baryogenesis are
fulfilled.  However, we must also take note of the theorem of 
Weinberg \cite{steve}, according
to which baryon number violating terms in the Hamiltonian must act 
twice in order to generate
an asymmetry.  In the decay of moduli, the first action of the 
Hamiltonian comes at no cost 
in amplitude, because the modulus must decay somehow and there is no 
reason for its baryon 
number
violating decays to be significantly smaller than those which conserve 
baryon number.
However the second baryon number violating interaction should not be 
highly suppressed if
we want to generate a reasonable baryon asymmetry.  Indeed, a $10$~TeV,
gravitationally coupled,  particle which produces 
a baryon asymmetry of order one in its decay, also produces of 
order $(10\,{\rm TeV}/ 3\,{\rm MeV})$
or $\sim 3 \times 10^6$ photons.  Thus a large suppression of the 
average baryon number per
decay would give too small a baryon asymmetry.  A way out of this 
difficulty is to admit
renormalizable baryon number violating operators in the supersymmetric 
standard model.
Discrete symmetries such as a $Z_2$ lepton parity \cite{irnir} can 
adequately suppress
all unobserved baryon and lepton number violating processes in the 
laboratory, while allowing
such operators with coefficients as large as $5 \times 10^{-3}$.
This might be large enough to produce the observed baryon asymmetry.
  
An unfortunate casualty of this mechanism is the lightest SUSY particle.  
The LSP is no longer
stable in the scenario described above and we have to look 
elsewhere for a dark matter candidate.
However, there are 
natural candidates for
dark matter.  Imagine a boundary modulus whose potential energy is 
substantially smaller
than the the estimate $\mu^3 / M^2$ coming from (\ref{supottwo}). We 
will call this the dark
modulus, because it will be our dark matter candidate. It has a 
potential of the form
$U = \Lambda^4 u(D/M)$. (In \cite{bdaxion}, where
this scenario was first proposed, the candidate was a QCD axion field.  
This model works, but the mechanism
is much more general and does not require energy densities as small as 
those of the axion.).

Now, briefly review cosmic history.  First we have inflation 
generated by bulk moduli fields
which are not on the true moduli space (which we have called
inflamoduli).
This period ends after of order $100$ $e$-foldings,
and the universe is heated by inflamoduli decay to a temperature of 
order $10^9$~GeV.  
The primordial plasma quickly redshifts away.  Furthermore, as soon as 
the inflamoduli
potential energy density falls to $\mu^6 / m_P^2$, the universe 
becomes dominated by the
coherent oscillations of the true bulk moduli.  
The dark modulus remains frozen at some generic point on its potential 
until the Hubble
parameter falls to the mass scale of this field.  At this point 
the energy density of the
universe is of order $\rho \sim m_P^2 \Lambda^4 / M^2$ which is of 
order $(m_P/M)^2 \sim 
10^{4}$ times larger than the energy density of the dark modulus.  
The important point now
is that this ratio is preserved by further cosmic evolution until the 
true bulk moduli decay.
After that time, the dark energy density grows linearly with 
the inverse temperature relative
to radiation, and
matter radiation equality occurs at $10^{-4}$~MeV.  This is close 
enough to the true value
for the observable universe that the factors of order one which 
we have neglected throughout 
might account for the difference.   $\Lambda$ must satisfy two 
constraints in order
for this scenario
to work:  the dark moduli must remain frozen until the true bulk 
moduli begin to oscillate, 
and 
the dark modulus must have a lifetime at least as long as the age of 
the universe.
The second constraint is by far the stronger, and leads to 
$\Lambda < 3 \times 10^6$~GeV.
Axions satisfy this constraint by a large margin.  Note that this 
scenario completely
removes the conventional cosmological constraint on the axion 
decay constant.  Axions will
be very weakly coupled and will escape all of the usual schemes 
for detecting them.

Another possible mechanism for baryogenesis in this scenario is that of
Affleck and Dine \cite{ad}\footnote{This was suggested to me by a student
at the school.  I thank M.Dine for detailed discussions of it and for
pointing out the reference below.}.  Indeed the authors of \cite{heavmod}
have investigated a scenario with a 10~TeV modulus and Affleck Dine
baryogenesis and found that it can account for all cosmological data.
In this scenario the dark matter can either be an LSP, or if we have
strong R parity violating interactions,  the dark modulus 
(or a combination).   

All in all, this seems to be the simplest solution of the cosmological
moduli problem, and has the added virtue of allowing an invisible axion
solution of the strong CP problem.  I am also fond of the way in which
the version of this scenario with a dark modulus
predicts the correct (within an order of magnitude) temperature for
matter radiation equality in terms of fundamental parameters.

For completeness, we should also discuss the possibility that SUSY breaking
itself is caused by gauge interactions which are weakly coupled at the 
fundamental scale.  This is required if we assume, with Dine 
\cite{mike} \cite{evadine} \cite{islands},
that moduli are fixed at some enhanced symmetry point.  Scenarios of
this sort are attractive because they allow us to use the idea of gauge
mediation \cite{fischlerdinenelson} to solve the SUSY flavor problem.
Gauge interactions generate superpotentials of the form
$\mu_1^3 {w_g}_1 (C_1/m_1) + \mu_2^3 {w_g}_2 (C_2/m_2)    $, where the
$C's$ are composite superfields and the $m_i$ the nonperturbative low energy
scales generated by asymptotic freedom.  Here, in order to cancel the
cosmological constant, we must introduce an R breaking gauge theory 
with scale $(m_1)$, 
which preserves SUSY and a SUSY breaking gauge theory, with scale
related by $m_1^6 = m_P^2 m_2^4$.   This is the price one must pay for
giving up the idea that true bulk moduli are the instigators of SUSY
breaking.  The ratio of scales between SUSY and R breaking no longer
comes out naturally, but must be put in by hand.  In compensation
there is no cosmological moduli problem in this picture, since all
moduli are assumed to be 
frozen by the initial superpotential. 

\subsection{The effects of a dynamical radius}

We now have to include the dynamics of the radial modulus $T$.
The R symmetry
violating superpotential has an expansion\footnote{It is important that,
as a consequence of our assumption of an R symmetry under which $T$ is
neutral, all terms in this expansion are proportional to the R breaking
scale $\mu^3$. This means that we cannot invoke mechanisms like that of
\cite{Raman} to explain the stabilization of the radius.}
\eqn{rvsupot}{W = \sum_{n=0}^{\infty} \mu^3 W_n (m/m_P) e^{-n T/m_P}}
At large radius the exponential terms are negligible.  We then have
no scale SUSY breaking even if all other bulk moduli have SUSY
minima\footnote{However, once we take into account the SUSY violating
potential coming from the $F$ term of $T$, there is no reason to assume
that the other fields sit at their SUSY minima.  The minimum of the
potential might be achieved with F terms for all the fields.}.
One can then hope, as in \cite{horwitc}, that the radius is stabilized by
higher order terms in the K\"ahler potential.  This would give a SUSY
breaking scale close to $\mu^3 /m_P^2 $.  The resulting scenario is similar
to that of the previous section.

There is a much more substantial difference in the case where (what we
previously called) the true moduli space is a point.  $T$ still plays
the role of a true modulus, and we again get no-scale SUSY breaking when
the low energy theory violates R symmetry without breaking SUSY.  
We can, if we wish, also add a low energy SUSY breaking sector, but to
leading order in $R$ this leads to a large positive cosmological
constant\footnote{It should be noted that a large positive cosmological
constant is not a disaster only for our ability to ``fit the data''.
The size of the event horizon for a De Sitter space with energy density
of scale $1$~MeV is about a light second in linear size, and for the
scale of SUSY breaking it is smaller by a factor of $10^{12}$.  
In a theory with
multiple late time attractors it is not hard to explain why we are not
there to observe such a universe.}.  
This is true no matter what we choose for the relative scales of low
energy SUSY breaking and R symmetry breaking (as long as we try to
be consistent with the lower bound on superpartner masses).  
Thus, once the radius is allowed to be dynamical there do not seem to be
consistent scenarios with gauge mediated SUSY breaking.

\subsection{Generalizing Ho\v rava-Witten}

As we have noted, the 
moduli space of 11 dimensional SUGRA compactifications which preserve 
${\cal N} =1$ SUSY in four Minkowski dimensions splits into three components.
These are Joyce sevenfolds, F theory limits of compactification on Calabi-Yau
fourfolds, and Heterotic limits of compactification on $K3\times CY_3$.  
These may be continuously connected when short distance physics 
is properly taken 
into account. In addition,
there may be many branches of moduli space which join onto these 
through generalized
extremal transitions.  The moduli space is thus highly complex.

The cosmological arguments of these lectures indicate that the 
phenomenologically relevant
compactifications may belong to a highly constrained submanifold 
of this complicated space.
Namely, they should preserve eight supercharges in the bulk.  
The breaking to ${\cal N} =1$ 
should occur only on branes.  SUGRA compactifications preserving eight SUSYs are much
more constrained.  The holonomy must be contained in $SU(3)$ 
which implies that the
manifold is the product of a Calabi-Yau threefold times a torus, 
modded out by a discrete
group $\Gamma$.  In order to obtain a smooth manifold with eight SUSYs, 
$\Gamma $ should act freely and the holonomy around the new cycles 
created by $\Gamma$
identification should be in $SU(3)$.  Clearly, a way to obtain 
Ho\v rava-Witten like
scenarios is to allow fixed manifolds of $\Gamma$, on which 
an additional SUSY is broken.
The original scenario of Ho\v rava and Witten was a 
$CY_3 \times S^1$ compactification in
which $\Gamma$ is a $Z_2$ reflection on the $S^1$.  The fixed planes 
carry $E_8$ gauge
groups, and one must also choose an appropriate gauge bundle.  
A further generalization
allows five branes wrapped on two cycles of $CY_3$ to live between the planes.

It seems likely that more complicated choices of $\Gamma$ might 
lead to a wider class
of scenarios.  The problem of classifying scenarios of this type 
seems quite manageable\footnote{Preliminary 
results on the classification problem have 
been obtained by
L.Motl.}.  The moduli space of 
compactifications of M-theory on $CY_3$ times a torus has a reasonably
complicated structure, replete with extremal transitions.  Nonetheless, 
it is considerably
simpler than the fourfold or Joyce manifold problem, and we know much 
more about its
structure.  Thus, if cosmology really points us in the direction of 
generalized Ho\v rava-Witten
compactifications, we have made real progress in the search for the 
true vacuum of
M-theory.

\subsection{Conclusions}

Witten's explanation of the discrepancy between the Planck and 
unification scales
in the context of Ho\v rava-Witten compactifications,
poses a challenge for inflationary cosmology and particularly for 
the notion that 
moduli are inflatons.  In fact, the enhanced bulk SUSY of these 
compactifications gives
us a clean definition of modular inflatons.  The scenario then makes an 
order of
magnitude prediction of the amplitude of primordial density fluctuations 
in terms
of the unification scale.  The major problem with this inflationary
scenario comes from stabilization of the radius of the Ho\v rava-Witten
orbifold.  In leading order in the large radius approximation, radial
dynamics appears
to destroy inflation.  We pointed out several sources of friction for
the radius field, which could restore inflationary solutions, but there
is more work to be done here and a mystery remains.  Assuming the radion
is slowly rolling during inflation, why is it near its vacuum value?

An alternative, which seems more compelling, is to recognize that the
Dine Seiberg problem for the radial field probably requires us to
contemplate the breakdown of the large radius expansion for its K\"ahler
potential near the true VEV of this field.  We argued that this meant
that the K\"ahler potential was rapidly varying (as a function of $T/m_P$)
near
the low energy VEV and that this implied that the radius would not be an
inflaton but instead would rapidly be driven to the minimum of its
potential during inflation.  It is not clear whether the inflationary
minimum is close enough to the VEV to salvage our explanation of the
size of density fluctuations.  This depends on properties of the K\"ahler
potential which are, at the moment, incalculable.  

In the context of this large class of inflationary scenaria,
arguments first discussed in \cite{prevb} then focus 
attention on the
true moduli space of M-theory, a locus of enhanced discrete R symmetry.  
Such a space
almost certainly exists \cite{bdmod}.  It is the attractor of postinflationary
cosmological evolution.  The further evolution of the universe then
depends on 
whether this space contains bulk moduli.  In the attractive scenario 
in which it does,
the initial Hot Big Bang generated by inflation, is soon dominated by 
the energy
density stored in coherent oscillations of true bulk moduli.  
By making optimistic
but plausible assumptions about coefficients of order one in order of 
magnitude
estimates, one obtains a reheat temperature above that required 
by nucleosynthesis.
The decay of true bulk moduli, rather than that of the inflaton, 
generates the Hot Big
Bang of classical cosmology.  The baryon asymmetry might also be 
generated in these
decays, and this is possible if the SUSY standard model 
contains renormalizable baryon
number violating interactions (compatible with laboratory 
tests of baryon and lepton
number conservation).  As a consequence of this, there is no 
LSP dark matter candidate.
Instead, boundary moduli with a suppressed potential energy act as a natural
source of dark matter.  Indeed, the ratio between the Planck 
and unification scales
appears again in this scenario, this time in explaining the temperature at 
which matter and radiation make equal contributions to the energy 
density of the
Universe.  This estimate comes out an order of magnitude too high, 
but given the
crudity of the calculation it seems quite plausible that this mechanism 
could be
compatible with observation.  The ``dark modulus'' which 
appears in this
scenario could be a QCD axion with decay constant of order the
unification scale.
Our unconventional origin for the Hot Big Bang completely removes the 
cosmological upper
bound on this decay constant.  Such a particle would be undetectable in 
presently
proposed axion searches.

An alternative is to postulate the Affleck-Dine mechanism as the source
of the baryon asymmetry in this late decaying modulus scheme.  Dark
matter could then be an LSP, a unification scale QCD axion, or some
combination of the two. 

If a cosmology like that outlined here turns out to be correct, one 
might be tempted
to revise Einstein's famous estimate of the moral qualities of a 
hypothetical Creator.
The current standard model of cosmology was constructed in the sixties.  
Since then there
has been much speculation about cosmology at times earlier than that at 
which the
primordial elements were synthesized.  Most of it has been based on an 
eminently reasonable
extrapolation of the
Hot Big Bang to energy densities orders of magnitude higher.  
If the present scenario
is correct, no such extrapolation is possible, and the conditions 
in the Universe in
the first fraction of the First Three Minutes were considerably 
different from those 
at any subsequent time.  There was a prior Big Bang after 
inflation, whose remnants may be 
forever hidden from us.  The dark matter which dominates our 
universe is so weakly coupled
to ordinary matter that its detection is far beyond the reach of 
currently planned
experiments.   The QCD and electroweak phase transitions 
never occurred.

The only dramatic prediction of this scenario for currently 
planned experiments is
the occurrence of renormalizable baryon number violation 
in the low energy SUSY
world\footnote{In the version of the scenario with Affleck Dine
baryogenesis, even this signature might be absent.}.  
The details of the baryogenesis scenario envisaged here 
should be worked out
more carefully, and combined with laboratory constraints, 
to nail down precisely
which kind of operators are allowed.  The scenario is thus easily 
falsifiable, but even
the discovery of renormalizable baryon number violating 
interactions among SUSY particles 
will not be a confirmation of our cosmology.  Similarly, 
any evidence for the existence
of more or less conventional WIMP dark matter will be a 
strong indication that the present
speculations are incorrect, but the failure to discover 
WIMPS will not prove that they
are correct.

Instead one will have to rely on the slow accumulation of 
evidence against alternatives:
ruling out vanishing up quark mass and spontaneous CP violation 
as solutions to the strong
CP problem, the failure of conventional axion and 
WIMP searches, the discovery of
renormalizable B violation.  These will be steps on 
the road to proving that this cosmology
is correct, but the end of that road is not in sight.

We have travelled a long road, from the exotic reaches of \mth\ to what
I hope have been glimpses of more practical applications of modular
physics to cosmology.  I hope I have convinced you that the moduli of
\mth\ are likely to play a crucial role in any inflationary cosmological
model and that many of the phenomenological and fundamental problems of
\mth\ are likely to be resolved in a cosmological context.  Perhaps the
somewhat unorthodox cosmological scenaria presented here will also prove
to be more than just a theorist's toys, and will play some role in the
future of cosmology.

\acknowledgments

I would like to thank Sean Carroll, Michael Dine, Andrei Linde, Markus
Luty, Raman Sundrum, Paul Steinhardt, Neal Turok, Willy Fischler,
and the students at the Les Houches school for
insightful comments on and questions about the material presented here.
I would like to thank the organizers, Pierre Binetruy, Richard
Schaeffer, and Joe Silk for giving me the opportunity to present these
ideas and for organizing a terrific school.  This work was supported in
part by the DOE under grant DE-FG02-96ER40559.


\end{document}